\documentstyle[preprint,aps,epsfig,tighten,floats]{revtex}

\newcommand{\ba}{\begin{array}}
\newcommand{\ea}{\end{array}}
\newcommand{\be}{\begin{equation}}
\newcommand{\ee}{\end{equation}}
\newcommand{\bea}{\begin{eqnarray}}
\newcommand{\eea}{\end{eqnarray}}
\newcommand{\beq}{\begin{equation}}
\newcommand{\eeq}{\end{equation}}

\def\Im{\mbox{Im}}

\def\GSW_sign{}

\def\bra{\langle}
\def\ket{\rangle}
\def\a{\alpha}
\def\b{\beta}
\def\g{\gamma}

\def\e{\epsilon}

\def\l{\lambda}
\def\m{\mu}
\def\n{\nu}
\def\G{\Gamma}

\def\to{\rightarrow}

\renewcommand\Im{\mbox{Im}}
\renewcommand\Re{\mbox{Re}}
\renewcommand\a{\alpha}

\newcommand{\cO}{{\cal O}}

\newcommand{\no}{\nonumber}

\newcommand{\mbMS}{\overline{m}_b}

\def\npb#1#2#3{    {\it Nucl. Phys. }{\bf B\,#1} (19#2) #3}

\def\prd#1#2#3{    {\it Phys. Rev. }{\bf D\,#1} (19#2) #3}

\def\zpc#1#2#3{    {\it Zeit. f\"ur Physik }{\bf C\,#1} (19#2) #3}

\def\npb#1#2#3{    {\it Nucl. Phys. }{\bf B #1} (#2) #3}

\def\prd#1#2#3{    {\it Phys. Rev. }{\bf D #1} (#2) #3}

\def\zpc#1#2#3{    {\it Z. Phys. }{\bf C #1} (#2) #3}

\def\bra{\langle}
\def\ket{\rangle}
\def\a{\alpha}
\def\b{\beta}
\def\g{\gamma}

\def\e{\epsilon}

\def\l{\lambda}
\def\m{\mu}
\def\n{\nu}
\def\G{\Gamma}

\def\mub{\mu_b}

\def\to{\rightarrow}

\def\o{\over}
\def\b{\begin{equation}}
\def\e{\end{equation}}

\begin{document}

\preprint{CERN-TH/2002-264, SLAC-PUB-9604}
\vspace{4cm}

\title{\large \bf 
Present Status of Inclusive Rare B Decays}


\author{Tobias Hurth\thanks{Heisenberg Fellow}}

\address{CERN, Theory Division\\ CH-1211 Geneva 23, Switzerland\\
tobias.hurth@cern.ch\\ and \\
SLAC, Stanford University\\
Stanford, CA 94309, USA\\
hurth@slac.stanford.edu}

\maketitle

\begin{abstract}
We give a status report on inclusive rare $B$ decays,
highlighting recent developments and open problems. We 
focus on the decay modes $B \rightarrow X_{s,d} \gamma$,
$ B \rightarrow X_s  \ell^+\ell^-$ and $B \rightarrow X_s \nu \bar \nu$
and on their role in the  search for new physics.
 
Most of the inclusive rare $B$ decays are important modes of 
flavour physics due to  the small  hadronic uncertainties. They 
can be regarded as laboratories to search for new physics.  

We collect the experimental data already available from 
CLEO  and the $B$ factories BABAR and BELLE. We review  
the NLL and NNLL QCD calculations of the inclusive decay rates
that were recently completed, and discuss future prospects,
especially           the issue of the charm mass scheme ambiguity. 
Finally, we analyse the phenomenological impact of these decay 
modes, in particular on the CKM phenomenology and 
on the indirect search for supersymmetry.

We also briefly discuss direct CP violation in inclusive rare $B$ decays,
as well as  
the rare kaon decays $K^+\rightarrow \pi^+\nu\bar{\nu}$ and 
\mbox{$K_L \rightarrow \pi^0 \nu \bar{\nu}$}, which offer 
complementary theoretically clean information.
\end{abstract}

\vfill
\begin{center}
\bf Invited Contribution to Reviews of Modern Physics\\
\end{center}
\vfill


\pagebreak
\tableofcontents
\pagebreak


\section{Introduction}
\label{intro}
\setcounter{equation}{0}

The precise test of the flavour structure and the mechanism of CP
violation of the standard model (SM) is at the centre of 
today's research in high-energy physics. 
By definition, flavour physics deals with that  part of the 
SM that  distinguishes between the three  generations of fundamental fermions.
It is still a mystery why there are exactly three generations.
Also the origin of the fermion masses  and their mixing is unknown; 
in particular, the SM does not explain the hierarchical  pattern of these 
parameters. Flavour physics can be regarded as the least tested part of the 
SM.  This is reflected in the rather large error bars of 
several flavour parameters such as the mixing parameters 
at the $20 \%$ level \cite{PDG2002}.

\medskip

However, the experimental situation concerning flavour physics
is drastically changing.  
Several $B$ physics experiments are successfully
running at  the moment and,  in the upcoming years, 
new facilities will start
to explore  $B$ physics with increasing 
sensitivity and within various    
experimental settings:
apart from the CLEO experiment (Cornell, USA),
located at the Cornell Electron--Positron Storage Ring (CESR) \cite{cleoexpo}, 
two $B$ factories, 
 operating at the  $\Upsilon (4S)$ resonance in an asymmetric mode, are 
successfully obtaining data:
the BABAR experiment at SLAC 
(Stanford, USA) 
\cite{babarexpo}  
and the BELLE experiment 
at KEK (Tsukuba, Japan) 
\cite{belleexpo}.
Besides the hadronic 
$B$ physics program at FERMILAB (Batavia, USA) 
\cite{fermilabexpo} 
there are $B$ physics experiments  planned at the
hadronic colliders. Within the LHC project at
CERN in Geneva \cite{lhcbexpo} all three experiments 
have strong $B$ physics programs. Also at FERMILAB  
an independent $B$ physics experiment, $B$TeV,  is planned 
\cite{btevexpo}. 
The main motivation for a $B$ physics program at hadron colliders 
is the huge $b$ quark production cross section with  
respect to  the one at 
$e^+ e^-$ machines.

\medskip

While the time of electroweak precision physics focusing on the 
{\it gauge} sector of the SM,  draws to 
a close with the completion of the 
LEP experiments at CERN and the SLC experiment  in Stanford, the era of 
precision flavour physics, focusing on the {\it scalar} sector of the SM, 
has just begun  with the start of the $B$ factories.

\medskip 

The $B$ system represents an ideal framework for the study of
flavour physics. Since the $b$ quark mass is much larger than the typical
scale of the strong interaction $\Lambda_{QCD}$, long-distance 
strong interactions are generally  
less important and are under better control than in kaon physics,
thanks to the expansion in that heavy mass.
Moreover, GIM suppression is not active in loop diagrams
involving the top quarks, which leads to experimentally accessible 
rare decays and to large CP violating effects within
$B$ physics. 
Thus, the CP violation  in the $B$ system 
represents  an important independent test of the SM description of 
CP violation (see \cite{BELLECP,BABARCP,Durham}). 
$B$ meson decays also allow for 
a rich CKM phenomenology and stringent tests of the 
unitarity.

\medskip 

\begin{figure}
\begin{center}
\epsfig{figure=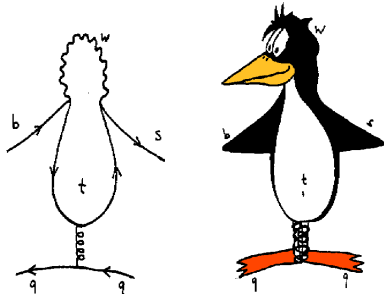,width=5cm}
\caption{Penguin decays of $B$ mesons.}
\label{penguin}
\end{center}
\end{figure}

The so-called rare decays  are of particular interest.
These processes represent flavour changing neutral currents (FCNCs)
and occur in the SM  only at the loop  level.
They also run under the name of `penguin decays'  
(see fig. \ref{penguin} \cite{lenz}), first introduced
in \cite{Ellis} as a result of a bet.

\medskip

In contrast to the exclusive rare $B$ decay modes, the inclusive ones 
are theoretically clean  observables, because no specific model is needed 
to describe the  hadronic final states. For instance 
the decay width $\Gamma (B \to X_s \gamma)$ 
is well approximated by the partonic decay rate
$\Gamma (b \to X_s^{parton} \gamma)$, which can be
analysed within the framework of renormalization-group-improved 
perturbation theory. Non-perturbative contributions play
only a subdominant role and can be calculated in 
a model-independent way by using the heavy-quark expansion.

\medskip

The role of inclusive rare $B$ decays is  twofold:  
on the one hand they are relevant to the determination of CKM matrix
elements.
On the other hand they are particularly sensitive
to new physics beyond the SM, since additional contributions to the 
decay rate, in which SM particles are replaced by new particles, such as the
supersymmetric  charginos or gluinos, are not suppressed by additional
factors $\alpha/(4\pi)$ relative to the SM contribution. This makes 
it possible to observe new physics indirectly -  a strategy 
\mbox{complementary} to the direct production of new (supersymmetric) 
particles. The latter production is reserved for the planned 
hadronic machines such as the LHC at CERN, while the indirect search of the 
$B$ factories  already implies significant restrictions for the parameter 
space of supersymmetric models and, thus, lead to important 
clues for the direct search of supersymmetric particles.

\medskip

It is even possible that these rare processes lead to the first 
evidence of new physics outside the neutrino sector by a significant 
deviation from the SM prediction,
for example in the observables concerning direct 
CP  violation within the $\Delta F =1$ sector;  such a measurement 
would definitely not be in conflict with the recent measurements of 
 CP violation in the $B_d$ system, which confirms 
the SM predictions at the $10\%$ level \cite{BELLECP,BABARCP}. 
But also in the long run,
after new physics has already been discovered, inclusive rare $B$  
decays will
play an important role in analysing in greater detail the 
underlying new dynamics.

\medskip 

The expression {\it inclusive rare $B$ decay} is loosely defined 
and calls for a  
precise definition. Within the present paper it is understood 
as 
a  FCNC process  $ B \rightarrow  X\,\, Y$, 
where $B$ denotes a  $B^\pm$,\, $B_d$ or $ B_s$ meson.  
$X$ is an 
inclusive hadronic  state containing no charmed particles, and 
$Y$ is a state built out of leptons, neutrinos and photons. 
The possibilities for $Y$ are for example $\gamma$ (one particle),
$\ell^+ \ell^-$, $\gamma\, \gamma$  or $\nu \bar\nu$ (two particles), etc. 
The  most  interesting ones are 
$B \rightarrow X_{s,d} \gamma$,\, 
$B \rightarrow X_{s} \ell^+ \ell^-$,\, $B \rightarrow X_{s} \nu \bar\nu$, 
on which we will focus in this paper. Clearly, the cases with  $X = \O$ 
are regarded as exclusive decay modes. Nevertheless,  
for example the rare decay $B_{s,d} \rightarrow \ell^+ \ell^-$ is 
also theoretically rather clean, 
in contrast to other exclusive $B$ rare modes.

\medskip

In 1993, the first evidence for a rare  $B$ meson decay 
was found by the CLEO collaboration. At CESR,
the exclusive  electromagnetic penguin process $B \rightarrow
K^* \gamma$ was measured  \cite{CleoFirst}.
Among inclusive rare $B$ decays, the $B \rightarrow X_s \gamma$ mode
is the most prominent, 
because it was already measured by several independent experiments
\cite{CLEOincl,CLEOneu,ALEPH,Bellebsg,Babarbsg} and  the stringent bounds 
obtained from  that mode on 
various non-standard scenarios
(see e.g.~\cite{Carena,Degrassinew,OUR,NEWNEW})
are a clear example of the importance of 
theoretically clean FCNC observables
in discriminating new-physics models. 
Also the inclusive $B \rightarrow X_s \ell^+\ell^-$ transition
is already  accessible at $B$ factories \cite{BELLEbsll2}. It 
represents a new source of theoretically clean observables, 
complementary to the $B \rightarrow X_s \gamma$ rate.
In particular, kinematic observables such as the invariant 
dilepton mass spectrum and the forward--backward (FB) asymmetry
in \mbox{$B \rightarrow X_s \ell^+\ell^-$}, provide 
clean information 
on short-distance couplings not accessible in 
$B \rightarrow X_s \gamma$
\cite{AliMannel}.

\medskip 

Although the general focus within flavour physics is 
at present on $B$ systems, kaon physics 
offers interesting complementary opportunities in the new 
physics search,  such as the exclusive rare
decays $K^+ \rightarrow \pi^+ \nu \bar{\nu}$ and 
 $K_L \rightarrow \pi^0 \nu \bar{\nu}$. 
They are  specifically interesting in view of the current experiments 
at the Brookhaven National Laboratory 
(USA) and suggested experiments at FERMILAB (USA) and at KEK (Japan).
They are also theoretically clean observables.

\medskip

This present paper  is meant as a status report
to  highlight recent developments and open problems;
for the technical tools 
the reader is often guided to excellent reviews that already
exist in the literature. 
The paper is organized as follows: in section II we briefly discuss 
the role of the strong interaction within flavour physics.
In section III the experimental status of rare $B$ decays is
summarized. In section IV and V we focus on the perturbative
calculations; in section VI we discuss the non-perturbative corrections. 
Phenomenological implications are discussed in section VII.  
In section VIII we explore the implications of these decays 
for our search of physics beyond the SM. 
In section IX we 
discuss direct CP violation  
and in  section X the 
complementary role of rare kaon decays in precision 
flavour physics. In section XI, we present our summary.

\newpage

\setcounter{equation}{0}
\section{Strong interaction in $B$ decays} 
\label{strong}

Flavour physics is governed by the interplay of strong and weak interactions. 
One of the main difficulties in examining the observables in 
flavour physics is the influence of the strong interaction.
As is well known, for matrix elements dominated by 
long-distance strong interactions, there is no
adequate quantitative solution available in quantum field theory.
The resulting hadronic uncertainties restrict the opportunities in
flavour physics significantly, in particular within the indirect 
search for new physics. 

\medskip

The present discussion on the new $g-2$ muon
data \cite{muondata}  
also reflects this issue (for a recent review, see \cite{Melnikov}): 
the hadronic 
self-energy contribution 
to the $g-2$ observable can be determined by experimental data, however, 
the results found from $e^+ e^-$-based data and from the $\tau$-based data 
differ from each other. Furthermore
the well-known
light-by-light scattering contribution 
can only be modelled at present.
It is obvious that these hadronic uncertainties  make it difficult to 
deduce  strict constraints on a 
new physics scenario from this measurement.

\medskip

There are several fundamental tools available,  which 
are directly based on QCD.
High hopes for precise QCD predictions are placed on lattice gauge theoretical 
calculations.  
While there are competitive predictions from lattice gauge theory for
form factors of semi-leptonic $B$ decays, pure hadronic decays  
are less accessible to these methods \cite{Latt}.

Another approach is the method of factorization \cite{Fakt1}. 
This method
has recently been systemized for non-leptonic decays
in the heavy quark limit \cite{Fakt2,Keum}.  
However, within this approach,   a 
quantitative method to estimate the  $1/m_b$ corrections to this limit 
is missing \cite{Fact3}.
A promising step  in this direction was recently presented in \cite{Diehl}.

Further well-known fundamental methods whose applications and precision 
are also  somewhat restricted are  chiral perturbation theory \cite{CHIRAL},
heavy quark effective theory (HQET) \cite{HQET}, QCD sum rules \cite{SUM}  
and the $1/N$ expansion \cite{1/N}.

In view of this, the goal must be to minimize theoretical 
uncertainties with the help of an optimized combination of different
fundamental methods  solely based on  QCD. 
This can only be done for a selected number of observables in flavour
physics. It is also clear that an active cooperation 
between theory and 
experiment is necessary in order to make progress on this issue.

\medskip

There are a few golden channels in which the hadronic physics 
can be disentangled and clean tests of the SM are possible.
Moreover, there are also observables, dominated 
by   perturbative contributions, which make precision
flavour physics possible in the near future. 
Among them inclusive rare $B$ decays play  
the most important role.
Inclusive decay modes are theoretically clean and represent 
a theoretical laboratory of perturbative QCD.
In particular,  the decay width $\G(B \to X_s \gamma)$ 
is well approximated by the partonic decay rate
$\G(b\to X_s^{parton} \gamma)$, which can be
analysed in renormalization-group-improved 
perturbation theory:
\begin{equation}
\Gamma ( B \rightarrow X_s \gamma) = \Gamma ( b \rightarrow X_s^{parton} \gamma ) +
\Delta^{nonpert.}  
\end{equation}
Non-perturbative effects, $\Delta^{nonpert.}$, play a subdominant role
and are under control thanks to the heavy mass expansion \cite{HME}
and the assumption of quark--hadron duality \cite{QHD}. 

\medskip

Thus, in general,  inclusive decay modes 
should be preferred to exclusive ones from 
the theoretical point of view.
The inclusive modes $B \rightarrow X_{s \, (d)} \gamma$
and $B \rightarrow X_{s\, (d)} \ell^+\ell^-$ can be measured by the 
electron--positron experiments ($B$ factories, CLEO) with their kinematic 
constraints and their 
controlled background, while they are more difficult to measure
 at hadronic machines.
Exclusive decay modes, however, are
more accessible to experiments, in particular at hadronic machines. 
But in contrast 
to the inclusive modes, they have in general large  non-perturbative 
QCD contributions, which makes it difficult to deduce valuable  information
on new physics from those decay modes.
However, as mentioned in the introduction, 
the exclusive decays $B_{d,s} \rightarrow \mu^+ \mu^-$
are distinguished observables at hadronic colliders.

\medskip

\begin{figure}
\begin{center}
\epsfig{figure=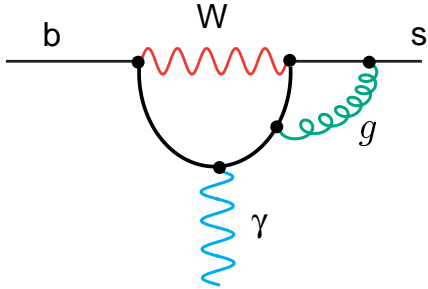,width=6cm}
\end{center}
\caption{QCD corrections to the decay $B \rightarrow X_s \gamma$.}
\label{QCDfigure}
\end{figure}

Within inclusive $B$ decay modes, 
short-distance QCD effects turn out to be very important. 
For example, in the decay $B \rightarrow X_s \gamma$ 
they lead  to a tremendous rate enhancement.
These effects  are induced by 
hard--gluon exchange between the quark lines of the one-loop electroweak
diagrams (fig. \ref{QCDfigure}).

The QCD radiative corrections  bring in large logarithms of the form 
$\alpha_s^n(m_b)$ \, $ \log^m(m_b/M)$,
where $M=m_t$ or $M=m_W$ and $m \le n$ (with $n=0,1,2,...$).
This is a natural  feature in any process where two different mass scales
are present.
In order to get a reasonable result at all, one has  to resum at least
the leading-log (LL) series 
\begin{equation}
\alpha_s^n(m_b) \,  \log^n(m_b/M), \,\, \mbox{(LL)}
\label{LLQCD}
\end{equation}
with the help of renormalization--group techniques.
Working to next-to-leading-log (NLL) precision means that one is also 
resumming all the
terms of the form 
\begin{equation}      
\a_s(m_b) \, \a_s^n(m_b) \, \log^n (m_b/M), \,\,\, \mbox{(NLL)}.
\label{NLLQCD}
\end{equation}

\begin{figure}
\begin{center}
\epsfig{figure=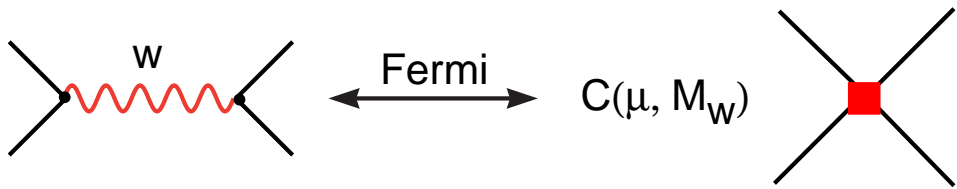,width=13cm}
\end{center}
\caption{Operator product expansion: full versus effective theory.}
\label{operatorproductexpansion}
\end{figure}

A suitable framework to achieve the necessary resummations 
of the large logs is an  effective 
low-energy theory with five quarks, obtained by integrating out the
heavy particles, which, in the SM, are the electroweak bosons and the
top quark. 
The standard method of the operator product expansion (OPE) 
allows for a separation
of the meson decay amplitude into two distinct parts, 
the long-distance contributions contained 
in the operator matrix elements and 
the short-distance physics described by the so-called Wilson coefficients
 (see fig. \ref{operatorproductexpansion}). 
In the case of $B$ decays, 
the $W$ boson and the top quark with mass 
larger than the factorization scale are 
integrated out, that is removed from the 
theory as dynamical fields. 
The effective Hamiltonian can be written
\begin{equation}
 H_{eff} = - \frac{4 G_{F}}{\sqrt{2}} \, 
\sum  {C_{i}(\mu, M_{heavy})}\,\, \, {\cal O}_i(\mu), 
\end{equation}
where ${\cal O}_i(\m)$ are the relevant operators and 
$C_{i}(\mu, M_{heavy})$ are the corresponding Wilson coefficients.
As the heavy fields are integrated out, the complete top and
$W$ mass dependence is contained in the Wilson coefficients.
Working out a convenient set of quantities, both in the effective
(low-energy) theory and in the full (standard model) theory, and 
requiring equality (matching)
up to terms suppressed by higher powers of $m_W$ or $m_t$,
these coefficients can be determined.
At the high scale $\mu_W \approx m_W, m_t$,
the matrix elements of the operators  in the 
effective theory lead to the  same logarithms  as the full theory
calculation. 
Consequently, the Wilson coefficients 
$C_i(\mu_W)$ only pick up small QCD corrections,
which can be calculated in fixed-order perturbation theory.

\medskip

Within this framework QCD corrections for the decay rates are 
twofold: the 
ingredients are  the order $\alpha_s$ corrections to the matrix elements
of the various operators and the order $\alpha_s$ 
corrections to the Wilson coefficients, of course both 
at the low-energy scale $\mu_b \approx  m_b$.
Only the sum of the two 
contributions is renormalization-scheme- and scale-independent; in fact, 
from the $\mu$-independence of the effective Hamiltonian,
one can derive a renormalization group equation 
(RGE) for the Wilson 
coefficients $C_i(\mu)$:
\be
\label{RGEa}
\mu \frac{d}{d\mu} C_i(\mu) = \gamma_{ji} \, C_j(\mu) \quad ,
\ee  
where the matrix $\gamma$ is the anomalous dimension
matrix of the operators ${\cal O}_i$, which describes the 
anomalous scaling of the operators with respect to the 
 one at the  classical level. 
At leading order, the solution is given by
\be
\label{Wilsonsummation}
\tilde{C}_i(\mu)= \left[ 
\frac{\alpha_s (\mu_W)}{\alpha_s(\mu)} 
\right]^{\frac{\tilde{\gamma}^{(0)}_{ii}}{2 \beta_0}}  \, \tilde{C}_i (\mu_W) =
\left[\frac{1}{1+\beta_0\frac{\alpha_s(\mu)}{4\pi}\ln\frac{\mu^2_W}{\mu^2}}
\right]^{\frac{\tilde{\gamma}^{(0)}_{ii}}{2 \beta_0}} \, \tilde{C}_i (\mu_W)
\ee
with $\mu d/d\mu \alpha_s = -2 \beta_0 \alpha^2_s / (4 \pi)$;
$\beta_0$ and $\tilde{\gamma}^0_{ii}$  correpond to  the leading anomalous 
dimension of the coupling constant and the operators, respectively. 
The tilde indicates that the diagonalized anomalous dimension matrix
is used. 
The formula (\ref{Wilsonsummation}) to LL precision  
makes the renormalization-group improvement transparent. 
It represents a summation of the form of the Eq. (\ref{LLQCD}).

\medskip 
There are  three principal calculational 
steps leading to the
leading-log (next-to-leading-log) result within the
effective field theory approach (for a pedagogical review see \cite{Buraslecture}):
\begin{itemize}
\item {Step 1:\, } The full SM theory has to be matched 
with the effective theory at the scale $\m=\mu_W$, where
$\mu_W$ denotes a scale of order $m_W$ or $m_t$. 
As mentioned above, 
the Wilson coefficients 
$C_i(\mu_W)$ only pick up small QCD corrections,
which can be calculated in fixed-order perturbation theory.
In the LL (NLL)  program, the matching has to be worked out at the 
$O(\a_s^0)$  ($O(\a_s^1)$) level. 
\item {Step 2:\, } Then the   
evolution of these Wilson coefficients from 
$\m=\mu_W$ down to $\m = \mu_W$ has to be performed 
with the help of the 
renormalization group, where $\mu_b$ is of the order of $m_b$.
As the matrix elements of the operators evaluated at the low scale
$\mu_b$ are free of large logarithms, the latter are contained in resummed
form in the Wilson coefficients. For a LL (NLL) calculation, this RGE step
has to be done  using the anomalous--dimension matrix up 
to order $\a_s^1$  ($\a_s^2$).
\item {Step 3:\, } To LL (NLL) precision, the 
corrections to the matrix elements 
of the operators $\bra s \g |{\cal O}i (\mu)|b \ket$ at the scale  $\mu = \mub$
have to be calculated to order $\a_s^0$ ($\a_s^1$) precision.
This includes also bremsstrahlung corrections.
\end{itemize}

Finally, we stress that the step from the leading (LL) to
the next-to-leading (NLL) order within the framework of the 
renormalization--group--improved perturbation theory is not only a
quantitative one, increasing the precision of the theoretical prediction,
 but also a qualitative one, which tests the validity of the perturbative
approach in the given problem.

\newpage

\setcounter{equation}{0}
\section{Experimental status}
\label{inclusivesection}

\subsection{Experimental data on $B \rightarrow X_{s} \gamma$}
\label{experimental}

\begin{figure}
\begin{center}
\epsfig{figure=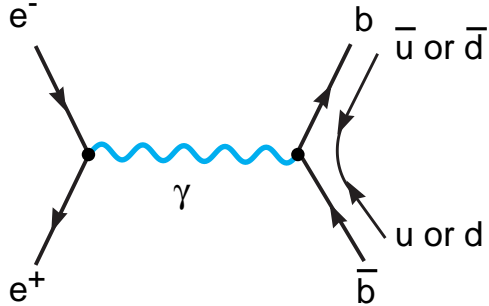,width=7cm}
\end{center}
\caption{$e^+ e^- \rightarrow \Upsilon (4 S) \rightarrow B^+ B^- , B^0 \bar{B}^0$.}
\label{e+e-}
\end{figure}

Among inclusive rare $B$ decays, the $B \rightarrow X_s \gamma$ mode
is the most prominent 
because it was already measured by several independent $e^+e^-$--experiments,
mostly at the $\Upsilon(4 S)$ resonance, fig. \ref{e+e-} 
 \cite{CLEOincl,CLEOneu,ALEPH,Bellebsg,Babarbsg} (see also 
\cite{experiment1,experiment3,experiment2}). 
In 1993, the first evidence for a penguin-induced $B$ meson decay was found
by the CLEO collaboration. At CESR,
they measured the exclusive 
electromagnetic penguin process $B \to K^* \gamma$ \cite{CleoFirst}.
The inclusive analogue $B \to X_s \gamma$,  
which is the quantity of 
theoretical interest, 
was also found by the CLEO 
collaboration through the measurement of its characteristic photon 
energy spectrum in 1994.
As this process is dominated by the two-body 
decay $b \to s \gamma$, its photon energy spectrum 
is expected to be a smeared delta function centred 
at $E_\gamma \approx m_b/2$, 
where the smearing is due to perturbative gluon bremsstrahlung and  to 
the non-perturbative motion of the $b$ quark within the $B$ meson.

\medskip

\begin{figure} 
\begin{center}
\epsfig{figure=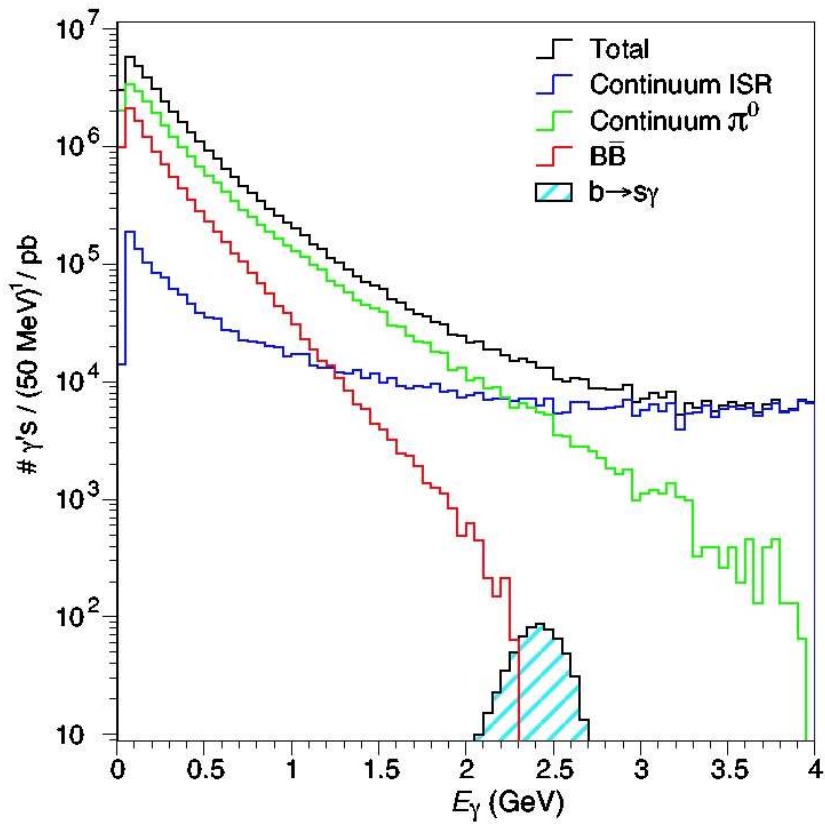,width=11cm}
\end{center}
\caption[]{Levels of inclusive photons from various background processes at 
$\Upsilon (4S)$ and the expected signal from $b \rightarrow s \gamma$:
ISR, $B \bar B$ and $\pi^0$ backgrounds are shown 
(from the bottom to the top at $E_\gamma = 0.5$), from \cite{experiment2}.}
\label{Cleo}
\end{figure}

Only the high part of  the $B \to X_s \gamma$ photon spectrum is observed.
Some lower cut-off in the photon energy was imposed in order to suppress 
the background from other $B$ decay processes. The $B \bar B$ background 
mainly arises from the processes $B \rightarrow \pi^0 X$
and $\pi^0 \rightarrow \gamma_1 \gamma_2$  or $B \rightarrow \eta X$
and $\eta \rightarrow \gamma_1 \gamma_2$, where $\gamma_1$ has high energy
and $\gamma_2$ either has energy too low to be observed or is not 
in the geometric
acceptance of the detector. Moreover, there is a small component ($\sim 5 \%$) 
from the process $B \rightarrow \bar n X$ or $B \rightarrow K_L X$, where 
the anti-neutron or the neutral kaon interacts hadronically with the electromagnetic
calorimeter, faking a photon.

\medskip

Therefore only the `kinematic' branching ratio for
$B \to X_s \gamma$ in the range between 
$E_\gamma=2.2$ GeV and the kinematic endpoint at $E_\gamma= 2.7$
GeV could be measured directly  within this first measurement. 
To obtain from this measurement the `total'  branching ratio, one has 
to know
the fraction $R$ of the $B \to X_s \gamma$ events with
$E_\gamma \ge 2.2$ GeV. 
This was first done  in \cite{AG91} where 
the motion of the $b$ quark in the $B$ meson was
taken into account by using a phenomenological model \cite{ACCMM}
and taking into account  a large systematic 
error for this model dependence. 
Using this {\it theoretical} input regarding the photon energy spectrum
the value  $R=0.87 \pm 0.06$ was used  
by the CLEO collaboration, leading to the CLEO branching ratio
\cite{CLEOincl}
\begin{equation} 
\label{cleoincl}
{\cal B}(B \to X_s \gamma) = (2.32 \pm 0.57_{stat} \pm 0.35_{syst}) \times 10^{-4}. 
\end{equation}
The first error is statistical and the second 
is systematic (including model dependence). 
This measurement was based on a sample of $2.2 \times 10^6 \,  B\bar B$
events. 

\medskip

\begin{figure} 
\begin{center}
\epsfig{figure=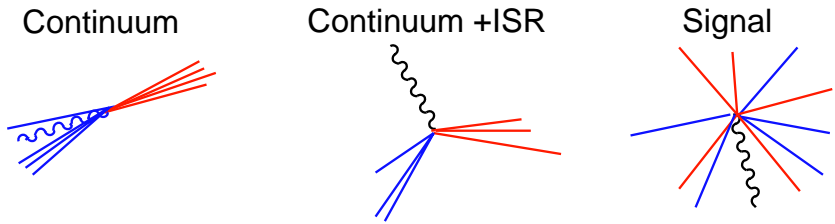,width=11cm}
\end{center}
\caption[]{Examples of idealized event shapes. The straight lines indicate hadrons and the wavy lines photons, from \cite{experiment3}.} 
\label{Shape} 
\end{figure}

Besides the high energy cut-off to suppress the 
background from other $B$ decays,  
two different techniques were used to suppress the continuum background
in this first CLEO measurement. 
In the first (semi-inclusive) technique  all products were reconstructed as in the 
exclusive measurement. 
The background in the
measurement of exclusive modes is naturally low,  because of
kinematical constraints 
and of the beam energy constraint. In order to reduce the combinatoric
 background, only 
$K (n \pi) \gamma$, with $n \leq 4$ and at most  
one $\pi^0$, were chosen as final states in this analysis, which accounts for 
$\sim 50 \%$ of the inclusive rate. 
In the second (fully inclusive) technique, only the photon was 
explicitly reconstructed.
As shown in fig. \ref{Cleo},  
there are very large backgrounds, both from the 
initial-state-radiation
(ISR) process $e^+e^- \rightarrow q \bar q \gamma$, where one of the
beam electrons radiates a hard photon before annihilation,  and from 
inclusive $\pi^0 / \eta$ production in which one of the photons from the decay is not detected.
Background suppression was therefore  more 
difficult with this technique. 
For this purpose, topological differences between the spherical 
$B\bar{B}$ events 
and the two jets $e^+e^- \rightarrow q\bar{q}$ as shown in fig. 
\ref{Shape} were used. While the signal events are spherical 
because the $B$ mesons are almost at rest at the $\Upsilon (4S)$ resonance, 
the continuum
events have a jet-like  structure.  
With the help of a neural network, several event-shape variables 
were combined into a single one,  which tends towards $+1$ 
for $b \rightarrow s \gamma$  and towards $-1$ for the ISR and $q\bar q$ 
processes; the signal was extracted from a one-parameter fit 
to that variable.

\medskip

The signal efficiency ($32 \%$) 
was very high with respect to the first technique ($9 \%$). However the first 
technique has a better signal-to-noise ratio, so that the two methods 
had nearly equal sensitivity. 
In the first CLEO measurement in 1994, they found ${\cal B}(B \to X_s \gamma) 
= (2.75 \pm 0.67_{stat})\times 10^{-4}$
with the first technique and ${\cal B}(B \to X_s \gamma) = (1.88 \pm 0.74_{stat}) \times 10^{-4}$ using the second technique. 
The branching ratio stated above (\ref{cleoincl}) represents the average 
of the two measurements, taking into account the correlation between
the two techniques.

\medskip

In 1999, CLEO had presented a preliminary improved  measurement
\cite{CLEOneualt},
which was based on $53\%$ more data ($3.3 \times 10^6$ events).
 They also used the
slightly wider $E_{\gamma}$ window starting at $2.1$ GeV. 
The relative error dropped by a factor of almost $\sqrt{3}$ already.
In 2002, CLEO published  a new  measurement
\cite{CLEOneu}, based on three times more data 
 ($10 \times 10^6$ events).
The spectrum down to $2.0$ GeV was used, which includes 
almost $90 \%$ of the $B \rightarrow X_s \gamma$ yield.
This also leads to a significant background from $B$ decay 
processes other than $B \rightarrow X_s \gamma$, 
located within $2.0$ -- $2.2$ GeV. This $B\bar B $ background
arises from two components. First the inclusive $\pi^0 / \eta$ 
decays which account for $\sim 90 \%$ of the background. 
This is estimated by Monte Carlo in which the inclusive
$\pi^0 / \eta$ spectra have been tuned with independent 
processes to replicate the data. Second, hadronic interactions of 
anti-neutrons and neutral kaons in the electromagnetic calorimeter 
may fake a photon 
candidate. However, their lateral profile is different from that
of real photons, which allows a background subtraction.  
The continuum background was suppressed with the same 
two approaches as in the first measurement,  
but within a fully integrated analysis. 
What remained of the continuum background was subtracted 
using off-resonance data. 

\medskip

In order to obtain  
the corrected branching ratio of $ B \rightarrow X_s \gamma$, 
two extrapolations  were  necessary. 
What was directly measured was  the branching fraction 
for $B \rightarrow X_s \gamma $ plus $B \rightarrow X_d \gamma$.
The $B  \rightarrow X_d \gamma$ part was subtracted by using the 
theory input that, according to the 
 SM expectation, the $B \rightarrow X_d \gamma$ and the
$B \rightarrow X_s \gamma$ branching fractions  are in the ratio 
$|V_{td}/V_{ts}|^2$. Therefore the branching ratio was corrected 
down by $(4.0 \pm 1.6) \%$ of itself - assuming the validity of the 
SM suppression factor $|V_{td}/V_{ts}|^2$.
Moreover, one has 
to know again the fraction $R$ of the $B \to X_s \gamma$ events with
$E_\gamma \ge 2.0$ GeV. In this measurement, the corresponding fraction  
was estimated  to be $R = 0.915 ^{+0.027}_{-0.055}$ using the model
of Kagan and Neubert  
(see also section \ref{photonspectrum}), which allowed for the 
extrapolation of  the measured branching ratio to the `total' 
$ B \rightarrow X_s \gamma$ branching 
ratio ($E_\gamma > 0.25$ GeV).  
With these two theoretical corrections, the present CLEO measurement 
for the $B \rightarrow X_s \gamma$ branching ratio is
\beq 
\label{cleoneu}
{\cal B}(B \to X_s \gamma) = (3.21 \pm 0.43_{stat} \pm 0.27_{syst} 
{^{+0.18}_{-0.10}}_{mod})\times 10^{-4}. 
\eeq

The errors represent statistics, systematics, and
the model dependence (due to the \mbox{extrapolation} below $E_{\gamma} = 2.0$ GeV) 
respectively.

\medskip

There are also data at the $Z^0$ peak from the LEP experiments.
The ALEPH collaboration \cite{ALEPH} has measured the
inclusive branching ratio based on $0.8 \times 10^{6}$ $b \bar b$ pairs.
\beq
\label{braleph} 
{\cal B}(H_b \to X_s \gamma) = (3.11 \pm 0.80_{stat} \pm 0.72_{syst}) \times 10^{-4}. 
\eeq
The signal was isolated in lifetime-tagged $b \bar b$ events by the presence of a hard photon associated with a system of high momentum and high rapidity 
hadrons. 
It should be noted that the branching ratio in (\ref{braleph}) involves 
a  weighted
average of the $B$ mesons and $\Lambda_b$ baryons produced in $Z^0$ decays
(hence the
symbol $H_b$) different from the corresponding one given by CLEO, 
which has been
measured at  the $\Upsilon (4S)$ resonance.
High luminosity is more difficult to obtain
at higher $e^+e^-$ collision energies. Thus, $B\bar{B}$ samples
obtained by the LEP experiments are rather small. 
The rate measured by ALEPH is consistent with the CLEO measurement,
with an error twice as large as the present CLEO measurement. 

\medskip

BELLE has also presented a measurement \cite{Bellebsg}
based on $6.07 \times 10^{-6}$ $ B \bar B$ events at the $\Upsilon (4S)$ resonance.
A semi-inclusive analysis was used to reconstruct 
the $B \rightarrow X_s \gamma$ decay 
from a primary photon, a kaon and multiple pions 
(no more than one  $\pi^0$). 
The background reduction includes an effective $E_\gamma > 2.24$ 
GeV photon energy cut-off which corresponds to a cut in 
the hadronic mass spectrum of $M_{X_s} = 2.05$ GeV as quoted 
in \cite{Bellebsg}; $E_{\gamma}=(M_B^2-M^2_{X_s})/(2 M_B)$:

\beq
\label{Bellebsg}
{\cal B}(B \to X_s \gamma) = (3.37 \pm 0.53_{stat} \pm 0.42_{syst} 
\pm 0.54_{mod}) \times 10^{-4}, 
\eeq
which is  consistent with previous  measurements. 

\medskip

BABAR presented two preliminary analyses on the $B \rightarrow X_s
\gamma$ branching ratio, a fully inclusive and a semi-inclusive 
one \cite{Babarbsg}. 
The fully inclusive BABAR measurement 
 has used the largest number of $B$ mesons, so far. It is  based on almost 
$60 \times 10^{6}$ $B \bar B$ events at the $\Upsilon (4S)$ resonance.
The method of extracting the signal from the data is similar 
to what was done for previous measurements:
the continuum background  was subtracted with the help of 
off-resonance data. The $B\bar B$ contribution was 
deduced from Monte Carlo predictions.  

Nevertheless, the high statistics available in this 
BABAR measurement allowed for 
additional techniques:
a lepton tag on a high-momentum electron or muon 
was also required to suppress continuum 
backgrounds. For the $ B \rightarrow 
X_s \gamma$ signal events,  the lepton arises from the semi-leptonic decay 
of the other $B$ meson. 
Leptons also occur in the continuum background, 
most notably from the semi-leptonic decays of charm hadrons, but their
production  
is significantly less frequent and their  momentum lower than
those from a $B$ decay. 
Because a lepton  tag is imposed on the other $B$ meson, not on the signal
$B$, one can reject the continuum background  without introducing any
model dependence because one does not impose any requirements on the 
signal decay. 
A $\times 1200$ reduction of the background was achieved by  
5$\%$ efficiency of the lepton tag. This effective method to suppress 
the continuum background was possible because of the high statistics
of the new BABAR measurement.


The systematic precision was limited by the size of the $B \bar B$
background control samples scaling in proportion to the signal sample.
The systematic precision limited the lower bound to $E_\gamma > 2.1$ GeV
(measured in the $e^+ e^-$ centre-of-mass system).  
The preliminary BABAR measurement is 
\beq
\label{BABARbsg}
{\cal B}(B \to X_s \gamma) = (3.88 \pm 0.36_{stat} \pm 0.37_{syst} 
  {^{+ 0.43}_{-0.23}}_{mod}   ) \times 10^{-4}.
\eeq

\medskip

\begin{figure} 
\begin{center}
\epsfig{figure=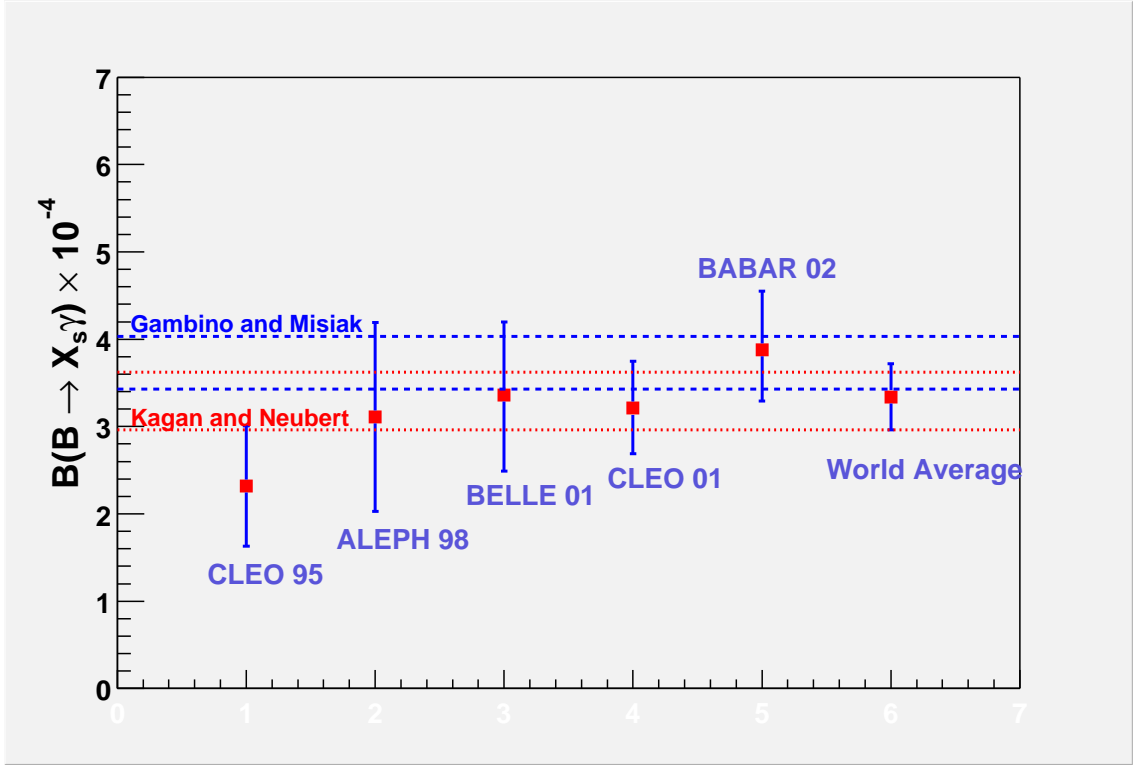,width=15cm}
\end{center}
\caption[]{$B \rightarrow X_s \gamma$ measurements 
versus theoretical predictions (see section \ref{phenobsg}), 
from \cite{Babarbsg}.}
\label{bsgall} 
\end{figure}

Besides this fully inclusive analysis, BABAR also presented a semi-inclusive
analysis where twelve exclusive $b \rightarrow s \gamma$ decays were fully 
reconstructed, which led to the following  measurement of the inclusive 
branching ratio:

\beq
\label{BABARbsg2}
{\cal B}(B \to X_s \gamma) = (4.4 \pm 0.5_{stat} \pm 0.8_{syst} 
  \pm 1.3_{mod}   ) \times 10^{-4}.
\eeq

The error is much larger than the one of the previous semi-inclusive
measurements, but includes also less final states; only 
states including $1-3$ pions rather than $1-4$ pions were 
reconstructed. 

\medskip 

When much more statistics is available, the fully-inclusive strategy 
using the lepton tag will get the priority in the future 
measurements of the $B \rightarrow X_s \gamma$ branching ratio
because model dependence and systematic errors can 
be reduced significantly compared to the semi-inclusive  method.

\medskip

As fig. \ref{bsgall} shows, all the measurements of the `total' 
$B \rightarrow X_s \gamma$ branching ratio available so far 
are consistent with  each other and also consistent with the SM 
predictions (see section \ref{phenobsg}).  
A weighted average of the available experimental measurements is 
problematic, because the model dependence errors (and also 
the systematic errors) are correlated and differ within  
the various measurements. A recent analysis taking into account
the correlations  leads to the following world average~\cite{Colin}:

\beq
\label{world}
{\cal B}(B \to X_s \gamma) = (3.34 \pm 0.38) \times 10^{-4}.
\eeq

\medskip 

With the expected high luminosity of the $B$-factories, 
the systematic uncertainty in the $B \bar B$ background will be reduced 
along with statistical uncertainties. This reduction in the systematic 
uncertainty will also allow for a lower photon energy cut-off, which will 
further reduce
the model dependence from the theory-based interpolation to the whole energy 
spectrum \cite{Babarbsg}. Thus, in the future the lower energy 
cut-off in a fully inclusive analysis has to balance  the 
systematic error due to the $B \bar B$ background and the 
model dependence due to the extrapolation. 
An experimental accuracy  below $10\%$ in the inclusive 
$B \rightarrow X_s \gamma$ mode is possible in the near future.

\subsection{Photon spectrum of $B \rightarrow X_s \gamma$}
\label{photonspectrum}

\begin{figure} 
\begin{center}
\epsfig{figure=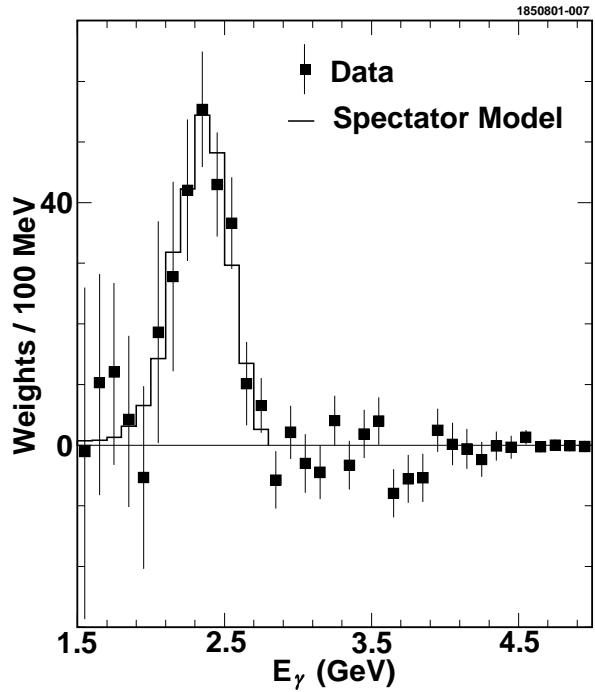,width=8cm}
\end{center}
\vspace{0.5cm}
\caption[]{Photon energy spectrum measured by CLEO and
spectrum from Monte Carlo simulation of the spectator model with
parameters $\hat m_b = 4.690$ GeV, $p_F = 410$ MeV/c, a good fit to
the data, from \cite{CLEOneu}.}
\label{photonspectrumcleo}
\end{figure}

The uncertainty regarding the fraction $R$ of the $B \to X_s \gamma$ 
events above the chosen lower photon energy cut-off 
$E_\gamma$ GeV quoted in the experimental measurement, also
cited as model dependence, should 
be regarded as a purely {\it theoretical} uncertainty:
in contrast to the `total' branching ratio of $B \rightarrow X_s \gamma$,  
the photon energy spectrum  cannot be calculated
directly using the heavy mass expansion, because the operator product
expansion breaks down in the high-energy part of the spectrum,  
where $E_\gamma \approx m_b/2$. 
Therefore, the fraction $R$ was calculated in \cite{AG91} using a 
phenomenological model \cite{ACCMM}, where the motion of the $b$ quark 
in the $B$ meson 
is characterized by two parameters, the
average momentum $p_F$ of the $b$ quark and the average 
mass $m_q$ of the spectator quark.

\medskip

\begin{figure} 
\begin{center}
\epsfig{figure=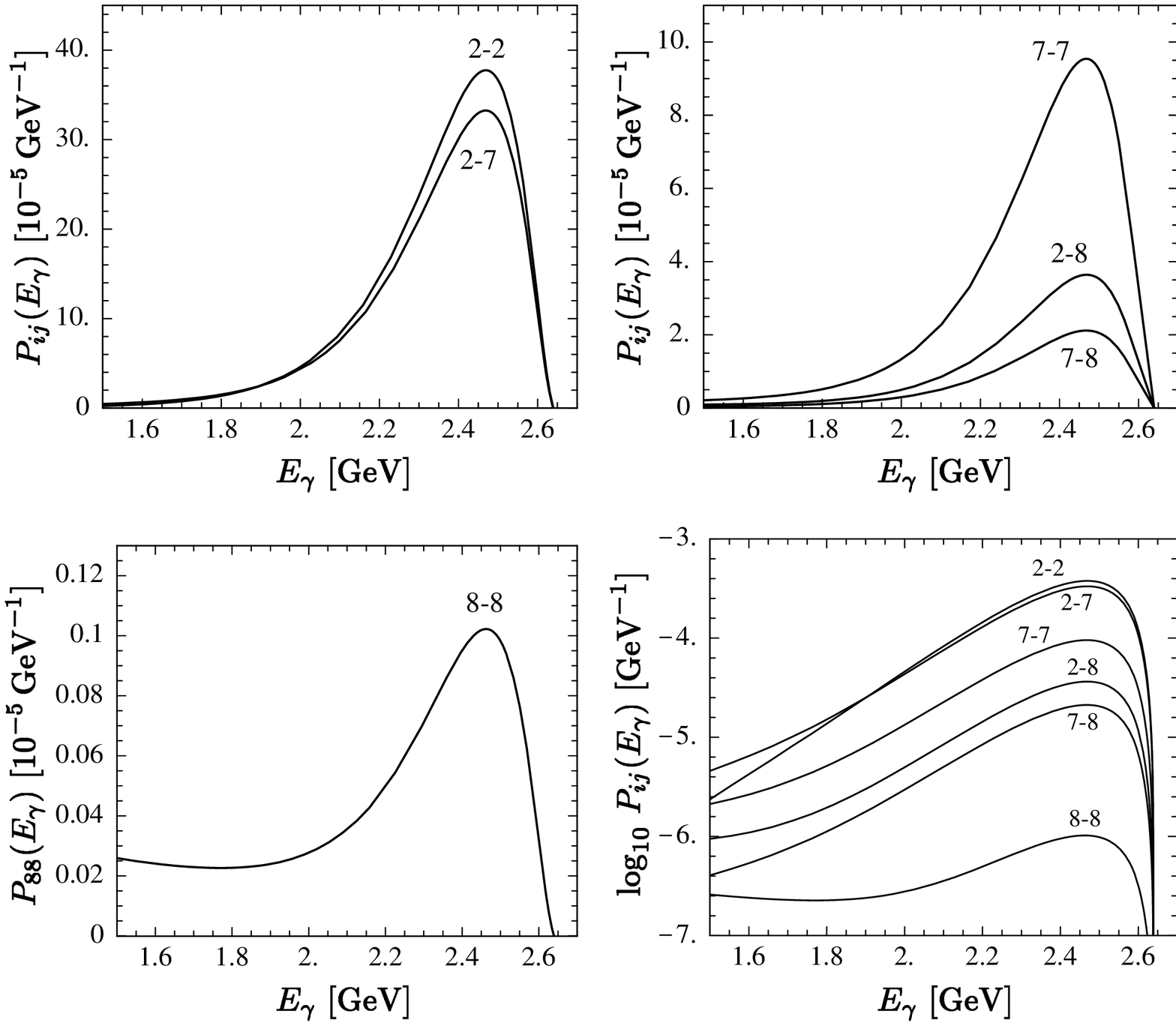,width=13.5cm}
\end{center}
\caption[]{Different components of the photon spectrum in the $B \to X_s \gamma$ decay, from \cite{Kagan}.}
\label{photonspectrumkagan}
\end{figure}

The error on the fraction $R$ is essentially obtained by varying 
the model parameters $p_F$ and $m_q$ in the range for which the 
model correctly describes the energy spectrum of the charged lepton in
the semi-leptonic decays $B \to X_c \ell \nu$ and $B \to X_u \ell \nu$,
measured by CLEO and ARGUS. 
In \cite{AG91} a first comparison between the calculated photon energy 
spectrum and the one measured by the CLEO collaboration was presented.
The (normalized) measured
photon energy spectrum and the theoretical one were  in agreement for those 
values of  $p_F$ and
$m_q$ that correctly describe the inclusive semi-leptonic CLEO data 
on $B \to X_c \ell \nu$ and $B \to X_u \ell \nu$.

\medskip

Besides this phenomenological model,
more fundamental theoretical methods 
are  available today to implement the bound-state effects, 
namely by making use 
of operator product expansion techniques in the framework of the heavy quark 
effective theory (HQET).
An analysis along these lines  was presented in \cite{Kagan}.
As mentioned above,  the operator product expansion breaks down 
near the endpoint of the photon energy  spectrum; therefore, an infinite 
number of leading-twist corrections  have to be resummed into
a non-perturbative universal shape function, which determines the
light-cone momentum distribution of the $b$ quark in the $B$ meson
 \cite{shapefunction}. The 
physical  decay distributions are then obtained from a convolution of 
parton model spectra 
with this shape function.
At present this function cannot be calculated, but there is at least some 
information 
on the moments of the shape function, which are related to the forward matrix 
elements 
of local operators. Ans\"atze  for the shape 
function, constrained by the latter information, are used. 
In contrast to the older analysis based on the phenomenological model
 proposed in \cite{ACCMM},  
the analysis of Kagan and Neubert \cite{Kagan}
includes the full NLL information. 

\medskip

In the latest CLEO measurement \cite{CLEOneu}, the phenomenological 
spectator model \cite{ACCMM,AG91}
 was used first. The momentum parameter $p_F$
and the $b$ quark average mass $\hat m_b$ were treated as free parameters, 
which allowed
the mean and the width of the photon energy spectrum to be varied:
see fig. \ref{photonspectrumcleo}. 
The Kagan--Neubert approach was also used by CLEO:
the  simple two-parameter shape function was fitted to the measured photon 
spectrum and very similar results to those obtained using the 
spectator model were obtained.

\medskip

An important observation is that the shape of the photon spectrum 
is practically insensitive to physics beyond the SM. 
As can be seen in fig. \ref{photonspectrumkagan},
all different contributions to the spectrum 
(corresponding to the interference
terms of the various operators involved, see section \ref{NLLQCDcorrections})
have a very similar
shape besides the small $8$--$8$ contribution. 
This implies that we do not have to assume the
correctness of the SM in the experimental analysis.

\medskip

A precise measurement of the photon spectrum  allows to 
determine the parameters of the shape function. 
The latter information is an  important input for the determination
of the CKM matrix element $V_{ub}$. 
One takes advantage of the universality of the shape 
function to lowest order in $\Lambda_{QCD}/m_b$. The same 
shape function occurs in the description of nonperturbative
effects in the endpoint region of the $B \rightarrow X_s \gamma$
 photon spectrum and of the $B \rightarrow X_u \ell \nu$ charged-lepton
spectrum up to higher $1/m_b$ corrections \cite{shapefunction}. 
Thus, 
 from the photon spectrum one can  determine the shape function;
with the help of the latter and of the measurement of the charged-lepton
spectrum of $B \rightarrow X_u \ell \nu$, one can extract a value for $V_{ub}$.  
This method represents one of the best ways to measure the CKM matrix
element $V_{ub}$. 
Following this strategy, CLEO has presented the following 
 measurement \cite{cleoVub}:
\begin{equation}
V_{ub} = ( 4.08 \pm 0.56_{exp} \pm 0.29_{th} ) . 
\end{equation}
The impact of the 
higher-order corrections in $1/m_b$ was quite recently 
investigated \cite{highermb}.

The  future aim should be to determine the shape function
 by using the high-precision measurements of the photon  energy
spectrum more precisely.

\medskip

Moreover, the first and the second moment
of the photon spectrum can be determined 
within the measurement of $B \rightarrow X_s \gamma$.
These results can be used to  
extract values for the HQET parameters $\bar\Lambda$ 
and $\lambda_1$ (see \cite{Ligeti}).
CLEO has measured these moments and extracted for example from the first
moment 
 $\bar \Lambda = 0.35 \pm 0.08 \pm 0.10 GeV$, where the first error 
is from the experimental error and the 
second error is from the theoretical expression, in particular from neglected
higher-order terms \cite{CLEOneu}.

\medskip

\begin{figure} 
\begin{center}
\epsfig{figure=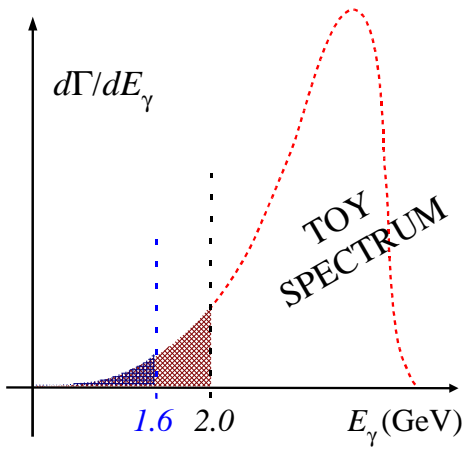,width=9.5cm}
\end{center}
\caption[]{Schematic photon spectrum of $B \rightarrow X_s \gamma$.}
\label{Toymodel}
\end{figure}

A lower experimental cut in the photon energy spectrum  within the 
measurement of 
$B \rightarrow X_s \gamma$ decreases the sensitivity 
to the parameters of the shape function 
and therefore the  model dependence. With respect to this, 
the ideal energy cut
would  be $1.6$ GeV (see fig. \ref{Toymodel}). But in this case a better 
understanding of the  $\psi$ background would be mandatory.
The intermediate $\psi$ background, namely 
${B} \to \psi X_s$ followed by $\psi \to X' \gamma$, is  
more than $4 \times 10^{-4}$ in the `total' branching ratio.  
With the present energy cut of $2.0$ GeV, this contribution is
suppressed and estimated to be less than $1.5 \%$; for 
$2.1$ GeV it is  $0.6 \%$ \cite{Misiakichep}.

\subsection{Experimental status of  $ B \rightarrow X_s  \ell^+\ell^- $
 and  $B \rightarrow X_d \gamma$ }
 \label{experibsll}

The inclusive $B \rightarrow X_s \ell^+\ell^-$ transition
also starts to be accessible at the $B$ factories.
BELLE and also BABAR have already established measurements of the
exclusive mode $B \rightarrow K \ell^+ \ell^-$ \cite{BELLEbsll1,BABARbsll1}.
The two measurements are compatible with each other.

\medskip

\begin{figure} 
\begin{center}
\epsfig{figure=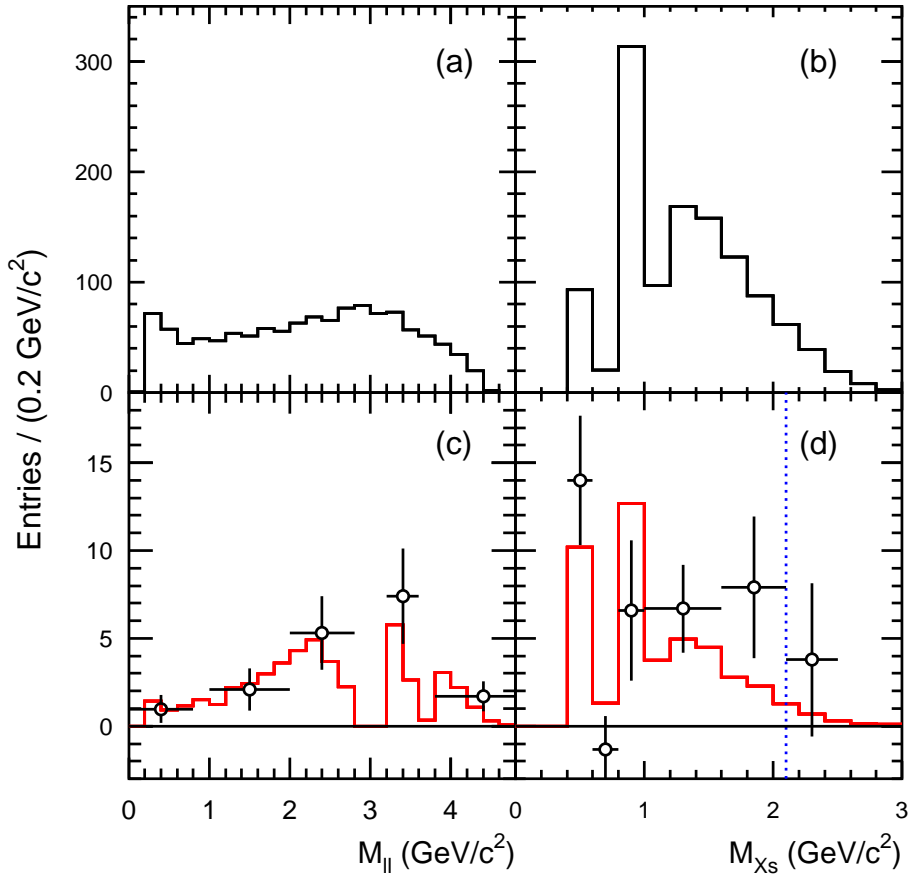,width=12cm}
\end{center}
\caption[]{SM expectations for the (a) dilepton and (b) recoil mass
 spectra; the observed (c) dilepton and (d) recoil mass
 spectra (circles).  The histograms in (c), (d) show the
 SM expectations after all the selections are applied;
 histograms are normalized to the expected branching fractions.  The
 dotted line in (d) indicates the $M_{Xs} <2.1$ GeV  requirement, 
from \cite{BELLEbsll2}.}
\label{bsllhigh}
\end{figure}

Quite recently, BELLE has also announced the first measurement of the inclusive
$B \rightarrow X_s \ell^+ \ell^-$ mode based  on a semi-inclusive analysis
\cite{BELLEbsll2,BELLEbsll3}.  
The hadronic system $X_s$ 
is reconstructed from a kaon
with $0$ to $4$ pions (at most one $\pi^0$). 
The used data sample contains 
$65.4 \times  10^6$ $B \bar B$ pairs. 

The signal characteristics within this semi-inclusive analysis 
is determined by modelling the invariant mass $M_{X_s}$ spectrum 
using the phenomenological model first proposed in \cite{ACCMM}. 
The reconstruction efficiencies of the signal are determined by the
MC samples based on this model, leading to a large part of the 
systematic uncertainty. 

\medskip 

The non-peaking backgrounds are estimated by sideband subtraction. 
But there 
are two peaking backgrounds:
the first one is the process $B \rightarrow
X_s \pi^+ \pi^-$, where the two pions are misidentified as leptons. 
This background is estimated explicitly by reconstructing the 
$B \rightarrow X_s \pi^+ \pi^-$ and multiplying the yield by
$( f_{\pi}^\mu )^2$, where $f_{\pi}^\mu$ is the probability 
of a pion faking a muon measured in an independent data set.
The second source are the charmonium decays
$B \rightarrow \psi ( \rightarrow \ell^+ \ell^-)\, X_s$ and
$B \rightarrow \psi' (\rightarrow \ell^+ \ell^-)\, X_s$. They are vetoed 
by excluding lepton combinations whose invariant mass falls within a window 
around the nominal $\psi$ and $\psi'$ mass.

The continuum background and the $B \bar B$ background, however, 
can be suppressed by the kinematical constraints.  Further suppression
is achieved with methods similar to those in the 
$B \rightarrow X_s \gamma$ analysis.

\medskip 

Moreover, there is a cut used in the $X_s$ invariant mass spectrum at 
$2.1$ GeV. This removes a large part of the combinatorial background
while a model calculation determines that
$(93 \pm 5) \%$ of the signal is within this experimental window
(leading to an additional model dependence). 
Events with a dilepton mass $M_{\ell^+\ell^-}$ less than $0.2$ GeV 
are also rejected in order to suppress electron pairs from 
$\pi^0 \rightarrow e^+ e^- \gamma$ and $\gamma \rightarrow e^+ e^-$ 
conversion.  

\medskip 

A comparison of the histograms in fig. \ref{bsllhigh}, (a) and (c), 
indicates that the efficiency within this measurement
is much higher in the high dilepton mass region. 

\medskip 

The uncertainty of this first measurement of the inclusive decay
is still at the $30 \%$ level and in agreement with the SM expectations.
One can expect much higher accuracy from the $B$ factories in the near 
future. 

\medskip

Also the inclusive decay $B \rightarrow X_d \gamma$ is within reach of  
the high-luminosity $B$ factories.
Such a measurement will rely  on high statistics and on powerful methods
for the kaon--pion--discrimination.
At present only upper bounds on corresponding exclusive modes
are available from CLEO \cite{CLEObdg}, BELLE \cite{BELLEbdg} and also from
BABAR  \cite{BABARbdg}.

\newpage

\section{Perturbative calculations in  $ B \rightarrow X_{s,d}\, \gamma$}
\label{NLLQCDcorrections}

The inclusive decay $B \rightarrow X_s \gamma$ is a
laboratory for perturbative QCD. 
Non-perturbative effects (see section \ref{sectionnonpert}) 
play a subdominant role and are well  under control thanks to 
the heavy quark expansion. 
The dominant short-distance  QCD corrections enhance the 
partonic decay rate $ \Gamma(b \to X_s^{parton} \g)$  by a factor of more than 
$2$.
The corresponding large  logarithms of the form 
$\alpha_s^n(m_b) \, \log^m(m_b/M)$,
where $M=m_t$ or $M=m_W$ and $m \le n$ (with $n=0,1,2,...$), have to
be summed with the help of the renormalization-group-improved perturbation 
theory,  as presented in section \ref{strong}.

\medskip

\begin{figure}
\begin{center}
\epsfig{figure=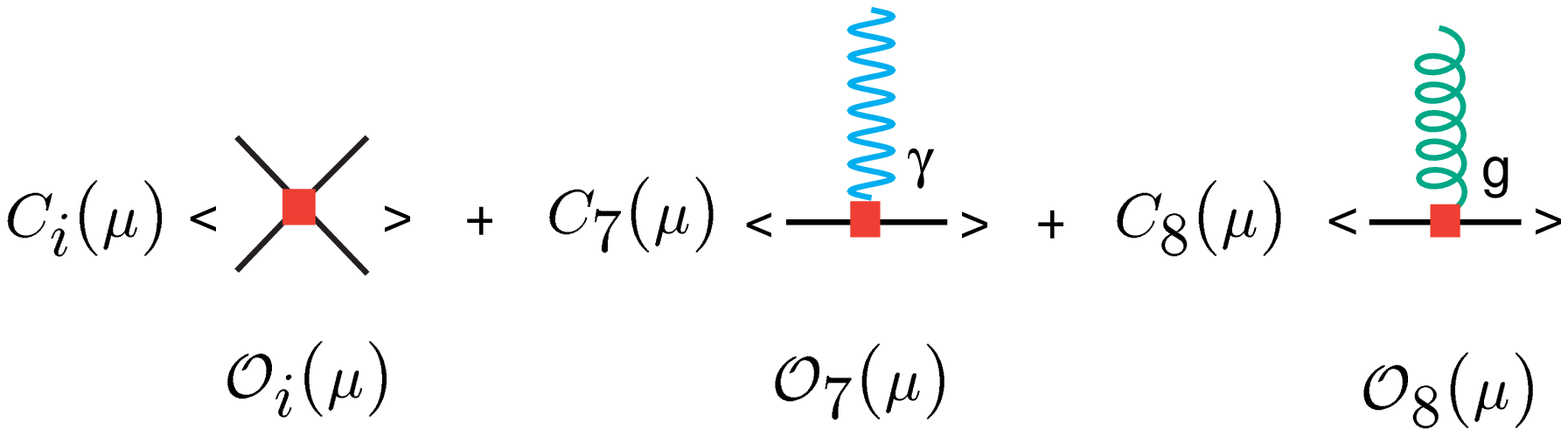,width=13cm}
\end{center}
\caption{Effective Hamiltonian in the case of $B \rightarrow X_{s,d} \gamma$.}
\label{SMhamiltonian}
\end{figure}

The effective Hamiltonian relevant to  $B \to X_s \gamma$ 
in the SM reads 

\begin{equation}
\label{heff}
H_{eff}(B \to X_s \gamma)
       = - \frac{4 G_{F}}{\sqrt{2}} \, \lambda_{t} \, \sum_{i=1}^{8}
C_{i}(\mu) \, {\cal O}_i(\mu) \quad, 
\end{equation}
where ${\cal O}_i(\m)$ are the relevant operators, 
$C_{i}(\mu)$ are the corresponding Wilson coefficients,
which contain the complete top- and $W$-mass dependence
(see fig. \ref{SMhamiltonian}),
and $\lambda_q=V_{qb}V_{qs}^*$ with $V_{ij}$,  the
CKM matrix elements. The CKM dependence globally factorizes,
if one  works in the approximation $\lambda_u=0$\,\footnote{
This approximation is not used within the recent theoretical
predictions.}. 
One neglects the operators with dimension $>6$, which are suppressed 
by higher powers of $1/m_{W}$. 

\medskip 

Using the equations of motion for the operators, one arrives at the 
following basis  of dimension-6 operators \cite{Zakharov,Grinstein}: 
\begin{equation}
\begin{array}{rlrl}
{\cal O}_{1} ~= &\!\!
(\bar{s} \gamma_\mu T^a P_L c)\,  (\bar{c} \gamma^\mu T_a P_L b)\,, & 
{\cal O}_{2} ~= &\!\!
(\bar{s} \gamma_\mu P_L c)\,  (\bar{c} \gamma^\mu P_L b)\,,  \\[1.2ex]
{\cal O}_{3} ~= &\!\!
(\bar{s} \gamma_\mu P_L b) \sum_q (\bar{q} \gamma^\mu q)\,,   &  
{\cal O}_{4} ~= &\!\!        
(\bar{s} \gamma_\mu T^a P_L b) \sum_q (\bar{q} \gamma^\mu T_a q)\,, \\[1.2ex]
{\cal O}_{5} ~= &\!\! 
(\bar{s} \gamma_\mu \gamma_\nu \gamma_\rho P_L b) 
 \sum_q (\bar{q} \gamma^\mu \gamma^\nu \gamma^\rho q)\,, &
{\cal O}_{6} ~= &\!\! 
(\bar{s} \gamma_\mu \gamma_\nu \gamma_\rho T^a P_L b) 
 \sum_q (\bar{q} \gamma^\mu \gamma^\nu \gamma^\rho T_a q)\,, \\[2.0ex]
{{\cal O}}_{7}   ~= &\!\!     
  \frac{e}{16\pi^2} \, \mbMS(\mu) \,
 (\bar{s} \sigma^{\mu\nu} P_R b) \, F_{\mu\nu}\,,    &
{{\cal O}}_{8}   ~= &\!\!  
  \frac{g_s}{16\pi^2} \, \mbMS(\mu) \,
 (\bar{s} \sigma^{\mu\nu} T^a P_R b)
     \, G^a_{\mu\nu}\,,        \\[2.0ex]                       
\end{array}
\label{operators}                                        
\end{equation}

In the dipole-type operators ${\cal O}_7$ and ${\cal O}_8$, 
$e$ and $F_{\m \n}$ ($g_s$ and $G^A_{\m \n}$)
denote the electromagnetic (strong)
coupling constant and field strength tensor,  respectively. 
$T^a$ ($a=1,8$) denote $SU(3)$ colour generators and  $P_{R,L} = (1 \pm \gamma_5)/2$. 

\medskip

The error of the leading logarithmic (LL) result \cite{counterterm}
was  dominated by a large renormalization scale dependence 
at the $\pm 25\%$ level, which already indicated
the importance of the NLL series.
By convention, the dependence on the renormalization scale $\mu_b$ is
obtained by the variation $m_b/2 < \mu_b < 2 m_b$.
The former measurement of the CLEO collaboration (see (\ref{cleoincl})) 
overlaps with the estimates based on LL  calculations,
and the experimental and 
theoretical errors are comparable.
In view of the expected increase in the experimental precision, 
it became clear that a systematic inclusion of 
the NLL corrections
was becoming necessary. Moreover, such a NLL program 
was also important in order to ensure the validity
of renormalization-group-improved perturbation theory in this specific 
phenomenological application.

\medskip 

\begin{figure} 
\centerline{
\psfig{figure=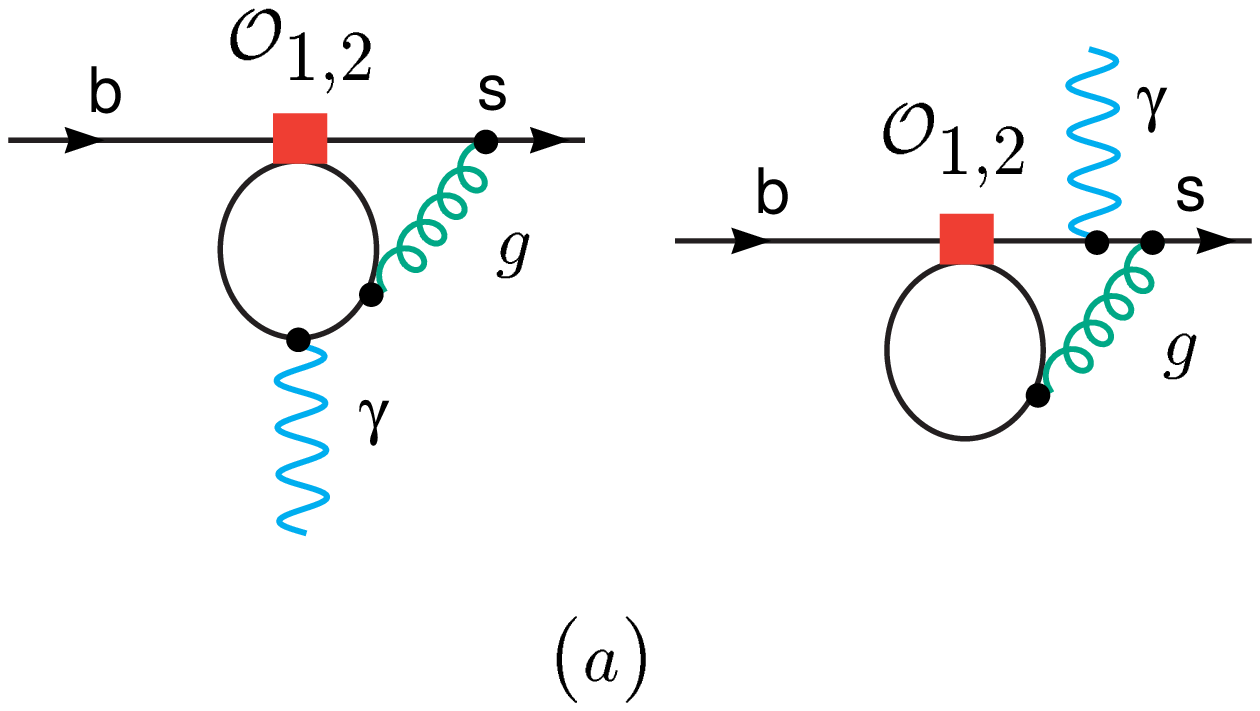,height=4.8cm} 
\hspace{1cm}\psfig{figure=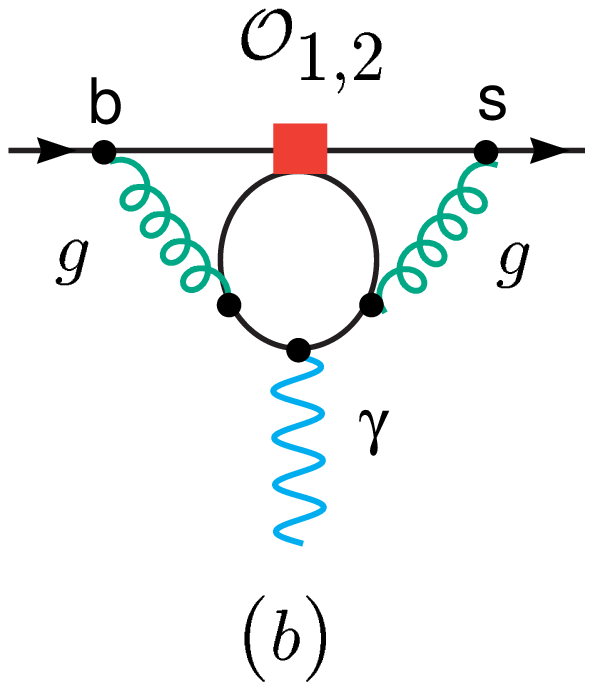,height=4.8cm}} 
\caption[]{(a) Typical diagrams (finite parts) contributing 
to the matrix element of the operator ${\cal O}_2$ at the NLL level, Step 3;
(b) typical diagram (infinite part) contributing to the NLL anomalous 
dimension matrix, Step 2;  see fig.\ref{QCDfigure} for 
a typical diagram (finite part) contributing 
to the NLL matching calculation, Step 1.
\label{feynman}}
\end{figure}

This ambitious NLL enterprise was completed some years ago.
This was a joint effort of many different groups 
(\cite{AG91}, \cite{GHW}, 
\cite{Adel}, \cite{Mikolaj}).
The theoretical error of the  previous 
LL result was substantially reduced, 
to $\pm 10\%$, and the central value of the partonic
decay rate increased by about $20\%$.
All three steps to NLL precision listed below (\ref{RGEa})
involve rather difficult calculations (see fig. \ref{feynman}).

\begin{itemize}

\item The most difficult part in Step 1 is the 
two-loop (or  order $\a_s$) matching of the dipole operators ${\cal O}_7$ and 
${\cal O}_8$. 
It involves two-loop diagrams both in the full and in the effective theory. 
It was first worked out by Adel and Yao \cite{Adel}. 
As this is a crucial step in the NLL program,
Greub and Hurth confirmed their findings in a detailed recalculation 
using a different method \cite{GH}.
Two further complete  
\cite{CDGG,mikolaj99} and one partial recalculations \cite{Infrared}
of this result were presented in the meanwhile, confirming
the original results in \cite{Adel}.  
In order to match the dimension-6 operators ${\cal O}_7$ and ${\cal O}_8$, 
it is sufficient to
extract the terms  of order $ \frac{m_b^2}{M^2} $ ($M=m_W,m_t$)
from the SM 
matrix elements for $b \to s \g$ and $b \to s g$.
Terms suppressed by additional powers of
$m_b/M$ correspond to higher-dimensional operators in the effective theory.
In \cite{GH} the finite parts of the 
two-loop diagrams in the SM were calculated by means of the 
well-known method of asymptotic mass expansions, which naturally leads 
to a systematic expansion of Feynman diagrams in inverse powers of
$M$.

\item 
The order $\a_s^2$ anomalous dimension matrix (Step 2) has been 
worked out
by Chetyrkin, Misiak and M\"unz \cite{Mikolaj}. 
In particular, the calculation of the elements $\gamma_{i7}$ 
and $\gamma_{i8}$ ($i=1,...,6$) in the $O(\a_s^2)$ anomalous
dimension matrix involves a huge number of three-loop diagrams
from which the pole parts (in the $d-4$ expansion) have to be extracted.
This extraction was
 simplified by a clever decomposition of the scalar propagator 
\cite{Misiakcalc1}.
Moreover, the number of necessary evanescent operators was reduced by 
a new choice of a basis of dimension-6 operators \cite{Misiakcalc2}.   
Using the matching result (Step 1), these authors obtained 
the NLL  correction to the Wilson coefficient $C_7(\m_b)$.
Numerically, the LL and  NLL values 
for $C_{7}(\m_b)$ 
turn out to be  rather similar; the NLL 
corrections to the Wilson coefficient $C_7(\m_b)$ 
lead to a change of the  $B \to X_s \gamma$
decay rate that does not exceed  6\% in the $\overline{MS}$ scheme
 \cite{Mikolaj}.

It should be stressed that the result of Step 2, in particular  
the entries $\gamma_{i7}$  and $\gamma_{i8}$ ($i=1,...,6$)
of the anomalous dimension matrix to NLL precision, is the only 
part of the complete  NLL enterprise that  has not been confirmed 
by an independent group. An independent check of this
important part of the NLL program is already on the way 
\cite{Gambinoprivate}.

\item
Step 3 basically consists of bremsstrahlung corrections and virtual
corrections. While the bremsstrahlung corrections
were worked out some time ago by Ali and Greub \cite{AG91}, and were
 confirmed and extended by Pott \cite{Pott}, a
complete analysis of the virtual two-loop corrections (up to the contributions 
of the four-quark operators with very small coefficients) was presented
by Greub, Hurth and Wyler \cite{GHW}. 
This  calculation involves two-loop diagrams, where the full charm 
dependence has to be taken into account.   
By using Mellin--Barnes techniques in the Feynman parameter integrals, 
the result of these two-loop
diagrams was obtained in the form
\be
\label{Mellin}
c_0 + \sum_{n=0,1,2,...;m=0,1,2,3} c_{nm} \left( \frac{m_c^2}{m_b^2}
\right)^n \, \log^m \frac{m_c^2}{m_b^2} \, ,
\ee
where the quantities $c_0$ and $c_{nm}$ are independent of $m_c$.
The convergence of the Mellin-Barnes series was proved;
the practical convergence of the series (\ref{Mellin}) was
also checked explicitly. 
Moreover, a finite result is obtained in the limit $m_c \to 0$,
as there is no naked logarithm of $m_c^2/m_b^2$. This observation
is of some
importance in the $b \to d \gamma$ process, where the $u$-quark
propagation
in the loop is not CKM-suppressed (see below).
The main result of Step 3  consists in a drastic reduction of the 
renormalization 
scale uncertainty from about $\pm 25\%$ to about $\pm 6\%$.
The central value was shifted by about $20 \%$. 

In \cite{GHW} these results are presented also in the 
't Hooft--Veltman scheme, which may be regarded
as a first step towards a cross-check 
of the complete  NLL calculation  prediction in a different 
renormalization  scheme.
Recently, the results of the ${\cal O}_{1,2}$ matrix elements in the
$\overline{MS}$ scheme calculated in  \cite{GHW}
were confirmed by an
independent group \cite{Burasnew} with the help of the method
of  asymptotic expansions. Also two further calculations with the help of 
a direct analytical \cite{Misiak2002} and  of a numerical method 
\cite{Adrian2} confirmed these results.  The direct analytical 
method also allowed control over the matrix elements 
of the penguin operators ${\cal O}_{3-6}$. As expected from the smallness 
of the corresponding Wilson coefficients, their  effect  
on the branching ratio  does not exceed $1 \%$.

\medskip

\end{itemize}

Combining the NLL calculations of the three steps, 
the first practically complete theoretical prediction to 
NLL  precision 
for the branching ratio of $B \rightarrow X_s \gamma$
was presented in \cite{Mikolaj} (see also \cite{Towards}):
\be
\label{mikolajend}
{\cal B}(B \to X_s \g)=(3.28 \pm 0.33) \times 10^{-4}.
\ee
The theoretical error had two dominant sources,  $\mu$ dependence,
which was reduced to about $6 \%$, and the $m_c/m_b$ dependence. 
This first theoretical NLL prediction already included the 
non-perturbative correction scaling with $1/m_b^2$, which are rather small
(at the $1\%$ level) (see section \ref{sectionnonpert}).
Surprisingly, these first NLL predictions, \cite{Mikolaj} and \cite{Towards},  
are almost identical to the present predictions 
(\ref{currentprediction}) using the charm pole mass,  
in spite of so many important additional refinements such as the 
electroweak two-loop corrections and the non-perturbative
corrections,  which will be discussed below.

\medskip

Detailed studies of the two-loop electroweak corrections in the 
decay $B \rightarrow X_s \gamma$ were
performed.   In   \cite{Marciano} 
part of the electroweak two-loop contributions, 
namely contributions from  fermion loops in gauge boson
propagators ($\gamma$ and $W$) and from short-distance photonic 
loop corrections, were calculated. Moreover, it was observed  that  
the on-shell value of the fine
structure constant  $1/\alpha_{em} =137$ is more appropriate for real
photon emission than 
the value $1/\alpha_{em}  = (130.3 \pm 2.3)$ used in previous analyses.
The QED loop calculations in \cite{Marciano} confirmed this expectation.
This change in $\a_{em}$ leads to 
a reduction of $5\%$ in the perturbative contribution .
In \cite{Strumia} a calculation 
of the heavy top and the heavy Higgs corrections in the 
gaugeless limit $m_W \rightarrow 0$ was presented. 
In \cite{Kagan} the QED analysis made in 
\cite{Marciano} was improved
by resumming the contributions of order 
$\alpha_{em} \log(\mu_b/M) (\alpha_s \log(\mu_b/M)^n$
to all orders (while in \cite{Marciano} only the $n=0$ contribution
was included). This resummation decreases the QED corrections.
In \cite{Baranowski} the same calculation was performed taking into
account  the complete relevant set of operators. It was explicitly 
shown that the truncation of the operator basis in \cite{Kagan} 
turns out to be a correct approximation and that these 
corrections lead to a $0.8 \%$ correction only.

The first (practically) 
complete analysis of the electroweak contributions to order 
$\alpha_{em} (\alpha_s \log(\mu/M))^n$  was performed   
in \cite{Gambino:2000fz}. This includes a two-loop
matching to order $\alpha_{em}$, the QED--QCD running of the 
Wilson coefficients down to the $b$ quark and one- and two-loop
QED matrix elements.  
While in \cite{Gambinofirst} only the so-called 
purely electroweak contributions were considered
where terms vanishing in the limit $ \sin \theta \rightarrow 0$ 
were neglected if they are not enhanced by powers of the top mass,  
in \cite{Gambino:2000fz} the complete two-loop matching 
conditions to order $\alpha_{em}$ were presented.
It was shown that 
the electroweak two-loop corrections of order
$\alpha_{em} (\alpha_s \log(\mu/M)^n$ lead,
because of  accidental cancellations, 
to a 
$1.6 \%$ reduction of the branching ratio of 
$B \rightarrow X_s \gamma$ only. 
Thus, the electroweak corrections are well under control and 
shown to play a subdominant role.

\medskip

It is clear that many parts of the perturbative calculations at the
partonic level in the  case of $B \rightarrow X_s \gamma$ 
can be taken over to the cases $B \rightarrow X_d \gamma$ and 
$B \rightarrow X_s \ell^+ \ell^-$;
the latter case, however, needs some 
modifications, in particular the operator basis gets enlarged 
as will be discussed in the next subsection. 

\medskip

The perturbative 
QCD corrections in the decay  $b \rightarrow d \gamma$ can be treated  
in complete analogy to the ones in the decay $b \rightarrow s \gamma$
\cite{AG7}:
the effective Hamiltonian  is the same 
in  the processes $b \to s \gamma$ and $b \to d \gamma$
up to the obvious
replacement of the $s$-quark field by the $d$-quark field.  
But as $\lambda_u = V_{ub} V^*_{ud}$
for $b \to d \gamma$ is not small with respect to
$\lambda_t = V_{tb} V^*_{td}$ and $\lambda_c = V_{cb} V^*_{cd}$, 
one also has to take into account the operators proportional
to $\lambda_u$.
The matching conditions $C_i(m_W)$
and the solutions of the RG equations, yielding $C_i(\mu_b)$, coincide
with those needed for the process $B \to X_s \gamma$.

\medskip

The perturbative calculations at the partonic level of 
$B \rightarrow X_s \gamma$ can also be used for the partonic
process $c \rightarrow u \gamma$. As FCNC process, it does not occur 
at the tree level in the SM either. Moreover, it is strongly
 GIM-suppressed at one-loop.  The leading QCD logarithms 
are known to enhance the one-loop amplitude by more than one 
order of magnitude. It was shown in \cite{HurthMisiak} 
that the amplitude increases further by two orders of magnitude 
after including the formally NLL QCD effects. So the $c \rightarrow u \gamma$
process is completely dominated by a two-loop term. However,  this is only
of theoretical interest, because the $\Delta S = 0$ radiative decays of
charmed hadrons remain dominated by the $c \rightarrow d \bar d u \gamma$
and $c \rightarrow s \bar s u \gamma$ subprocesses.

\newpage 

\section{Perturbative calculations  in $B \rightarrow X_{s} \ell^+\ell^-$
and $B \rightarrow X_s \bar{\nu} \nu$ }
\label{NNLLQCDbsll}

\begin{figure}
\begin{center}
\epsfig{figure=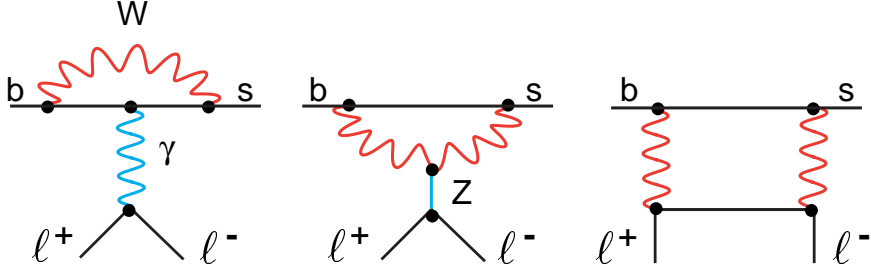,width=12cm}
\end{center}
\caption{One-loop contributions to the decay 
$B \rightarrow X_s \ell^+ \ell^-$.}
\label{llpicture}
\end{figure}

In comparison with the 
$B \rightarrow X_s \gamma$ decay, 
the inclusive $ B \rightarrow X_s \ell^+ \ell^-$ decay
presents a complementary and also more complex test of the SM, since different
contributions add to the decay rate (fig. \ref{llpicture}).

It is particularly attractive 
because of kinematic observables such as the invariant 
dilepton mass spectrum and 
 the forward--backward (FB) asymmetry. It is  also dominated 
by perturbative 
contributions, if one eliminates $c\bar{c}$ resonances with 
the help of kinematic cuts (see section \ref{bsllpheno}).

\medskip

The effective Hamiltonian relevant to  
$B \to X_s \ell^+ \ell^-$ in the SM reads 

\begin{equation}
\label{heffll}
H_{eff}(B \to X_s \ell^+ \ell^-)
       = - \frac{4 G_{F}}{\sqrt{2}} \, \lambda_{t} \, \sum_{i=1}^{10}
C_{i}(\mu) \, {\cal O}_i(\mu) \,  .
\end{equation}

\begin{figure}
\begin{center}
\epsfig{figure=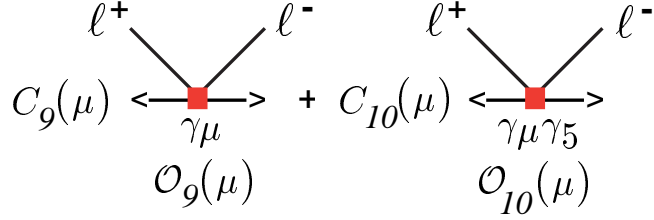,width=9cm}
\end{center}
\caption{Additional operators 
in the effective  Hamiltonian in the case of $B \rightarrow X_{s,d} \ell^+ \ell^-$.}
\label{SMhamiltonianbsll}
\end{figure}

Compared with the decay $B \rightarrow X_s \gamma$ (see (\ref{heff})),  
the effective Hamiltonian~(\ref{heffll}) contains two 
additional operators $O(\alpha_{em})$  (see fig. \ref{SMhamiltonianbsll}):

\begin{equation}
\begin{array}{ll}
{\cal O}_{9\phantom{a}}                 \,= &\!
  \displaystyle{\frac{e^2}{16\pi^2}} \,
 (\bar{s} \gamma_\mu  P_L b)\, (\bar{l} \gamma^\mu l)\,, \\[2.0ex]
{\cal O}_{10}                           \,= &\! 
  \displaystyle{\frac{e^2}{16\pi^2}} \,
 (\bar{s} \gamma_\mu P_L b)\,  (\bar{l} \gamma^\mu \gamma_5 l)\,.    
\end{array}
\label{leptonop}
\end{equation}

\begin{figure}
\begin{center}
\epsfig{figure=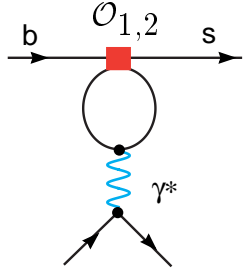,width=3.8cm}
\end{center}
\caption{Mixing of the operator  ${\cal O}_2$ into ${\cal O}_9$ at one loop.} 
\label{oneloopmixing}
\end{figure}

It turns out that 
the first large logarithm 
of the form $\log(m_b/M) \quad (M=m_W)$ already arises  without 
gluons,
because the operator ${\cal O}_2$ mixes into ${\cal O}_9$ at one loop 
via the diagram given in fig. \ref{oneloopmixing}.

This possibility, which has no equivalent  
in the $b \to s \gamma$ case, leads
to the following ordering of contributions to the decay amplitude 
(which should be compared with (\ref{LLQCD}) and (\ref{NLLQCD})):

\bea
&& \left[ \alpha_{em}\, \log(m_b/M) \right]\,
   \a_s^{n}(m_b)\, \log^n(m_b/M) \qquad [\mbox{LL}]~, \no \\  
&& \left[ \alpha_{em}\, \log(m_b/M) \right]\,
   \a_s^{n+1}(m_b)\, \log^n (m_b/M) \quad [\mbox{NLL}]~, \ldots
\eea

Technically, to perform the resummation, it is convenient
to transform these series into the standard form (\ref{LLQCD}).
This can be achieved by redefining magnetic, chromomagnetic and lepton-pair
operators  as follows~\cite{MM,BurasMuenz}:
\begin{equation}
\label{reshuffle}
{\cal O}_i^{new} = \frac{16\pi^2}{g_s^2} {{\cal O}}_i^{old}~, 
\quad C_i^{new} = \frac{g_s^2}{(4\pi)^2} {C}_i^{old}~, \quad \quad (i=7,...,10). \end{equation}

This redefinition
enables one to proceed in the standard way, or as in 
$b \to s \gamma$, in the three 
calculational steps discussed at the end of section \ref{strong}.
At the high scale, the $new$ Wilson coefficients can be computed at a
given order in perturbation theory and expanded in powers of 
$\alpha_s$:
\begin{equation} 
\label{Wilson}
C^{new}_i = C^{(0)}_i + \frac{\alpha_s}{(4 \pi)} C^{(1)}_i  
+ \frac{\alpha_s^2}{(4\pi)^2} C^{(2)}_i  + ...
\end{equation}
Obviously, the Wilson coefficients of the $new$ operators 
${\cal O}_{7-10}$ at the high scale start at order $\alpha_s$ only. 
Then the anomalous-dimension matrix has the canonical 
expansion in $\alpha_s$ and starts with a term
proportional to $\alpha_s$:
\begin{equation}
\label{dimmatrix}
{\gamma}^{new}
= \frac{\alpha_s}{4\pi} {\gamma}^{(0)}
+ \frac{\alpha_s^2}{(4\pi)^2} {\gamma}^{(1)}
+ \frac{\alpha_s^3}{(4\pi)^3} {\gamma}^{(2)} + ... 
\end{equation}
In particular,  after the reshufflings in (\ref{reshuffle}),
the one-loop mixing of the operator ${\cal O}_2$ 
with  ${\cal O}_9$ formally appears 
at order $\alpha_s$.

The last of the three steps, however, 
requires some care: among the $new$ operators with a non-vanishing 
tree-level matrix element, only  ${\cal O}_9$ has a non-vanishing 
coefficient at the LL level. Therefore, at this level, only the tree-level 
matrix element of this operator ($\langle {\cal O}_9 \rangle$)
has to be included. At NLL accuracy
the QCD one-loop contributions to 
$\langle {\cal O}_9 \rangle$, the tree-level contributions to 
$\langle {\cal O}_7 \rangle$ 
and $\langle {\cal O}_{10}\rangle$, and the electroweak 
one-loop matrix elements of the four-quark operators have to be 
calculated. Finally, at NNLL precision, one should  in principle 
take into account the QCD two-loop corrections 
to $\langle {\cal O}_9 \rangle$, the QCD one-loop corrections to
$\langle {\cal O}_7\rangle $ and  $\langle{\cal O}_{10}\rangle$, 
and the QCD corrections to the electroweak one-loop matrix elements 
of the four-quark operators. 
                       
\medskip 

The present status 
of these perturbative contributions to decay rate and 
FB asymmetry of $B \to X_s \ell^+\ell^-$ is the following: the 
complete NLL contributions to the decay amplitude
have been  found in \cite{MM,BurasMuenz}. 
Since the LL contribution to the rate turns out to be numerically 
rather small, NLL terms represent an $O(1)$ correction 
to this observable. On the other hand, since a non-vanishing 
FB asymmetry is generated by the interference of vector  
($\sim {\cal O}_{7,9}$) 
and axial-vector ($\sim {\cal O}_{10}$) leptonic currents,
the LL amplitude leads to a vanishing result and NLL terms represent 
the lowest non-trivial contribution to this observable.

In view of the forthcoming precise measurements at the $B$ factories, 
a computation of NNLL terms
in $B \to X_s \ell^+\ell^-$ is needed 
if one aims at the same numerical accuracy as achieved 
by the NLL analysis of $B \to X_s \gamma$. 
Large parts of the latter can be taken over 
and used in the NNLL calculation of
$B \to X_s \ell^+\ell^-$. But this is not the full story.

\begin{itemize} 
\item (Step 1) The full computation of initial conditions 
to NNLL precision was presented in Ref.~\cite{MISIAKBOBETH}.
The authors did the two-loop matching 
for all the operators relevant to $b \to s \ell^+\ell^-$
(including a confirmation of the $b \to s \gamma$ NLL matching 
results of \cite{Adel,GH}).
The inclusion of this  NNLL contribution removes the large
matching scale uncertainty (around $16 \%$) of the 
NLL calculation of the $b \to s \ell^+\ell^-$ 
decay rate.

\item (Step 2) Thanks to the reshufflings of the LL series,
most of the NNLL contributions to the anomalous-dimension matrix
can be derived from the NLL analysis of $b \to s \gamma$.
In particular the three-loop mixing of the four-quark operators 
${\cal O}_{1-6}$ into ${\cal O}_7$ and 
${\cal O}_8$ can be taken over from Ref.~\cite{Mikolaj}, which
allows an evaluation of the matrix element $U_{72}^{(2)}$  (using the 
usual convention  $C_i(\mu_b) = U_{ij} C_j (\mu_W)$). 
The only  missing piece for a full NNLL analysis of  
the $b \rightarrow s \ell^+\ell^-$
decay rate is the matrix element $U_{92}^{(2)}$ (see
fig. \ref{threeloopll}). 

In \cite{MISIAKBOBETH} an estimate was made, which suggests that the numerical 
influence of $U_{92}^{(2)}$ 
on the branching 
ratio of $b \rightarrow s \ell^+\ell^-$ is small.
Interestingly, since the FB asymmetry has no contributions 
proportional to $|\langle \cO_9 \rangle|^2$, 
this missing term is not needed for a NNLL analysis of this observable.

\item (Step 3) Within the $B \rightarrow X_s \gamma$ calculation
at NLL, the two-loop matrix elements of the four-quark operator ${\cal O}_2$
for an on-shell photon were calculated in \cite{GHW}, using Mellin--Barnes 
techniques. This calculation 
was extended in \cite{Asa1} to the case of an off-shell photon
(see fig. \ref{twoloopll})
\begin{figure}
\begin{center}
\epsfig{figure=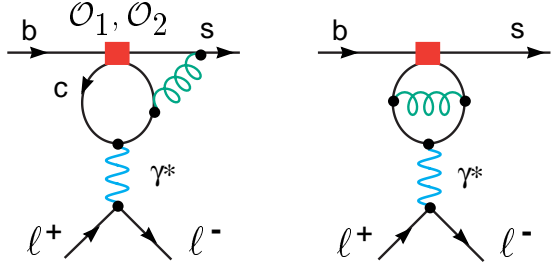,width=7.9cm}
\end{center}
\caption{Typical diagrams (finite parts) contributing to the matrix element 
of the operator ${\cal O}_2$ at the NNLL level (Step\,3).}
\label{twoloopll}
\end{figure}
with the help of a double Mellin--Barnes representation
which corresponds to a NNLL contribution relevant to the decay
$B \rightarrow X_s \ell^+\ell^-$.
The validity of these analytical results given in \cite{Asa1} 
is restricted to small
dilepton masses $q^2_{\ell^+ \ell^-}/m_b^2 < 0.25$.  
An independent 
numerical check of these results has been performed in \cite{Adrian2}. 
Moreover, the NNLL calculation in \cite{Adrian2}
is also valid for high dilepton masses for which 
the experimental methods have an efficiency much higher than
the one at low dilepton masses. 
Step 3 also includes 
the bremsstrahlung contributions which were calculated 
for the dilepton mass spectrum (symmetric part) in 
\cite{Asa3,Adrian1} and for the FB asymmetry in \cite{Adrian1,Asa2}.
In the low dilepton spectrum, 
these matrix element calculations reduce the error 
corresponding to the uncertainty of the low-scale dependence 
from $\pm 13\%$ down to $\pm 6.5\%$.

\begin{figure}
\begin{center}
\epsfig{figure=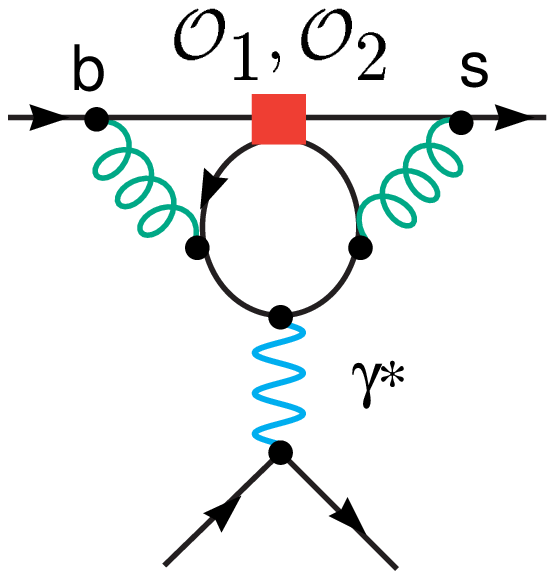,width=4.3cm}\, \epsfig{figure=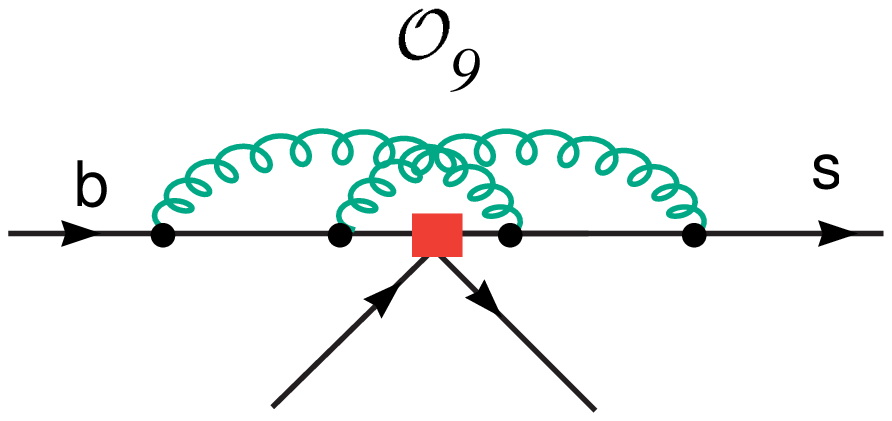,width=6.7cm}
\end{center}
\caption{Parts still  missing in a complete canonical NNLL analysis of the 
dilepton mass
 spectrum. Left: typical diagram (infinite part) contributing to the NNLL 
mixing $U_{92}^{(2)}$, Step 2. Right:  typical diagram (finite piece)
contributing to the NNLL matrix element of the operator ${\cal O}_9$, \mbox{Step 3}.}
\label{threeloopll}
\end{figure}
 
In principle, a complete NNLL calculation
of the $B \rightarrow X_s \ell^+\ell^-$ rate would require also 
the calculation of the renormalization-group-invariant 
two-loop matrix element of the operator ${\cal O}_9$ 
(see fig. \ref{threeloopll}).
But its impact to the dilepton mass spectrum is expected to 
be small. Similarly to the missing piece of 
the anomalous-dimension matrix, also this (scale-independent)
contribution does not enter the FB asymmetry at NNLL accuracy.

\end{itemize}

As anticipated, the canonical LL expansion is numerically not well 
justified, since the formally-leading $O(1/\alpha_s)$
term in $C_9$ is numerically  close to the  $O(1)$ term.  
For this reason,  it has been proposed in Ref.~\cite{Asa1} to use 
a different counting rule, where the $O(1/\alpha_s)$
term of $C_9$ is treated as $O(1)$.  
This approach, although it cannot be consistently 
extended at higher orders, seems to be well justified 
at the present status of the calculation. Within this 
approach, the two missing ingredients for a NNLL
calculation of the dilepton mass spectrum (see fig. \ref{threeloopll})
would be  of higher order.


\medskip

\begin{figure}
\begin{center}
\epsfig{figure=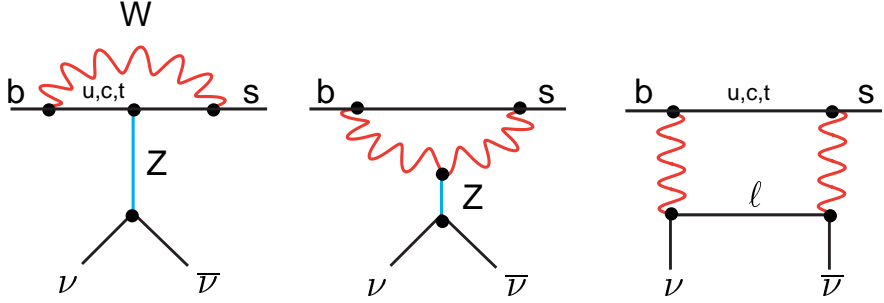,width=12cm}
\end{center}
\caption{One-loop contributions to the decay $B \rightarrow X_s \nu \bar \nu$.}
\label{oneloopnunu}
\end{figure}

The decay $B \rightarrow X_{s,d} \nu \bar \nu$
is induced by $Z^0$ penguin and box diagrams 
(see fig. \ref{oneloopnunu}). The main difference to the
semi-leptonic decay $B \rightarrow X_{s,d} \ell^+  \bar \ell^-$
is the absence of a photon penguin contribution.  
The latter implies 
only a logarithmic GIM suppression, while the former 
contributions  have a quadratic GIM suppression. 
As a consequence, the decay  $B \rightarrow X_{s,d} \nu \bar \nu$
is completely dominated by the internal top contribution. 

The effective Hamiltonian reads 
\begin{equation}
{H}_{eff}(B \rightarrow X_s \nu \bar \nu) = 
- \frac{4 G_F}{\sqrt{2}}  \frac{\alpha}{2 \pi \sin^2\Theta_W}\,  
 V_{tb} V^*_{ts}\,  C (m_t^2/ m_W^2)\,
(\bar{s} \gamma_{\mu} P_L b)(\bar{\nu} \gamma^{\mu} P_L \nu) +  h.c.
\end{equation}
For the decay $B \rightarrow X_d \nu \bar \nu$ the obvious changes have to be made. 

The hard (quadratic) GIM mechanism leads to $C (m_c^2 /m_W^2)/ C (m_t^2 /m_W^2) \approx O( 10^{-3})$. Moreover, the corresponding 
CKM factors in the top and the charm contribution are both of 
order $\lambda^2$. Therefore the charm contribution (and also the up quark contribution) can safely be neglected.   

The NLL QCD contributions to the partonic decay rate were 
presented in \cite{Buras93}. 
The perturbative error, namely the one due to the  renormalization scale, was 
reduced from $O(10 \%)$ at the LL level  to $ O(1 \%)$ at the NLL level. 

\newpage

\section{Non-perturbative contributions} 
\label{sectionnonpert}

\subsection{Inclusive decay rates of $B$ mesons}

In contrast to the exclusive rare $B$ decays, the inclusive ones
are theoretically clean observables and dominated by the 
partonic contributions. 
Bound-state effects of the final states are 
eliminated by averaging over a specific sum of hadronic states.  
Moreover, also long-distance effects 
of the inital state are accounted for, through the heavy mass 
expansion in which the inclusive decay 
rate of a heavy $B$ meson is calculated using
an expansion in inverse powers of the $b$ quark mass \cite{HME}. 

\medskip 

This heavy-mass expansion  is now a well-known method 
to  calculate the inclusive decay rates of a hadron
containing a heavy  quark, especially a $b$ quark.  
The optical theorem relates the {\it inclusive} decay rate of a hadron $H_b$ 
to the  imaginary part of certain forward scattering amplitudes
\begin{equation}
\Gamma (H_b \rightarrow X) = \frac{1}{2 m_{H_b}} \Im \, \langle 
H_b \mid {\bf T} \mid H_b \rangle\, ,  
\end{equation}
where the transition operator ${\bf T}$  is given by 
${\bf T} = i \int d^4 x \,  T [ H_{eff} (x) H_{eff} (0)]$.
The insertion of a complete set of states, $X\rangle\langle X$,
leads to the standard formula for the decay rate:
\begin{equation}
\Gamma (H_b \rightarrow X) = 
\frac{1}{2 m_{H_b}} \sum_X (2 \pi)^4 \delta^4 ( p_i
- p_f) \mid \langle X \mid H_{eff} \mid H_b  \rangle \mid^2 \, . 
\end{equation}

It is then  possible to construct an 
operator product expansion (OPE) of  the operator ${\bf T}$, which 
gets expressed as a series of {\it local} operators - suppressed by powers of 
the $b$ quark mass and written in terms of the $b$ quark field:
\be
T [ H_{eff}  H_{eff} ]  \stackrel{OPE}{=}  \frac{1}{m_b} \big( {\cal O}_0 + \frac{1}{m_b}
 {\cal O}_1 + \frac{1}{m_b^2} {\cal O}_2 + ... \big)\, . 
\ee
This construction is based on the parton--hadron duality
\cite{QHD}, using the facts that the sum is done over all exclusive final 
states and that the energy release in the decay is large with respect to  
the QCD scale, $m_b\ll\Lambda_{QCD}$.

\medskip

With the help of the HQET, namely the 
new heavy-quark spin-flavour symmetries arising in the heavy quark limit 
$m_b \rightarrow \infty$ \cite{HQET}, 
the hadronic matrix elements within the OPE, $\langle H_b \mid {\cal O}_i \mid H_b \rangle$, can 
be further simplified.  

\medskip 

The crucial  observations within this well-defined procedure, which are important for the 
application to the inclusive rare $B$ decays, are the following:
the free quark model   
turns out to be the first term in the constructed expansion 
in powers
 of $1/m_b$ and therefore the dominant contribution. 
This contribution can be calculated in perturbative QCD. 
 Second, in the applications to inclusive rare $B$ decays one finds 
no correction of order $1/m_b$ to the free quark model approximation, 
and the corrections to the partonic  decay rate start with $1/m_b^2$ only.
The latter  fact implies the rather small numerical impact of the non-perturbative 
corrections to the decay rate of inclusive modes.

\subsection{Non-perturbative corrections to  $B \rightarrow X_{s,d} \gamma$ and
$B \rightarrow X_s \ell^+ \ell^-$}

These techniques can directly be used in the decay $B \rightarrow X_s \gamma$,
in order to single out  non-perturbative corrections 
to the branching ratio. They are also applicable to the case
of $B \rightarrow X_d \gamma$ and, with some modifications, 
to the case of $B \rightarrow X_s \ell^+ \ell^-$.

\medskip

\begin{figure} 
\centerline{
\psfig{figure=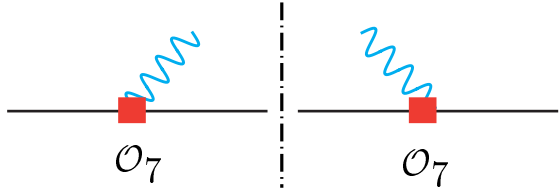,height=2.8cm}} 
\caption[]{Relevant cut diagram for the $({\cal O}_7,{\cal O}_7)$ contribution.}
\label{0707}
\end{figure}

If one neglects perturbative QCD corrections and assumes that the decay  
$B \to X_s \gamma$ is due to the operator ${\cal O}_7$ only,
then one has to consider  the time-ordered product
$T {\cal O}_7^+(x) \, {\cal O}_7(0)$ (see fig. \ref{0707}).
Using the OPE for $T {\cal O}_7^+(x) \, {\cal O}_7(0)$
and heavy quark effective theory methods as discussed above, the decay width
$\Gamma(B \to X_s \gamma)$ reads \cite{Falk,Alineu} 
(modulo higher terms in the $1/m_b$ expansion):
\bea
\label{width}
\Gamma_{B \to X_s \gamma}^{({\cal O}_7,{\cal O}_7)} &=&
\frac{\a G_F^2 m_b^5}{32 \pi^4} \, |V_{tb} V_{ts}|^2 \, C_7^2(m_b) \,
\left( 1 + \frac{\delta^{NP}_{rad}}{m_b^2} \right) \quad , \nonumber \\
\delta^{NP}_{rad} &=& \frac{1}{2} \lambda_1 - \frac{9}{2} \lambda_2 \quad ,
\eea
where $\lambda_1$ and $\lambda_2$ are the HQET parameters 
for the kinetic and the chromomagnetic
energy.  Using $\lambda_1=-0.5 \, \mbox{GeV}^2$ and 
$\lambda_2=0.12 \, 
\mbox{GeV}^2$, one gets $\delta_{rad}^{NP} \simeq -3\%$.

\medskip

The $B \rightarrow X_s \gamma$ decay width    is usually 
normalized by the semi-leptonic one.
The semi-leptonic decay width gets $1/m_b^2$ corrections,
which are also negative: 
\be
\delta^{NP}_{semilept} = \frac{1}{2} \lambda_1 - \big( \frac{3}{2} - \frac{6 (1-z)^4}{g(z)}  \big) \lambda_2, 
\ee
where $g(z)$ is a phase-space factor and $z$ the ratio 
$m_{c,pole}^2/m_{b,pole}^2$.

The non-perturbative corrections scaling with $1/m^2_b$ tend to cancel
in the branching ratio 
${\cal B}(B \to X_s \gamma)/{\cal B}(B \to X_c \ell \nu)$, and only about $1\%$
remains: 
the HQET parameter $\lambda_1$ cancels out in the ratio and 
the HQET parameter $\lambda_2$ is known from $B^*-B$ mass splitting, 
\be 
\lambda_2 = \frac{1}{4} ( m_{B^*}^2 - m^2_{B} ).
\ee

\medskip

\begin{figure} 
\centerline{\psfig{figure=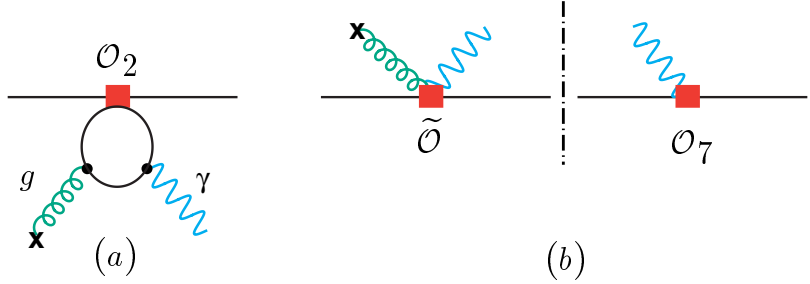,height=4.0cm}} 
\caption[]{a) Feynman diagram from which the operator $\tilde{\cal O}$
arises. b) Relevant cut diagram for the $(\tilde{\cal O},{\cal O}_7)$ interference.
\label{Voloshinfig}}
\end{figure}

Voloshin \cite{Voloshin} considered the non-perturbative
effects that arise  when including also the operator ${\cal O}_2$. 
This effect is generated by the diagram in fig. \ref{Voloshinfig}a 
(and by the one, not shown, where the gluon and the photon
are interchanged); $g$ is a soft gluon interacting with the
charm quarks in the loop. Up to a characteristic Lorentz structure,
this loop is given by the integral
\be
\label{volloop} 
\int_0^1 dx \, \int_0^{1-x} dy \, \frac{xy}{m_c^2-k_g^2 x(1-x) -2xy k_g k_\g}
\quad .
\ee
As the gluon is soft, i.e. $k_g^2,k_g k_\g \approx \Lambda^{QCD} \, m_b/2
\ll m_c^2$, the integral can be expanded in $k_g$. The (formally)
leading operator, denoted by $\tilde{\cal O}$, is
\be
\label{tildeo}
\tilde{\cal O} = \frac{G_F}{\sqrt{2}} V_{cb} V_{cs}^* C_2 \,
\frac{e Q_c}{48 \pi^2 m_c^2} \, \bar{s} \g_\mu (1-\g_5) g_s 
G_{\nu \lambda} b \, \epsilon^{\mu \nu \rho \sigma} \partial^\lambda
F_{\rho \sigma} \quad .
\label{extraop}
\ee 
Then working out the cut diagram shown in fig. \ref{Voloshinfig}b,
one obtains the non-perturbative 
contribution $\Gamma^{(\tilde{\cal O},{\cal O}_7)}_{B \to X_s \gamma}$
to the decay width,
which is due to the $({\cal O}_2,{\cal O}_7)$ interference.
Normalizing this contribution by the LL partonic width, one 
obtains \cite{Rey}
\be
\label{voleffect}
\frac{\Gamma^{(\tilde{\cal O},{\cal O}_7)}_{B \to X_s \gamma
}}{\Gamma_{b \to s \g}^{LL}} = -\frac{1}{9} \, \frac{C_2}{C_7} 
\frac{\l_2}{m_c^2} \simeq +0.03 \quad .
\ee

As the expansion parameter is $m_b \Lambda_{QCD}/m_c^2 \approx 0.6$
(rather than $\Lambda^2_{QCD}/m_c^2$), it is not a priori clear
whether, formally, higher order terms in the $1/m_c$ expansion are
numerically suppressed. More detailed investigations 
\cite{Wise,Peccei,Rey}
have shown that higher order terms are indeed suppressed, because
the corresponding expansion coefficients are small.

\medskip

The analogous $1/m_c^2$ effect 
has been found independently in the exclusive mode
$B \to K^* \gamma$ in ref. \cite{Wyler}. 
Numerically, the effect
there is also at the few percent level.

\medskip

As was emphasized by Misiak \cite{Misiak:1999jh}, 
an analogous  systematic analysis  of terms like 
$\Gamma^{{({\cal O}_2,{\cal O}_2)}}_{B \to X_s \gamma}$ at  
first order in $\alpha_s(m_b)$ is still missing. 
Rigorous  techniques such as OPEs do not seem to be applicable in this case. 

\medskip

The analysis of the $1/m_b^2$ and  $1/m_c^2$ effects was 
extended to the decay $B \rightarrow X_s \ell^+ \ell^-$ in \cite{Falk,Alineu,Rey,Savagenew,buchallanewnew,Bauer}. They  can be calculated 
quite analogously to  those in the decay $B \rightarrow X_s \gamma$. 
As was first noticed in \cite{Alineu}, 
the relative $\Lambda_{QCD}^2/m_b^2$ correction diverges in the 
high-$q_{\ell^+\ell^-}$ endpoint, which indicates the breakdown of the 
heavy mass expansion.

There are also  on-shell $c\bar{c}$ resonances,
which have to be taken into account.
While in the decay $B \rightarrow X_s \gamma$ (on-shell photon) 
the intermediate $\psi$ background for example, namely 
${B} \to \psi X_s$ followed by $\psi \to X' \gamma$, is 
suppressed for a high energy cut $E_\gamma$ and can be subtracted 
from the $B \rightarrow X_s \gamma$ decay rate
(see section \ref{experimental}),
the $c\bar{c}$ resonances show up as large peaks in the dilepton 
invariant mass spectrum 
in the decay $B \rightarrow X_s \ell^+ \ell^-$ (off-shell photon) 
\cite{Lim}.   
These resonances can be removed by appropriate  kinematic cuts
in the invariant mass spectrum.

\medskip

Also the non-perturbative contributions 
in the decay  $B \rightarrow X_d \gamma$ can be treated 
in analogy to the ones in the decay $B \rightarrow X_s \gamma$.
The power corrections in $1/m_b^2$ and $1/m_c^2$ (besides the CKM factors)
are the same for the two modes.
But the long-distance 
contributions
from the intermediate $u$-quark in the penguin loops are different.
These are  suppressed in the $B \rightarrow X_s \gamma$ mode by  the 
unfavourable CKM matrix elements. In $B \rightarrow X_d \gamma$, there 
is  no CKM suppression  and one
has to include the long-distance intermediate $u$-quark contributions.

Naively, one could expect that these contributions
from up-quark loops scale with $1/m_u^2$. However, 
following  the approach of \cite{Rey}, one easily shows that 
the general vertex function   
cannot in that case be expanded in the parameter $t = k \cdot q/m_u^2$ 
(where
$k$ and $q$ are the gluon and photon momentum respectively).  
But the expansion in inverse powers of $t$ is reasonable. 
The leading term in this expansion scales like $t^{-1} \sim m_u^2/k_g k_\gamma$ and therefore cancels the factor $1/m_u^2$ 
in the prefactor
(see the analogous  $1/m_c^2$ factor in (\ref{tildeo}))
and one gets  a  suppression factor $(\Lambda_{QCD}^2/m_u^2) \cdot 
(m_u^2/{k_g k_\gamma})$. Thus, although 
the  expansion in inverse
powers in $t$ induces  non-local operators, one explicitly finds 
that the leading term scales with $\Lambda_{QCD} / m_b$. This indicates no 
large long-distance intermediate $u$-quark contributions.

Model calculations, based on vector meson
dominance, also suggest this conclusion \cite{LDUP}. 
Furthermore, estimates of the long-distance constributions in
exclusive decays $B \rightarrow \rho \gamma$ and $ B \rightarrow 
\omega \gamma$ in the light-cone sum rule approach do not exceed 
15\% \cite{stollsum}.

Finally, it must be stressed that
there is no spurious enhancement  of the form $\log (m_u/\mu_b)$ 
in the perturbative contribution to the matrix elements
of the four-quark operators, 
as shown by the explicit calculation in \cite{GHW} (see 
also \cite{STERMAN}). In other words, the limit $m_u
\to 0$ can be taken.

All these observations exclude very large long-distance intermediate 
$u$-quark contributions in the decay $B \rightarrow X_d \gamma$.

\newpage

\section{Phenomenology}

\subsection{SM prediction of $B \rightarrow X_s \gamma$}
\label{phenobsg}

The theoretical prediction for the partonic $b \rightarrow X_s^{parton} \gamma$ decay 
rate  is usually 
normalized by the semi-leptonic decay rate in order to get rid of uncertainties 
related to the CKM matrix elements and the fifth power of the $b$ quark mass.
Moreover,  an explicit lower cut on the photon energy in the
bremsstrahlung correction has to be made: 
\be
R_{quark}(\delta) = 
\frac{\Gamma[ b \to s \gamma]+\Gamma[ b \to s \gamma  gluon]_\delta}{\Gamma[ b \to X_c e \bar{\nu}_e ]}, 
\ee
where the subscript $\delta$ means that only photons with energy 
$E_{\gamma}>(1-\delta) E_{\gamma}^{max} = (1-\delta) \frac{m_b}{2}$ are counted. 
The ratio $R_{quark}$ is divergent in the limit $\delta \rightarrow 1$, 
owing  to the 
 soft photon divergence in the subprocess $b \rightarrow s \gamma gluon$. 
In this limit only the sum of $\Gamma[b \to s \gamma]$,
$\Gamma[b \to s  gluon]$ and $\Gamma[b \to s \gamma gluon]$ is a reasonable 
physical quantity, in which all divergences cancel out. 
In \cite{Kagan} it was shown that 
the theoretical result is
rather sensitive to the unphysical soft-photon divergence;
the choice $\delta=0.90$ was suggested as the definition 
of the `total' decay rate. 

\medskip

It is suggestive to  give up the concept of a 
`total' decay rate of $b \rightarrow s \gamma$ 
and compare theory and experiment 
using the same energy cut. Then also  
the theoretical uncertainty  regarding the photon energy spectrum mentioned 
in  section  \ref{photonspectrum}   
would occur naturally in the theoretical prediction. 

\medskip

Using the measured semi-leptonic branching ratio ${\cal B}^{sl}_{exp.}$,
the  branching ratio ${\cal B}(B \to X_s \gamma)$ is given by 
\be
{\cal B}(B \to X_s \gamma) = R_{quark}  \times {\cal B}^{sl}_{exp.} (1 + \Delta_{nonpert}),
\ee
where the non-perturbative corrections scaling with $1/m^2_b$ and 
$1/m^2_c$, summed in $\Delta_{nonpert}$ (see section \ref{sectionnonpert}),
have a numerical effect of $+1\%$ and 
$+3\%$,  
respectively, on the branching ratio only.
For a comparison with the ALEPH measurement (\ref{braleph}),  
the measured semi-leptonic branching ratio ${\cal B}(H_b \to X_{c,u} \ell 
\nu)$ should be used consistently.
This leads to a slightly larger theoretical prediction for the LEP experiments.

\medskip

Including only the resummed QED corrections and the non-perturbative
corrections discussed in section \ref{sectionnonpert},
and using the on-shell value of 
$\alpha_{em}$ and the charm pole mass,  
one ends up with  the following theoretical prediction for the
$B \rightarrow  X_s \gamma$ branching ratio \cite{greubhurthQCD}: 
\beq
{\cal B}(B \rightarrow X_s \gamma) = (3.32 \pm 0.30)\times 10^{-4},
\label{currentprediction}
\eeq
where the error has two sources, the uncertainty 
regarding the scale dependences and the uncertainty due to 
the input parameters. In the latter 
the uncertainty due to the parameter $m_c/m_b$ is dominant. 
This prediction almost coincides with the prediction of Kagan and 
Neubert \cite{Kagan}.

\medskip 

In \cite{GambinoMisiak} two important observations were made. First it 
was shown that the charm-loop contribution to 
the decay $B \rightarrow  X_s \gamma$ 
is numerically dominant, and very stable under logarithmic
QCD corrections,  and that  the strong enhancement of the branching ratio 
by QCD logarithms is mainly due to the $b$-quark mass evolution in the
top-quark sector. This leads to a better control over the residual 
scale dependence at NLL.

\begin{figure}
\begin{center}
\epsfig{figure=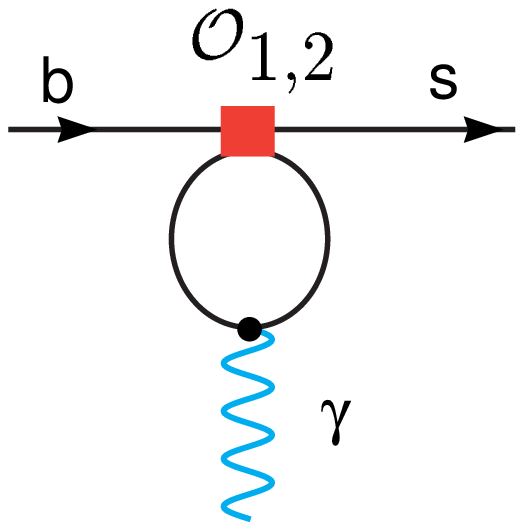,width=4.4cm}\epsfig{figure=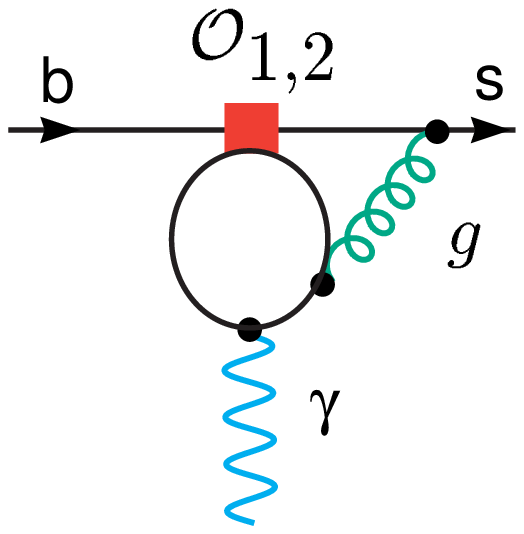,width=4.5cm}
\end{center}
\caption[]{The one-loop matrix element    of the four-quark operators ${\cal O}_{1,2}$ vanishes  (left), therefore the charm dependence starts with the 
two-loop (NLL) contribution (right).}
\label{mcharmbsg}
\end{figure}

Secondly, quark mass effects within the decay 
$B \rightarrow X_s \gamma$ were further analysed,
in particular the definitions of the quark masses $m_c$ and $m_b$ 
in the two-loop  matrix element of the four-quark operators ${\cal O}_{1,2}$
(see fig. \ref{mcharmbsg}).
 Since  the charm quark in the matrix element $\bra{{\cal O}}_{1,2} \ket$ are 
dominantly off-shell, it is argued 
that the running charm mass should be chosen instead
of the pole mass.                 
The latter choice was used  
in all previous analyses \cite{GHW,Mikolaj,CDGG,Kagan,greubhurthQCD}.  
\begin{equation}
 m_c^{pole}/m_b^{pole} \qquad 
{{\Rightarrow}}\, \qquad   
m_c^{\overline{MS}}(\mu)/m_b^{pole}, \,  \,\, \mu \in [m_c,m_b].
\end{equation}
Numerically, the shift from $m_c^{pole}/m_b^{pole} =  0.29 \pm 0.02$
to $m_c^{\overline{MS}}(\mu)/m_b^{pole} = 0.22 \pm 0.04$
is rather important and leads to a $+ 11 \%$ shift of the central 
value of the $B \rightarrow X_s \gamma$ branching ratio.  
The error in the charm mass within the $\overline{MS}$ scheme, 
is due to the uncertainty resulting from the scale variation and  
due to the uncertainty in $m_c^{\overline{MS}}(m_c^{\overline{MS}})$. 
With their new choice of the charm mass renormalization scheme 
and with $\delta = 0.9$, their theoretical prediction 
for the `total' branching ratio  is\footnote{Actually,  
  the theoretical prediction in \cite{GambinoMisiak}
is given  for the 
energy cut $E\gamma = 1.6$ GeV:\\ 
${\cal B}(B \to X_s \gamma)_{E\gamma > 1.6 GeV} = (3.60 \pm 0.30) \times 10^{-4}$. The theoretical error in (\ref{totalbr}) might be larger 
 due to nonperturbative corrections (see section \ref{inclusivesection}).}

\begin{equation} 
\label{totalbr}
{\cal B}(B \to X_s \gamma) = (3.73 \pm 0.30) \times 10^{-4}.
\end{equation}

Since the  matrix element starts at NLL order (see fig. \ref{mcharmbsg}) 
and, thus, 
the renormalization scheme for $m_c$ is an NNLL issue,
one should regard the choice of  the $\overline{MS}$ scheme 
as an educated guess of the NNLL
corrections. Nevertheless, the new choice
is guided by the experience gained from many higher order calculations
in perturbation theory.
Moreover, the  $\overline{MS}$ mass of the charm quark is also a 
short-distance quantity which does not suffer from non-perturbative 
ambiguities,  in contrast to its  pole mass. 
Therefore the central value 
resulting within this scheme, \ref{totalbr}, is definitely favoured
and should be regarded as the present theoretical prediction. 

One should note that the scheme 
ambiguity regarding the $b$ quark mass is under control.

Because the choice of the renormalization scheme 
for $m_c$ is a NNLL effect, 
one has to  emphasize that regarding the charm mass scheme 
the theoretical prediction (\ref{currentprediction}) using the
pole mass scheme 
is in principle as good as the new prediction (\ref{totalbr}) using 
the $\overline{MS}$ scheme; both predictions do not take into 
account the full impact of charm mass scheme ambiguity.
Thus, one has to argue for a  larger theoretical uncertainty  
in $m_c^{\overline{MS}}(\mu)/m_b^{pole}$,  
which includes also the value of $m_c^{pole}$.
This leads to a more appropriate error above $10 \%$ 
in \ref{totalbr}. 

\medskip 

One should emphasize that this present dominant uncertainty is due 
to a charm mass scheme ambiguity at the NLL level, i.e.   
to the question if the 
$\overline{MS}$ or the pole mass is more appropriate to estimate the 
unknown NNLL contributions. This uncertainty 
is of perturbative origin. The second uncertainty is due to the 
numerical value of $m_c$ within the specific choice of the 
charm mass scheme and is a parametrical uncertainty.  
The Particle Data Review in its latest edition \cite{PDG2002}
gives the following range for the latter uncertainty within the 
$\overline{MS}$ scheme: $ 1.0\,  \mbox{GeV}  \leq  
m_c^{\overline{MS}}(m_c^{\overline{MS}}) \leq 1.4 \, \mbox{GeV}$ 
and has therefore doubled
the uncertainty with respect to the one quoted previously  \cite{PDG2000}.
Nevertheless, there are determinations using the sum rule technique
\cite{summc} and also one that use the lattice technique (in the
quenched approximation)  \cite{sint}
which indicate much smaller uncertainties.

\medskip 

The only way to resolve this charm mass scheme ambiguity 
within the  present theoretical NLL prediction and  to
reduce the uncertainty below $10 \%$, would be a 
NNLL calculation,  which is  {\it not} beyond the power
of present technical tools in perturbation theory.   
Such an ambitious program makes sense only if one is 
sure that the associated  non-perturbative effects are 
under control. It was argued \cite{radcor} that the 
numerical behaviour of the branching ratio of  
$B \rightarrow X_s \gamma$ as a function of $m_c$ suggests 
that the dominant charm mass dependence originates from 
distances much smaller than $\Lambda_{QCD}$. In such a case,
the associated non-perturbative effects would be under control 
and negligible.

\subsection{CKM phenomenology with  $B \rightarrow X_{s,d} \gamma$} 
\label{phenoCKM}

Instead of making a theoretical prediction for the branching ratio
${\cal B}(B \to X_s \gamma)$,
one can use the experimental data  and theory in order to determine
the combination $|V_{tb} V_{ts}^*|/|V_{cb}|$
of CKM matrix elements; in turn, one can determine 
$|V_{ts}|$ by making use of the relatively well  known
 CKM matrix elements $V_{cb}$ and $V_{tb}$. 
But having used the unitarity constraint in the theory prediction
already, the $B \rightarrow X_s \gamma$ constraint  on $V_{ts}$ does not add 
 much  to what is already  known from the unitarity fit
\cite{aliokt97,greubhurthQCD}.
If one does not use the CKM unitarity,  the sensitivity of $B \rightarrow
X_s \gamma$ to  $V_{ts}$ gets significantly reduced, because the charm
quark contribution is twice as large as the top quark contribution.

\medskip

A  future measurement of 
the $B \rightarrow X_d \gamma$ decay rate will  help 
to reduce the currently allowed region
of the CKM Wolfenstein parameters $\rho$ and $\eta$.
It is also of specific interest with respect
to new physics, because its CKM 
suppression by the factor $|V_{td}|^2/|V_{ts}|^2$ in the SM may 
not be true in extended models.

Most of the theoretical improvements carried out in the
context of the decay $B \rightarrow X_s \gamma$ can 
straightforwardly be adapted
for the decay $B \rightarrow X_d \gamma$ and, thus,   
the NLL-improved and power-corrected decay rate for 
$B \rightarrow X_d \gamma$ has
much reduced theoretical uncertainty, which would allow an
 extraction  of  the CKM parameters from the 
measured branching ratio.

The predictions for the 
$ B \rightarrow X_d \gamma$ decay given in \cite{AG7}
show that, for  the central values of the input
 parameters, the difference between the LL and NLL results is 
$\sim 10\%$, increasing the branching ratio in the NLL case.
Of particular theoretical interest is the ratio of the
branching ratios, defined as
\begin{equation}
\label{dsgamma}
R(d\gamma/s\gamma) \equiv \frac{{\cal B}(B \to X_d \gamma)}
                           {{\cal B}(B \to X_s \gamma)},
\end{equation}
in which a good part of the theoretical uncertainties cancels. 
This suggests that 
a future  measurement of $R(d\gamma/s\gamma)$ will have a large impact on
the CKM phenomenology:
varying the CKM Wolfenstein parameters $\rho$ and $\eta$ in the ranges
$-0.1 \leq \rho \leq 0.4$ and $0.2 \leq \eta \leq 0.46$ and taking into
account other parametric dependences stated above, the 
results (without electroweak corrections) are
\begin{eqnarray}
\label{summarybrasy}
6.0 \times 10^{-6} &\leq &
 {\cal B}(B \rightarrow X_d \gamma)   \leq 2.6 \times 10^{-5}~, \nonumber\\
0.017 &\leq & R(d\gamma/s\gamma) \leq 0.074~.\nonumber
\end{eqnarray}
In these predictions \cite{AG7}
it is assumed that the long-distance intermediate $u$-quark contributions 
play only a subdominant role (see the discussion at the end of section
\ref{sectionnonpert}). 

\medskip

As mentioned in section \ref{photonspectrum}, the photon spectrum of 
$B \rightarrow X_s \gamma$ plays an important role in the determination 
of the CKM matrix element $V_{ub}$.

\subsection{Role of $b \rightarrow s \,  {gluon}$ for  $B \rightarrow X_{no\, charm}$}

Some remarks on the decay mode  $b \rightarrow s \, gluon$ are in order. 
The effective Hamiltonian in the decay mode $b \rightarrow s \, gluon$
coincides with the one   in the decay
$b \rightarrow s \gamma$. By replacing the photon
by the gluon, 
the NLL QCD calculation of $b \rightarrow s \gamma$  
can also be used.
But in the calculation of 
the matrix element  of the operator ${\cal O}_2$,  further diagrams 
with the nonabelian three-gluon coupling had to be calculated
\cite{Liniger}.
The NLL calculation \cite{Liniger} 
leads to 
${\cal B}(b \rightarrow s \, gluon)=(5.0 \pm 1.0) \times 10^{-3}$, which is 
a factor of more than 2 
larger than the former LL result 
${\cal B}(b \rightarrow s \, gluon)=(2.2 \pm 0.8) \times 10^{-3}$ 
\cite{counterterm}.
The mode $b \rightarrow s \, gluon$
represents one component of the  
inclusive charmless hadronic decays, $B \to X_{no charm}$, 
where $X_{no charm}$ denotes any hadronic 
charmless final state. 
The corresponding branching ratio allows for  the extraction of
the ratio $|V_{ub}/V_{cb}|$ \cite{LenzNierste}. 
At the quark level, there are decay modes with three-body final states,
$b \to q' \overline{q}' q$ ($q'=u,d,s$; $q=d,s$) and the modes
$b \to q g$, with two-body final-state topology.
The component $b \to s g$
of the charmless hadronic decays is expected to manifest itself in 
kaons with high momenta (of order $m_b/2$), owing to its two-body nature
\cite{Rathsman}. 
The impact of the NLL corrections in $b \rightarrow s g$ 
on the inclusive charmless hadronic decays, $B \to X_{no charm}$, 
turns out to be  as big 
as the NLL corrections to the $b$ quark decay modes with three quarks 
\cite{Liniger}.

There is still 
only marginal overlap between theory and experiment for the 
inclusive semi-leptonic branching ratio and the charm multiplicity 
in $B$ meson decays \cite{Golutvin},  
if usual scale variations are used in the 
theoretical predictions \cite{Liniger}.
A possible reinforcement of the decay $b \rightarrow s g$ 
due to new physics through the chromomagnetic (${\cal O}_8$)
contribution would 
lead to a natural explanation of these effects \cite{Rathsman}. There is
still room for such a scenario,  which would be also compatible 
with the present $B \rightarrow X_s \gamma$ constraint \cite{NEWNEW}.

\subsection{Phenomenology  of  $B \rightarrow X_s \ell^+ \ell^-$}
\label{bsllpheno}

\begin{figure}
\begin{center}
\vspace{-2.5cm}
\hspace{-3cm}\epsfig{figure=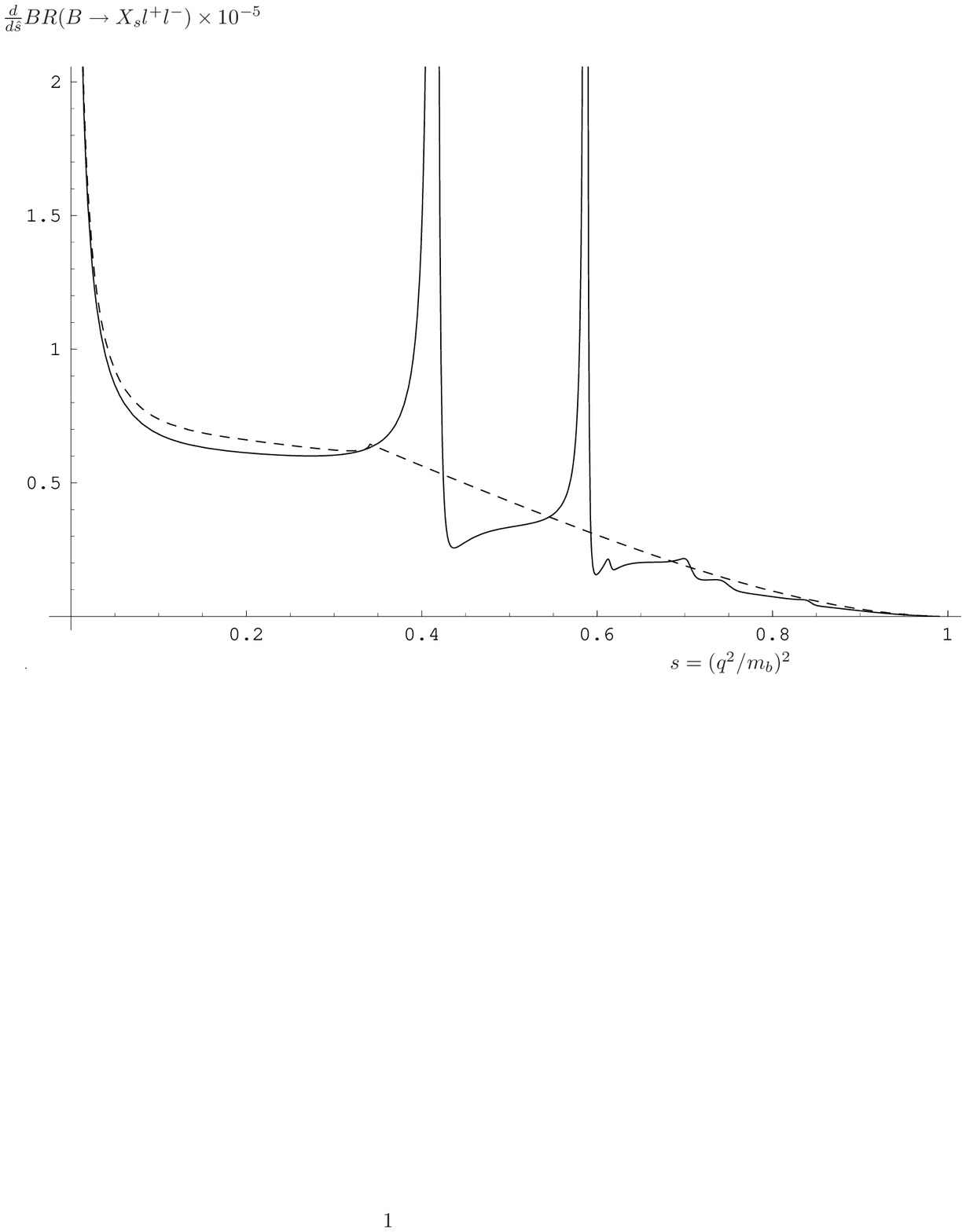,width=17.2cm}
\end{center}
\vspace{-10cm}
\caption{Schematic dilepton mass spectrum of $B \rightarrow X_s \ell^+ \ell^-$,
the dashed line corresponds to the perturbative contribution.}
\label{dileptonspectrum}
\end{figure}

In comparison with the 
$B \rightarrow X_s \gamma$, the inclusive $ B \rightarrow X_s \ell^+\ell^-$ 
decay presents a complementary and also more complex test of the SM. 
As mentioned above, also this decay  is  dominated 
by perturbative  contributions if the  $c \bar c$ resonances 
that show up as large peaks in the dilepton invariant mass spectrum 
(see fig. \ref{dileptonspectrum} \cite{courtesymikolaj}) are removed 
by appropriate  kinematic cuts. In the 'perturbative windows', namely  
in the low-$s$ region $ 0.05 <  s = q^2 / m_b^2 < 0.25 $ and 
also in the high-$s$ region with $ 0.65 < s < s_{max}$ (for $s_{max}$ see 
section  \ref{sectionnonpert}),
theoretical predictions for the invariant mass spectrum
are dominated by the purely perturbative contributions, 
and a theoretical precision comparable with  the one reached  
in the decay $B \rightarrow X_s \gamma$ is in principle possible. 
Regarding the choice of precise cuts in the dilepton mass 
spectrum, it is important that  one directly compares theory and  
experiment using the same energy cuts and avoids any kind of
extrapolation.

\medskip

In the high-$s$ region, one should encounter the breakdown of the 
heavy mass expansion at the endpoint (see section \ref{sectionnonpert}). 
This fact leads to sizeable $\Lambda^2_{QCD}/m_b^2$ non-perturbative 
corrections in this region. In \cite{Bauer} rather large
Wilson coefficients  to order 
$\Lambda_{QCD}^3/m_b^3$ were found. The latter can be used to give
an estimate of these corrections while the corresponding matrix
elements are unknown. 
Following an argument in \cite{Neubertold}, one can show that 
this $\Lambda_{QCD}/m_b$ expansion is effectively a
$\Lambda_{QCD}/m_c$ one in the high-$s$ region.

\medskip

The  kinematic observables,  
the invariant dilepton mass spectrum and the forward--backward 
(FB) asymmetry  are  usually normalized by the semi-leptonic
decay rate in order to reduce the uncertainties due 
to bottom quark mass and CKM angles. 
The normalized dilepton invariant mass spectrum 
and the FB asymmetry 
are defined as
\begin{equation}\label{decayamplitude}
R(s)= \frac{d}{d s}\Gamma(  B\to X_s\ell^+\ell^-) /
 \Gamma( B\to X_ce\bar{\nu}),
\end{equation}
\vspace{-0.3cm}
\begin{eqnarray}\label{forwardbackward}
A_{FB}(s)= \frac{1}{\Gamma( B\to X_ce\bar{\nu})} 
\times \int_{-1}^1 d\cos\theta_\ell ~
 \frac{d^2 \Gamma( B\to X_s \ell^+\ell^-)}{d s ~ d\cos\theta_\ell}
\mbox{sgn}(\cos\theta_\ell), 
\end{eqnarray}
where $\theta_\ell$ is the angle between $\ell^+$ and $B$ momenta 
in the dilepton centre-of-mass frame~\footnote{
The so-called `normalized' FB asymmetry, which is also often used, is given by \\$\overline{A}_{FB}(s)= 
\int_{-1}^1 d\cos\theta_\ell ~
 \frac{d^2 \Gamma( B\to X_s \ell^+\ell^-)}{d s ~ d\cos\theta_\ell}
\mbox{sgn}(\cos\theta_\ell) / \int_{-1}^1 d\cos\theta_\ell ~
 \frac{d^2 \Gamma( B\to X_s \ell^+\ell^-)}{d s ~ d\cos\theta_\ell}$}

\medskip

It was  shown in \cite{Alineu}
that $A_{FB}(s)$ is  equivalent to the energy asymmetry introduced in 
\cite{ChoMisiak}:
\begin{equation}
A_{energy}\, = \, \frac{N(E_{\ell^-} > E_{\ell^+}) - N(E_{\ell^+} > E_{\ell^-})}{N(E_{\ell^-} > E_{\ell^+}) + N(E_{\ell^+} > E_{\ell^-})}
\end{equation}
where $N(E_{\ell^-} > E_{\ell^+})$ denotes the number of lepton pairs whose
negatively charged member is more energetic in the $B$ meson rest frame than
its positive partner. 
From the experimental point of view, 
there is no significant  difference either between
measuring the inherent asymmetry  using the lepton energy 
distribution or using the
three-momentum information. 

\medskip

For the low-$s$ region the present partonic NNLL prediction is given in 
\cite{Asa3,Adrian2}:
\begin{equation}
\int_{0.05}^{0.25}  \, d \hat{s} \,  R^{\ell^+ \ell^-}_{quark}(\hat{s}) 
\, = \, (1.28 \pm 0.08_{scale} \,)\, \times  10^{-5} 
\label{partonicfinal}
\end{equation}
The error quoted in (\ref{partonicfinal}) 
reflects only the renormalization  scale uncertainty and is
purely perturbative.  
There is no additional problem due to the charm mass renormalization
scheme ambiguity within the decay $B \rightarrow X_s \ell^+ \ell^-$ 
because the charm dependence starts already at one loop, in contrast 
to the case of the decay $B \rightarrow X_s \gamma$ (see fig. 
\ref{oneloopmixing}). 
The charm dependence itself leads to an additional uncertainty
of $\sim 7.6 \%$ within the partonic quantity (\ref{partonicfinal}), 
if the pole mass is varied,  $m_c^{pole}/m_b^{pole} = 0.29 \pm 0.02$.

\medskip

As discussed in section \ref{NNLLQCDbsll}, the impact of the NNLL contributions
is significant. The large matching scale $\mu_W$ uncertainty of $16 \%$ 
of the NLL result was removed; the low-scale uncertainty $\mu_b$ 
of $13 \%$ was cut in half; and also the central value of the integrated
low dilepton spectrum (\ref{partonicfinal})   
was significantly changed by $\approx - 14 \%$ due to NNLL corrections.

\medskip 

These small uncertainties in the inclusive mode should be compared 
with the ones of the corresponding exclusive mode 
$B \rightarrow K^* \mu^+ \mu^-$  given in \cite{BallAli};
$\Delta BR 
= (^{+26}_{-17}, \pm 6, ^{+6}_{-4}, ^{-0.7}_{+0.4}, 
\pm 2)\% $.  The first dominant error represents the hadronic 
uncertainty due to the 
form factors.

\medskip

Using the measured semi-leptonic branching ratio ${\cal B}^{sl}_{exp.}$,
the prediction for the corresponding  branching ratio 
is given by 
\begin{eqnarray}
&& {\cal B}(B \rightarrow X_s \ell^+ \ell^-)_{Cut:\,\,\hat{s}  
\in [0.05,0.25]} = \nonumber\\
&=&\, {\cal  B}^{sl}_{exp.} \, \int_{0.05}^{0.25}  \, d \hat{s} \, \big[ R^{\ell^+ \ell^-}_{quark}(\hat{s}) + \delta_{1/m_b^2} R(\hat{s}) +
\delta_{1/m_c^2} R(\hat{s})  \big]
\nonumber\\
&=&\, 0.104 \, [ (1.27 \pm 0.08_{scale} \,) + 0.06 - 0.02] \times 10^{-5} \nonumber\\&=&\, ( 1.36
  \pm 0.08_{scale} \,) \times  10^{-6}.
\label{finalll}
\end{eqnarray}
$\delta_{1/m_b^2} R(\hat{s})$ and $\delta_{1/m_b^2} R(\hat{s})$ 
are the non-perturbative contributions discussed in section  
\ref{sectionnonpert}. 
The recent first measurement of BELLE, with a rather large
uncertainty \cite{BELLEbsll2}, is compatible with this SM prediction.

\medskip

\begin{figure}
\begin{center}
\epsfig{figure=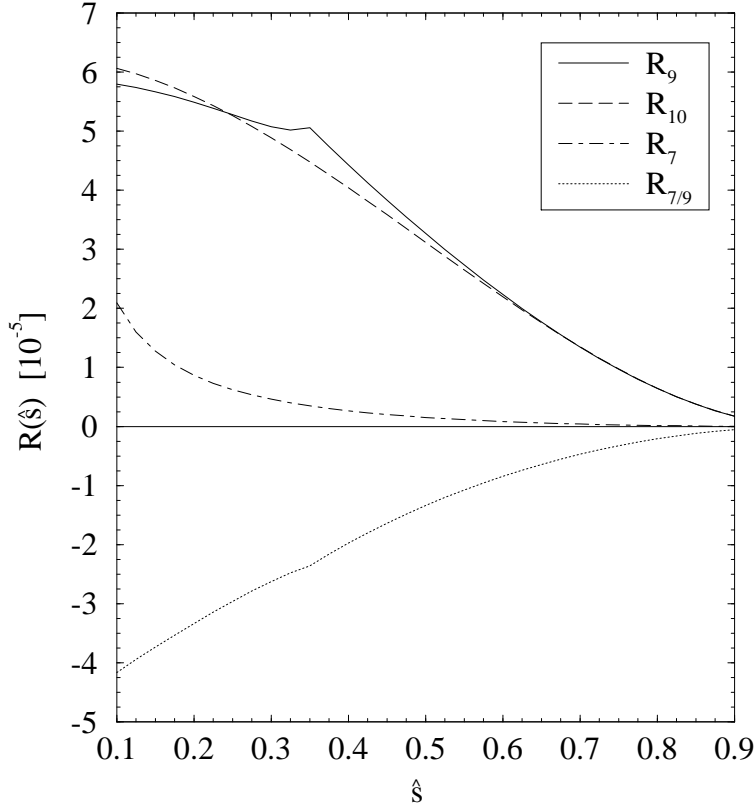,width=10cm}
\end{center}
\caption[]{Comparison of the different short-distance contributions to $R_{quark}(\hat s)$ (NLL precision), from \cite{BurasMuenz}.}
\label{Rpartiell}
\end{figure}

One could think that within the perturbative window at  low 
$\hat{s} \in [0.05,0.25]$, 
one is only sensitive to $C_7$, which would be redundant
information, since we already know it 
from the decay $B \rightarrow X_s \gamma$. 
But, as was explicitly shown  in \cite{MM,BurasMuenz}, one is also sensitive 
to  the new Wilson coefficients $C_9$ and $C_{10}$ and interference terms 
in the low $\hat{s}$ regime with $\hat{s} = m_{\ell^+\ell^-}/m_b^2 \in [0.05,0.25]$
(see fig. \ref{Rpartiell} where the various perturbative contributions
to $R_{quark}$ (with NLL precision) are plotted).

\medskip

A phenomenological  NNLL analysis  
including also the high dilepton mass region 
will be  presented in \cite{Adrian3}. Clearly, the kinematical cuts 
should be adjusted to the experimental choices.

\medskip

The phenomenological impact of the NNLL 
contributions on the FB asymmetry is 
also significant \cite{Adrian1,Asa2}. 
The position of the zero of the FB asymmetry, defined by 
$A_{FB}(s_0)=0$, is particularly interesting to determine 
relative sign and  magnitude of the Wilson coefficients 
$C_7$ and $C_9$ and it is therefore extremely sensitive to 
possible new physics effects.

\medskip 

Employing the counting rule proposed in \cite{Asa1}, i.e. treating the  
formally $O(1/\alpha_s)$ term of $C_9$
as $O(1)$ (see discussion at the end of section \ref{NNLLQCDbsll}), 
the lowest-order value of $s_0$ -- formally derived by the NLL 
expression of $A_{FB}$ --  is determined by the solution of 
\begin{equation}
s_0            C_9^{eff}(s_0) + 2            C_7^{eff} = 0~,
\end{equation} 
where the effective coefficients $C_i^{eff}$ encode also all dominant matrix element 
corrections, which leads to an explicit $s$ dependence
(see \cite{Adrian1}, (A.1)) 
One arrives at 
\begin{equation}
s^{NLL}_0 = 0.14 \pm 0.02~,
\label{eq:s0LO}
\end{equation}
where the error is determined by the  
scale dependence. 
That  NLL result is now modified by 
the  NNLL contributions to \cite{Adrian1,Asa2}
\begin{equation}
s^{NNLL}_0 = 0.162 \pm 0.008~.
\label{eq:s0NLO}
\end{equation}
In this case the variation of the result induced by the scale dependence is 
accidentally very small 
(about $\pm 1\%$) and cannot be 
regarded as a good estimate of missing higher-order effects. 
Taking into account the separate scale variation of both Wilson 
coefficients $C_9$ and  $C_7$,
and the charm-mass dependence, one estimates a conservative overall error
on $s_0$ of about $5 \%$ \cite{Adrian1}. 
In this $s$ region the non-perturbative  
$1/m_b^2$ and $1/m_c^2$ corrections to $A_{FB}(s)$ 
are very small and also under control (see section \ref{sectionnonpert}). 

\medskip 

\begin{figure}
\begin{center}
\epsfig{file=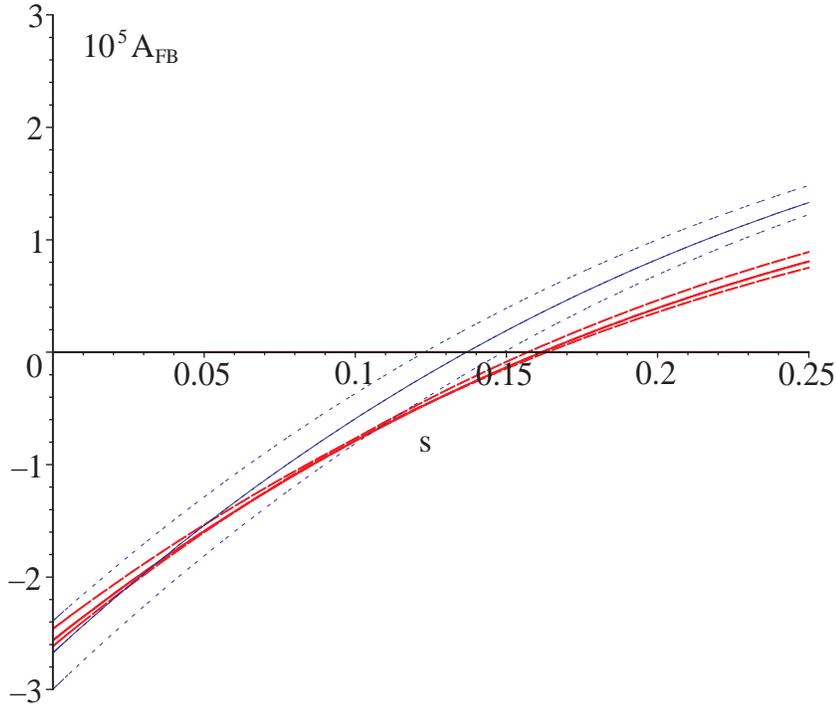,height=11cm}
\end{center}
\caption{Comparison between NNLL and  NLL results for 
$A_{FB}(s)$ in the low $s$ region. 
The three thick lines are the NNLL  
predictions for $\mu=5$~GeV (full),
and $\mu=2.5$ and 10 GeV (dashed); the dotted  curves 
are the corresponding NLL results. All curves for $m_c/m_b=0.29$.
Note that $A_{FB}(s)$ is normalized by the semi-leptonic 
decay rate, see (\ref{forwardbackward}).}
\label{fig:AFB}
\end{figure}

An illustration of the shift of the 
central value and the reduced scale dependence 
between NNL and NNLL expressions of $A_{FB}(s)$, in the low $s$
region, is presented in fig.~\ref{fig:AFB}. 
The complete effect of NNLL
contributions to the FB asymmetry  adds up to a $16\%$ shift compared 
with the  NLL result, with a residual error due to higher-order terms  
reduced at the 5\% level. 
Thus, the zero of the FB asymmetry in the inclusive mode turns out 
to be one of the most sensitive tests for new physics beyond the SM.

\medskip

The $B$ factories will soon provide statistics and resolution
needed for the measurements of $B \rightarrow X_s \ell^+\ell^-$
kinematic distributions. Precise theoretical estimates of 
the SM expectations are therefore needed in order to perform new 
significant tests of flavour physics.
The recently calculated new (NNLL) contributions 
\cite{MISIAKBOBETH,Asa1,Adrian2,Asa3,Adrian1,Asa2,Adrian3}
have significantly 
improved the sensitivity of the inclusive $B \rightarrow X_s \ell^+ \ell^-$ 
decay in  testing extensions of the SM in the sector of flavour 
dynamics. 
Together with the decay $B \rightarrow X_s \gamma$, 
the inclusive $B \rightarrow X_s \ell^+ \ell^-$ decay  will make 
precision flavour physics possible, if one can measure 
the kinematic  variables in the $B \rightarrow X_s \ell^+ \ell^-$ decay 
precisely.

\subsection{Golden mode $B \rightarrow X_s \bar{\nu} \nu$}

The decay $B \rightarrow X_{s} \nu \bar \nu$ is the theoretically 
cleanest rare $B$ decay, but also the most difficult experimentally. 

\medskip

As discussed in section \ref{NNLLQCDbsll}, the partonic  decay is completely
dominated by the internal top contribution due to the hard
GIM mechanism. The perturbative scale uncertainty is $O(1 \%)$.
Moreover, the non-perturbative contributions scaling with $1/m^2_b$ are under control
and small \cite{Falk,Alineu,buchallanewnew}. Because of the absence of the
photon-penguin contribution, the non-perturbative contributions scaling with
$1/m^2_c$ can be estimated to be at the level of $10^{-3}$ at most \cite{Rey}.

\medskip 

After normalizing to the semi-leptonic branching ratio
and summing over the three neutrino flavours, 
the branching ratio of the decay $B \rightarrow X_s \nu \bar \nu$  
is given by \cite{Buras93,Buraslecture}:

\begin{equation}
{\cal B} (B\rightarrow  X_s \nu \bar \nu) = {\cal B}_{exp.}^{sl}
\frac{12 \alpha^2}{\pi^2\sin^4 \Theta_W}
\frac{|V_{ts}|^2}{|V_{cb}|^2}\frac{C(m_t^2/m_W^2)\, \bar \eta}{f(m_c^2/m_b^2)\, \kappa(m_c^2/m_b^2)}.
\label{bsnn}
\end{equation}

Using the measured semi-leptonic branching ratio and 
the phase-space factor of the semi-leptonic decay $f$, the corresponding  
QCD correction $\kappa$, the QCD correction of the matrix element of 
the decay 
$B\rightarrow  X_s \nu \bar \nu$, namely $\bar \eta = \kappa(0)$, and 
scanning the input parameters,
one ends up with the theoretical prediction  \cite{Buraslecture}:

\begin{equation}
{\cal B} (B\rightarrow  X_s \nu \bar \nu) = (3.5 \pm 0.7) \times 10^{-5}. 
\end{equation}

The replacement of $V_{ts}$ by $V_{td}$ in (\ref{bsnn}) leads to the 
case of the decay $B\rightarrow X_d\nu\bar\nu$. Obviously all uncertainties
cancel out in the ratio of the two branching ratios of 
$B\rightarrow X_d \nu\bar\nu$ and  $B\rightarrow X_s \nu\bar\nu$.
Thus, it allows for the cleanest direct determination of the ratio 
of the two corresponding CKM matrix elements.  

\medskip

The measurement of these decay modes is rather difficult. The neutrinos 
escape detection; 
one, thus, has to search for the decays $B \rightarrow X_{s,d} \nu \bar \nu$
through large missing energy associated with the two neutrinos. 
Clearly, background control is more than  difficult in these channels. 
Hopefully, the $B$  factories, with their high statistics 
and their clean environment,  will be able to measure these decays
in the future. 

\medskip

However, the lack of an excess of events with large missing energy 
in a sample of $0.5 \times 10^{6}$ $b \bar b$ pairs at LEP already allowed
ALEPH to establish an upper bound on the branching ratio
of $B \rightarrow X_{s} \nu \bar \nu$ \cite{AlephWarsaw,Nardi}
at $90 \%$ C.L.,  
\begin{equation}
{\cal B} (B\rightarrow  X_s \nu \bar \nu) < 7.7 \times 10^{-4}\, .  
\end{equation}
This upper bound is still an order of magnitude above the SM prediction, 
but it already leads to constraints on new physics models \cite{Nardi}. 
For this purpose,  the QCD corrections to the decays $B \rightarrow X_{s,d} \nu \bar \nu$ in supersymmetric theories (MSSM) 
have recently been presented \cite{BobethQCD}.

\newpage

\setcounter{equation}{0}
\section{Indirect search for supersymmetry}

\subsection{Generalities}
\label{generalities}

Today supersymmetric models are given priority in our search for 
new physics beyond the SM. This is primarily suggested by theoretical 
arguments related to the well-known hierarchy problem. Supersymmetry 
eliminates the sensitivity for the highest scale in the theory and, thus, 
stabilizes the low energy theory.   
The precise mechanism of the necessary supersymmetry breaking is unknown. A 
reasonable approach to this problem is the inclusion of the most general 
soft breaking term consistent with the SM gauge symmetries in the 
so-called unconstrained minimal supersymmetric standard model
(MSSM). This leads to a proliferation of free parameters in the theory. 

\medskip

In the MSSM there are new sources of  FCNC transitions. 
Besides
the CKM-induced contributions,
which are brought about by  a charged Higgs or a chargino,  
there are  generic supersymmetric  contributions that arise from  
flavour mixing  in the squark mass matrices.
 The structure of the MSSM does not explain 
the suppression of 
FCNC processes, which is observed in experiments; the gauge symmetry
within the supersymmetric framework does not protect
 the observed strong suppression of the FCNC transitions. 
This is the crucial point of the well-known supersymmetric 
flavour problem. Clearly, the origin of flavour violation is 
highly model-dependent.

\medskip

Within the framework of the MSSM there are at present three favoured 
supersymmetric models that solve the supersymmetric 
flavour problem by a specific mechanism through which the sector of 
supersymmetry breaking 
communicates with the sector accessible to experiments: in the 
minimal supergravity model (mSUGRA)~\cite{MSUGRA},  
supergravity is the corresponding
mediator; in the other two models, this is achieved by gauge 
interactions (GMSB)~\cite{GMSB}
 and by anomalies (AMSB)~\cite{AMSB}. Furthermore, there are other 
classes of models in which the flavour problem is solved by particular 
flavour symmetries~\cite{FLAVOUR}.

\medskip

The decay  $B \rightarrow X_s \gamma$ is sensitive to the 
mechanism of supersymmetry breaking because,  
in the limit of exact supersymmetry, the decay rate would
be just zero:
\begin{equation}
{\cal B} (B \to X_s \gamma)_{Exact\, Susy} = 0.  
\end{equation}
This follows from an argument first given by Ferrara and Remiddi in 1974
\cite{Ferrara}. 
In that work  the absence of the anomalous 
magnetic moment in a supersymmetric abelian gauge theory was shown.

\medskip

Flavour violation thus originates
from the interplay  between the dynamics of flavour and the mechanism of  
supersymmetry breaking.  FCNC processes
therefore yield important (indirect) information on the construction 
of supersymmetric extensions of the SM and can contribute 
to the question of which mechanism ultimately breaks supersymmetry
and will thus yield important (indirect) information on the 
construction of supersymmetric extensions of the SM. In this context it is 
important to analyse the correlations between the different sets of information from 
rare $B$ and $K$ decays. Moreover, 
tight experimental bounds on some flavour-diagonal transitions, such
as the electric dipole moment of the electron and of the neutron, as
well as $g-2$, help constraining the soft terms that induce chirality
violations.

\medskip

As was already emphasized in the introduction, 
inclusive rare  decays, as loop-induced processes, are particularly 
sensitive to new physics and theoretically clean.
Neutral flavour transitions involving third-generation quarks,
typically in the $B$ system, do not yet pose serious threats to specific
models. However, 
the rare decay $B \rightarrow X_s \gamma$
has already carved out large  regions in the space of free
parameters of most of the models in the classes mentioned above. 
Once more precise data from the $B$ factories are  available, 
this decay will undoubtedly gain even more efficiency
in selecting the viable regions of the parameter space in the various
classes of models. This indirect search for new physics 
is a model-dependent issue;  especially in the MSSM with its
43 new CP-violating phases. Simplifying assumptions about the parameters
often introduce model-dependent correlations between different observables.  
Thus, flavour physics will also help in discriminating between  
the models that will be proposed by then.
In view of this, it is important to calculate the rate of the rare $B$ decays,
 with theoretical uncertainties as reduced 
as possible and general enough for generic 
supersymmetric models. 

\medskip

In the analysis of FCNC processes within supersymmetry, the additional 
assumption of 
minimal flavour violation (MFV) is often introduced.  Minimal 
flavour violation is then loosely defined as 
'the flavour violation that is completely dictated by the CKM angles'. 
In a top/bottom approach, one starts with a specific model of 
supersymmetry breaking and then one can try to justify 
the simplifying assumption of MFV explicitly within the specific model.
In a bottom/top approach, the naive assumption of MFV
is problematic, since it is not stable, within the MSSM,  
under radiative corrections and calls for a more 
precise  concept. In \cite{Giannew}, a consistent definition was
presented, which essentially requires that all flavour and CP-violating 
interactions be linked to the known structure of Yukawa couplings. 
The constraint within an  effective field approach  is introduced  
with the help of a symmetry concept and can be shown to be 
renormalization-group-invariant; it is also a valid concept beyond 
supersymmetric models \cite{Giannew}. 
This consistent MFV assumption for example is valid if  the 
 soft terms of the scalar mass   are universal and the trilinear 
soft terms are 
proportional to Yukawa couplings, at an
arbitrary high scale. Then the physical squark massses are not equal, 
but the induced flavour violation is described in terms of the 
usual CKM parameters.

\medskip 

Perhaps this MFV-based effective field theory approach is too 
pessimistic from the current  point of view. 
One of the key predictions of the MFV is the direct link 
between the  $b \rightarrow s$, $b\rightarrow d$ and $s \rightarrow d$
transitions. This prediction within the $\Delta F = 1$ sector is definitely 
not well-tested at the moment.

\medskip

In contrast to the scale of the electroweak
symmetry breaking, there is no similarly strong argument 
that new flavour structures have to appear at the electroweak scale.

\subsection{Constraints from $B \rightarrow X_s \gamma$}

While in the SM, the rate for $B \to X_s \gamma$
is known up to NLL in QCD, the calculation of this decay rate 
within supersymmetric models 
is still far from this level of sophistication.
There are several contributions to the decay amplitude: 
besides the 
$W\,t$-quark and the $H\,t$-quark contributions, there are also
the chargino, gluino and neutralino contributions.
The first systematic MSSM analysis of the decay $B \rightarrow X_s \gamma$
was presented in \cite{francesca1}.

The  phenomenological analyses of the decay 
$B \rightarrow X_s \gamma$ in the mSUGRA model \cite{francesca2,Baer,Goto}
already excluded large parts of the parameter space of this model.
However, within many  analyses the non-standard contributions   
were often not investigated  with NLL precision  
as the SM contribution.  
Besides the large uncertainties in the LL predictions,
the step from the LL to the NLL precision is also necessary 
in order to check  the validity of the perturbative 
approach in the model under consideration. 
Moreover, it was already shown in specific new-physics scenarios 
that bounds on the parameter space of  non-standard 
models are very sensitive to NLL contributions (see below).

\medskip

\begin{figure}
\begin{center}
\epsfig{figure=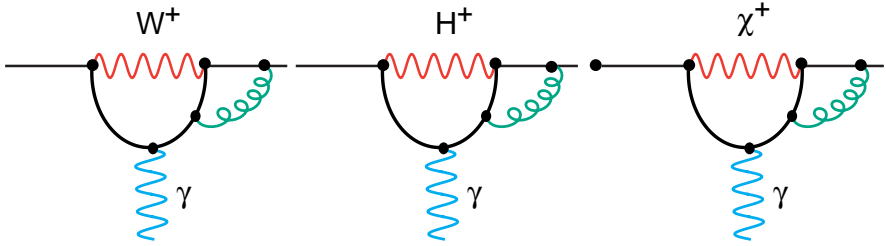,width=12cm}
\end{center}
\caption{SM, charged Higgs and chargino contribution at the matching scale.}
\label{susycalculation}
\end{figure}

\begin{figure}
\begin{center}
\epsfig{figure=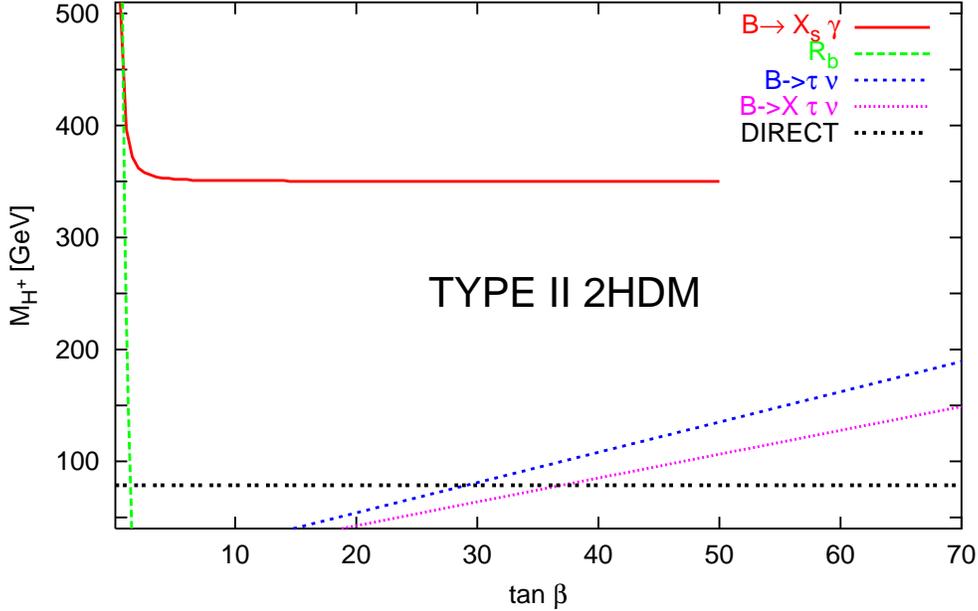,width=13cm}
\end{center}
\caption[]{ Direct and indirect lower bounds on $M_{H^+}$ from different 
processes in the 2HDM of Type II  as a function of $\tan\beta$. 
The $B\to X_s \gamma$ bound is based on the latest CLEO 
measurement (\ref{cleoneu}), from \cite{GambinoMisiak}.}
\label{paolo2}
\end{figure}

Nevertheless, within supersymmetric models, partial 
NLL results are available. 
The gluonic NLL two-loop matching contributions were 
presented some time ago \cite{mikolaj99}.  
A complete NLL calculation
of the $B \rightarrow X_s \gamma$ branching ratio in the simplest
extension of the SM, namely the two-Higgs-doublet model (2HDM),
is
already available \cite{CDGG,BG,GambinoMisiak}: 
in the 2HDM of Type II (which already represents 
a good approximation for 
gauge-mediated
supersymmetric models with large $\tan\beta$, where the charged Higgs
contribution dominates the chargino contribution), 
the $B \rightarrow X_s \gamma$ is only sensitive to two
parameters of this model, the mass of the charged Higgs boson 
and $\tan \beta$. 
Thus, the  experimental data of 
the decay $B \rightarrow X_s \gamma$ allow for stringent bounds on these
two  parameters,  which are much more restrictive 
than the lower bound on the charged Higgs mass found in the direct 
search at LEP (see fig. \ref{paolo2}).

One finds that these
indirect bounds are very sensitive to NLL QCD corrections and  even
to the two-loop electroweak contributions (see \cite{CDGG,BG}). 
Using  the latest theoretical NLL prediction (\ref{totalbr}) 
and  the latest CLEO measurement (\ref{cleoneu}),  
one finds the $\tan \beta$ independent bound $M_{H} > 350$ GeV
 \cite{GambinoMisiak}.
But this bound gets weakened if the charm mass renormalization 
scheme ambiguity of the present NLL prediction (see section \ref{phenobsg})
 is taken into account.
For example, if the pole mass  scheme is adapted as in the theoretical
prediction (\ref{currentprediction}), then the weaker bound  $M_H > 280$ GeV
 is  found \cite{GambinoMisiak}.

\medskip

In \cite{guidice}  
a specific  supersymmetric scenario is presented, 
where in particular the possibility of 
destructive interference of the chargino 
and the charged Higgs contribution is analysed. 
The analysis has been  done under two assumptions.
First, that the only source of flavour violation 
at the electroweak scale is that of the SM, 
encoded in the CKM matrix
(MFV).
Therefore, the analysis  applies to mSUGRA, GMSB and
AMSB models (in which the same features are assumed  at the 
messenger scale) only when the sources of flavour violation,  
generated radiatively between the supersymmetry-breaking scale and 
the electroweak scale, can be neglected with respect to those induced
by the CKM matrix. 
The second assumption is that there exists  
a specific mass hierarchy, in particular 
the heavy gluino limit.
Indeed, the NLL calculation has been done in the limit 
\begin{equation}
\mu_{\tilde{g}} 
\sim O(m_{\tilde{g}},m_{\bar{q}},m_{\tilde{t}_1}) \gg 
\mu_W \sim  O(m_{W},m_{H^+},m_{t},m_{\chi},m_{\tilde{t}_2}).
\label{masshierachy}
\end{equation} 
The mass scale of the 
charginos ($\chi$) and of the lighter stop ($\tilde t_2$) is
the ordinary electroweak scale $\mu_W$, while
the scale $\mu_g$ is 
characteristic of all other strongly interacting supersymmetric
particles (squarks and gluinos) and is 
assumed to be of the order of $1$ TeV. 
NLL QCD corrections have been  calculated up to first order
in  $\mu_W/\mu_{\tilde{g}}$, including the important 
 non-decoupling effects \cite{guidice}.

\medskip 

At the electroweak scale $\mu_W$, the new contributions do not induce any new 
operators in this scenario. Thus, the only  step in the new NLL calculation 
beyond the one within the SM is Step 1, 
the matching calculation at the scale
$\mu_W$,  where we encounter the two new CKM-induced contributions
of the charged Higgs and the chargino
(see fig. \ref{susycalculation}):
\begin{equation}
 C_{{NLL}} (\mu_W) = C_{{NLL}}^{{SM}} (\mu_W) + 
C_{{NLL}}^{{H^+}} (\mu_W) + C_{{NLL}}^{{\chi}} (\mu_W). 
\end{equation}

\begin{figure}
\begin{center}
\epsfig{figure=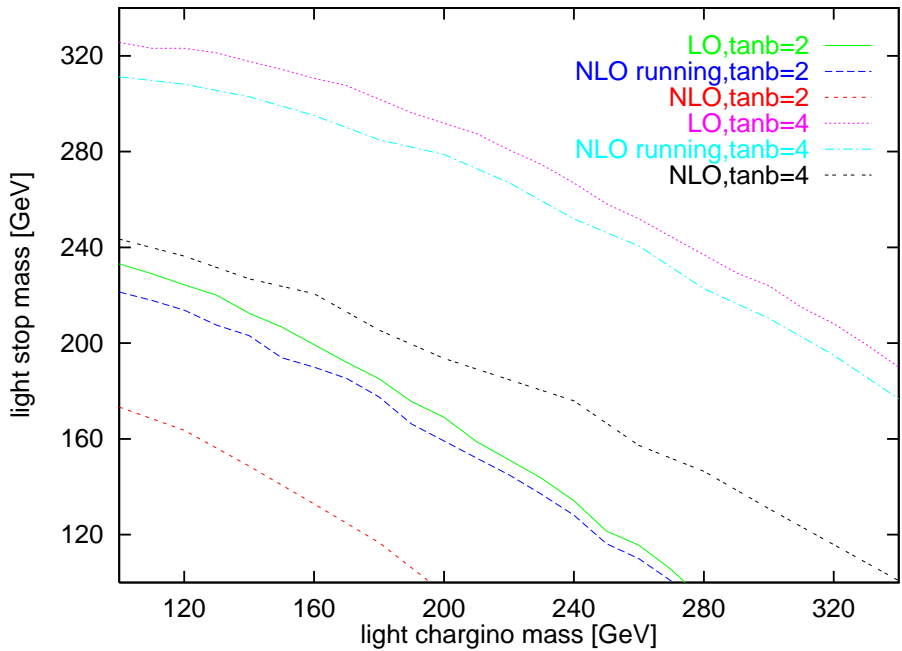,width=13cm}
\end{center}
\caption[]{Upper bounds on the lighter chargino and stop masses 
from $B \rightarrow X_s \gamma$ data  in the scenario
(\ref{masshierachy}) if a light charged Higgs mass  
is assumed;
for  $\tan\beta=2$ (three lower curves) and $4$ 
(three upper plots)
the LL, NLL-running and NLL results
(from the top  to the bottom) are shown (see text),
from \cite{guidice}.}
\label{paolo}
\end{figure}

It was found  \cite{guidice} 
that,  in this  specific supersymmetric scenario,  
bounds on the parameter space are rather sensitive 
to NLL contributions 
and they lead to a significant reduction of the stop-chargino 
mass region,  where the supersymmetric contribution has
a large destructive interference with the charged-Higgs boson 
contribution. In fig. \ref{paolo} the 
upper bounds on the lighter chargino and stop masses 
from $B \rightarrow X_s \gamma$ data  in the scenario of
(\ref{masshierachy}) are illustrated if a light charged Higgs mass 
of $m_{H^\pm}=100$ GeV is assumed.
The stop mixing is set to $|\theta_{\tilde t}|<\pi/10$, which 
corresponds to the assumption of a mainly right-handed light stop.
Moreover, $|\mu|<500$ GeV and all heavy masses are around 1 TeV. 
For  $\tan\beta=2$  and $4$ 
the results of the LL, `NLL running' and NLL
calculations are given. 
The result of
neglecting the new NLL supersymmetric contributions to the 
Wilson coefficients is labelled as `NLL running' and illustrates 
the importance of the NLL chargino contribution \cite{guidice}. 

\medskip 

\begin{figure}
\begin{center}
\epsfig{figure=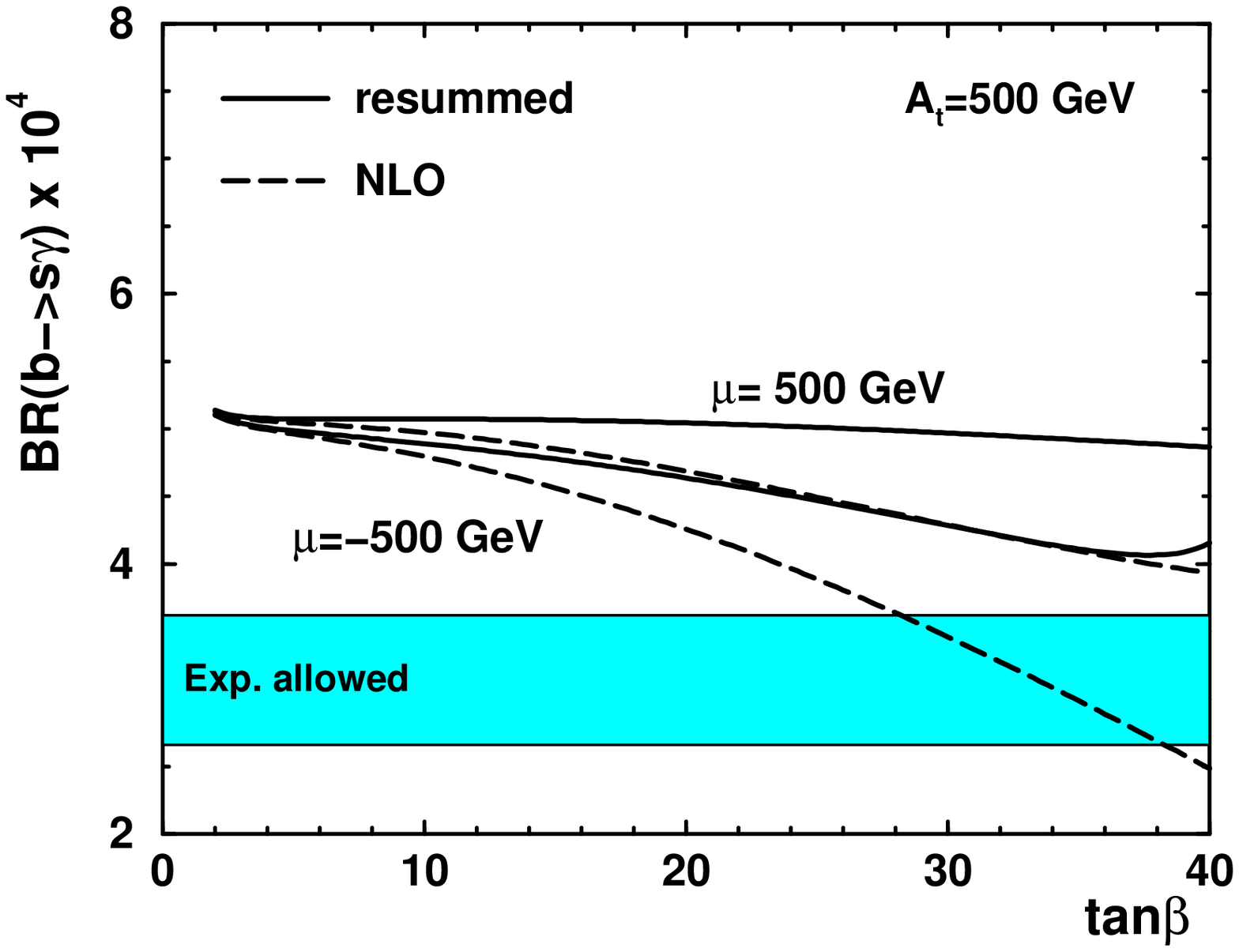,width=8cm} \epsfig{figure=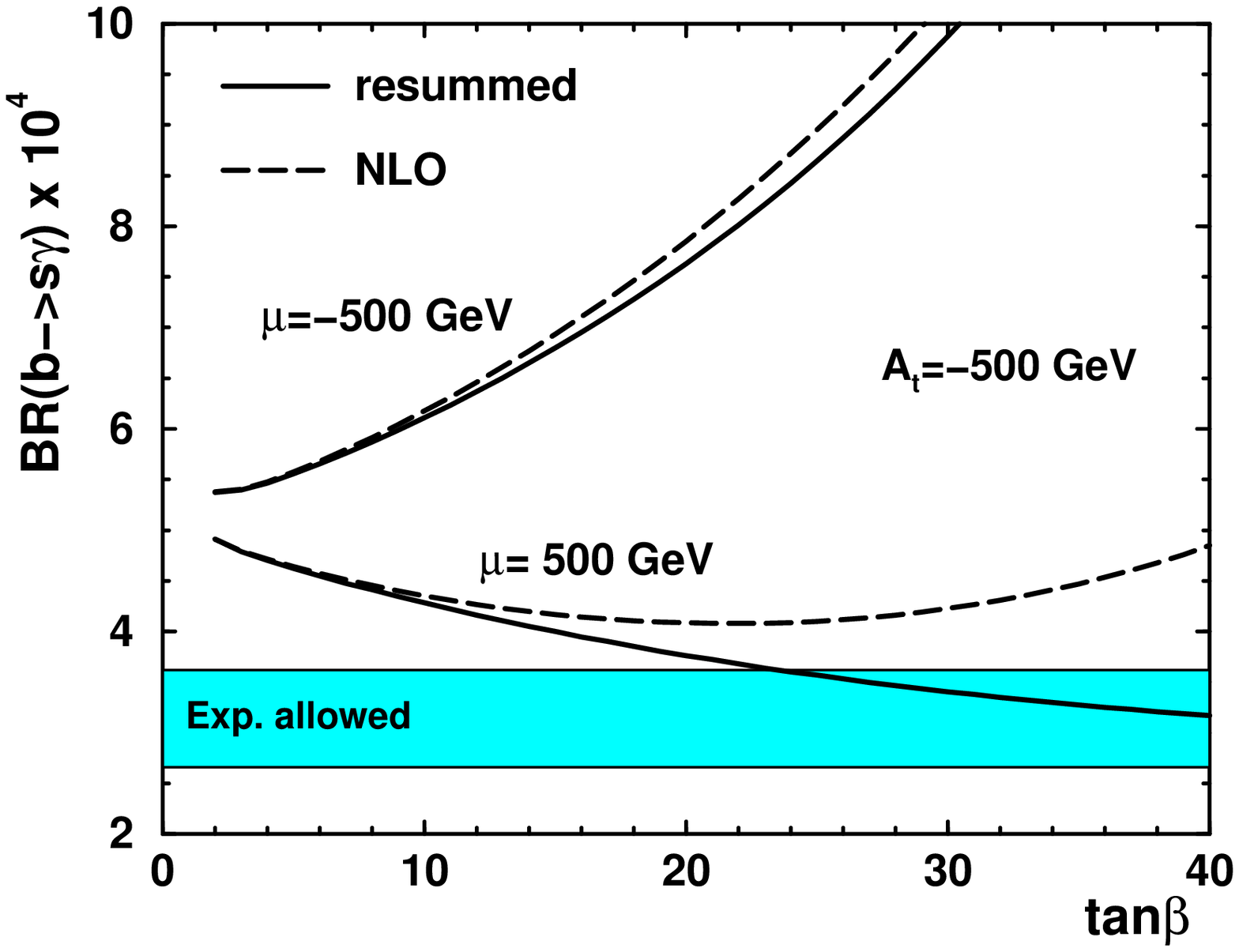,width=8cm} 
\end{center}
\caption[]{Comparison of the theoretical NLL predictions within
a special MSSM scenario (similiar to \ref{masshierachy})
{\it with} and  {\it without} the resummed large $\tan \beta$ terms;
 the charged Higgs boson mass is $200$ GeV and the light stop mass is 
$250$ GeV.  
The values of $\mu$ and $A_t$ are indicated in the plot, while  
the gluino, heavy stop and down-squark masses are set at $800$ GeV; from
 \cite{Carena}.}
\label{Carenap}
\end{figure}

This specific MFV
scenario was refined and extended to the 
large $\tan \beta$ regime by the resummations of terms of the form 
$\alpha_s^n \, \tan^{n+1} \beta$ \cite{Carena,Degrassinew}.
Additional $\tan \beta$ terms, which have to be summed in the 
large $\tan \beta $ regime were singled out in \cite{Giannew}. 
The stability of the renormalization--group--improved perturbation theory 
was re-assured for this specific scenario: the resummed NLL results 
in the large $tan \beta$ regime show constraints similar to the
LL results (see \cite{Deboernew}). 
For example, it is a well-known feature in the mSUGRA model
that,  depending on the sign of $A_t \cdot \mu$ (where $A_t$ denotes 
the stop mixing parameter) the chargino contribution 
can  interfere constructively ($A_t \cdot \mu > 0$)
or destructively ($A_t \cdot  \mu <  0$)
with the SM and the charged Higgs contribution. 
Therefore, the scenario
$A_t \cdot \mu > 0$ within this model 
requires very heavy superpartners in order to
accommodate the $B \rightarrow X_s \gamma$ data. But also
the case $A_t \cdot \mu <  0$ is constrained in the large $\tan \beta$ 
regime where the chargino contribution is strongly enhanced
\cite{Carena,Degrassinew,Deboernew} (see fig. \ref{Carenap}).

\medskip 

\begin{figure}
\begin{center}
\epsfig{figure=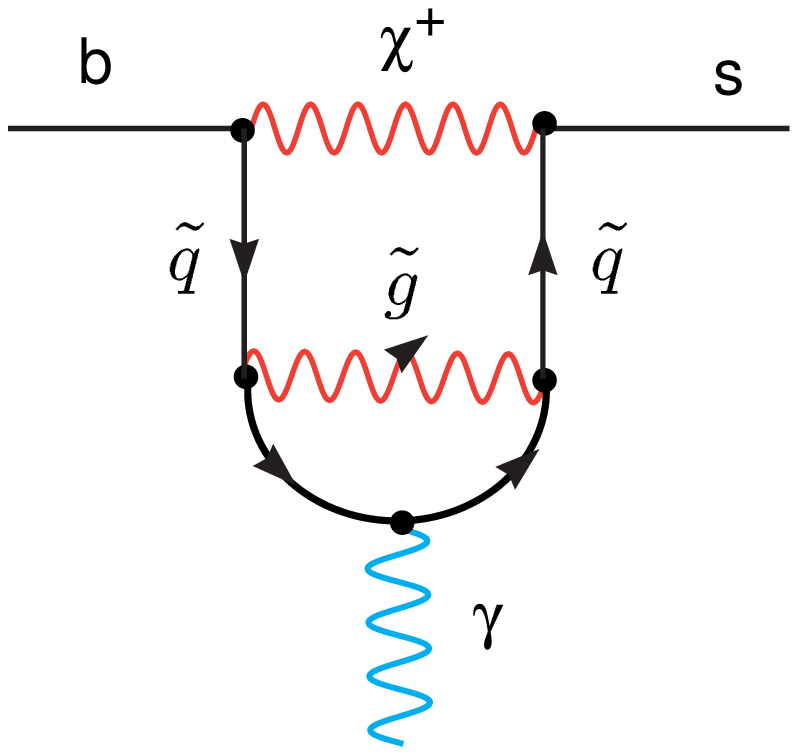,height=4.4cm} \epsfig{figure=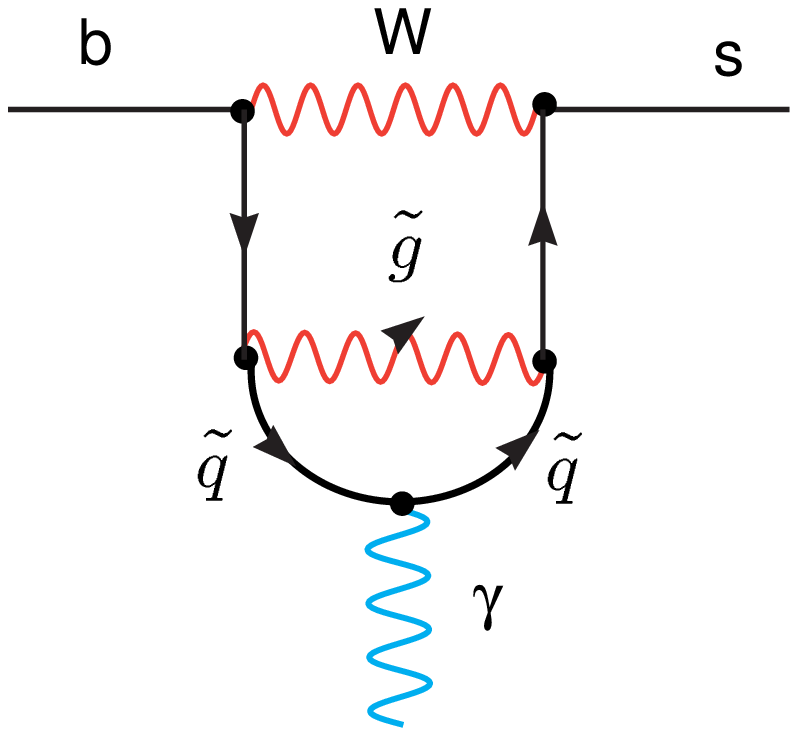,height=4.4cm} \epsfig{figure=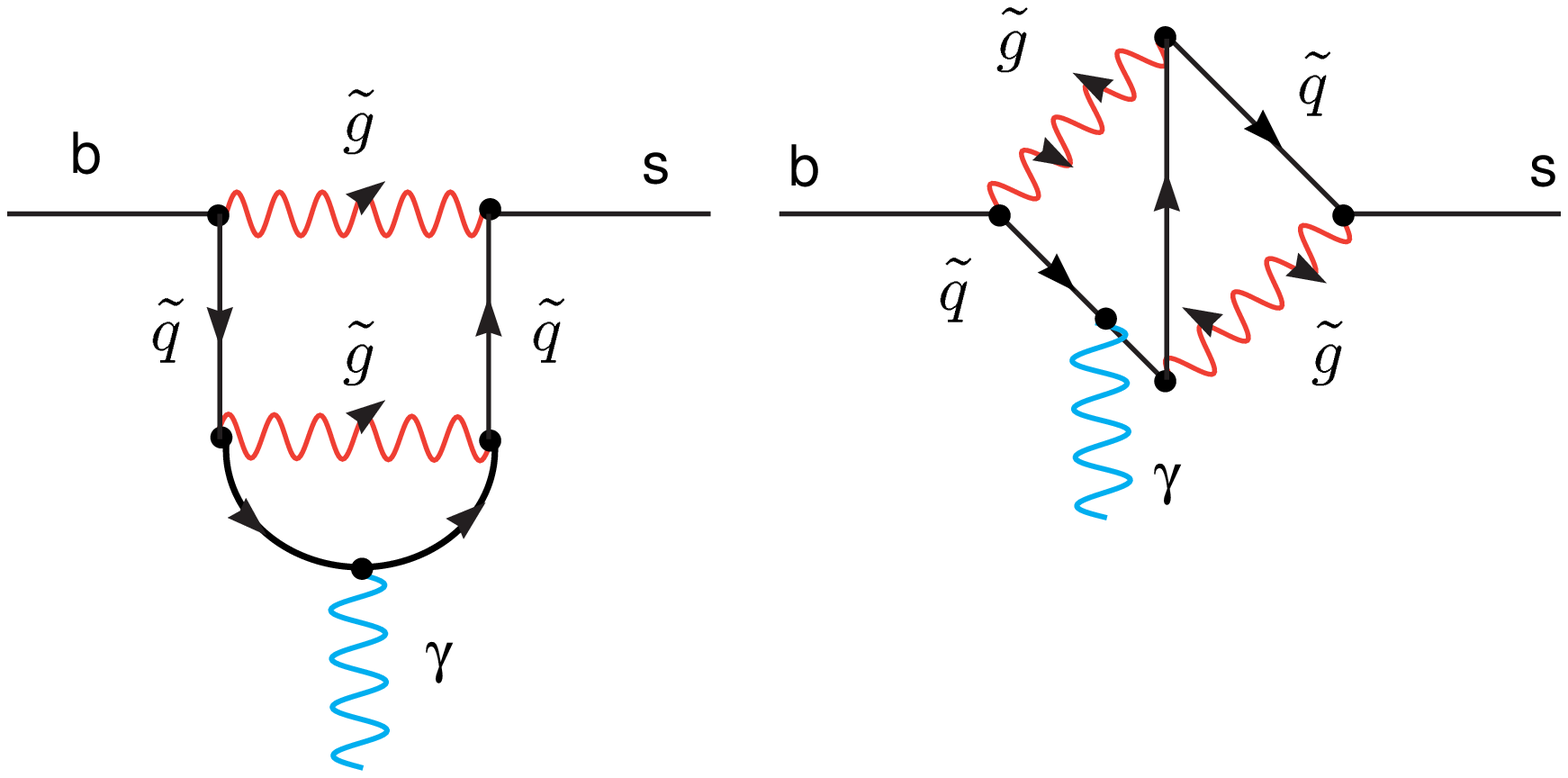,height=4.4cm} 
\end{center}
\caption[]{NLL gluino contributions to the decay $B \rightarrow X_s \gamma$.}
\label{MFVgluino}
\end{figure}

However, all these NLL analyses  are valid only in 
the heavy gluino regime. 
Thus, they 
cannot be used in particular directions of the parameter 
space of the above-listed models in which quantum effects induce a 
gluino contribution as large as the chargino or the SM contributions. 
Nor can it be used as a model-discriminator tool, able to constrain 
the potentially
large sources of flavour violation typical of generic 
supersymmetric models. Therefore, a 
complete NLL calculation, also  within the MFV approach, should  
include contributions where the gluon is replaced by its superpartner   
gluino (see fig. \ref{MFVgluino})~\cite{Matthias}.

\medskip

The flavour non-diagonal gluino--quark--squark vertex induced by
the flavour violating scalar mass term and trilinear terms
is particularly interesting. This vertex is generically assumed to induce the 
dominant contribution to quark flavour transitions, as this vertex is weighted
by the strong coupling constant $g_s$.
Therefore, it is often taken as the 
only contribution 
to these transitions and in particular to the $B \rightarrow X_s \gamma$
decay, when attempting to obtain order-of-magnitude upper bounds
on flavour violating terms in the scalar potential. 
Once the experimental constraints are imposed, 
however, the gluino contribution is reduced to values such that the SM 
and the other supersymmetric contributions can no longer  be neglected. 
Any LL and NLL calculation of the $B \rightarrow X_s \gamma$
rate in generic supersymmetric models, therefore, should then include
all possible contributions.

\medskip

\begin{figure}
\begin{center}
\epsfig{figure=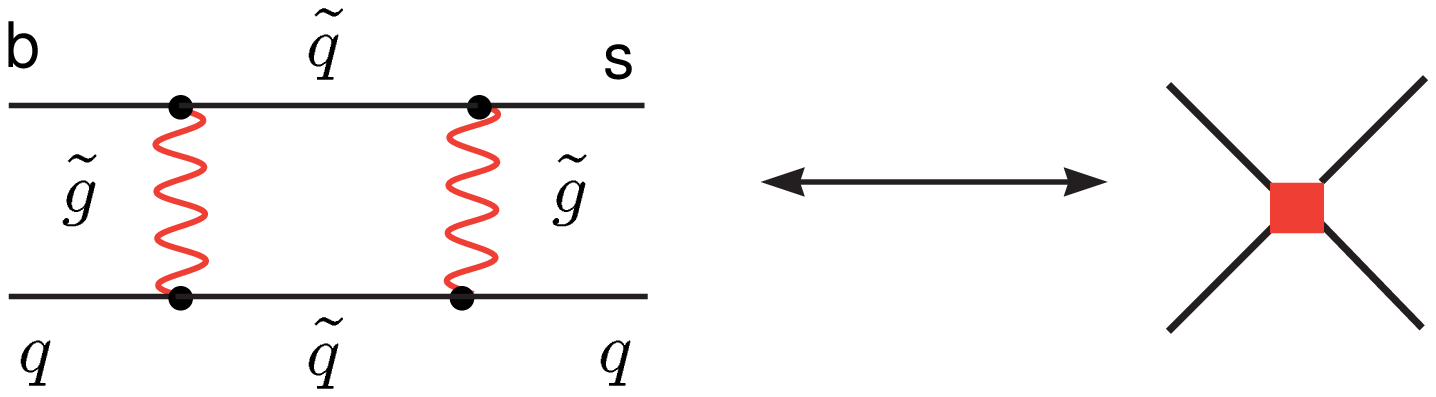,width=8.3cm}
\hspace{0.5cm}
\epsfig{figure=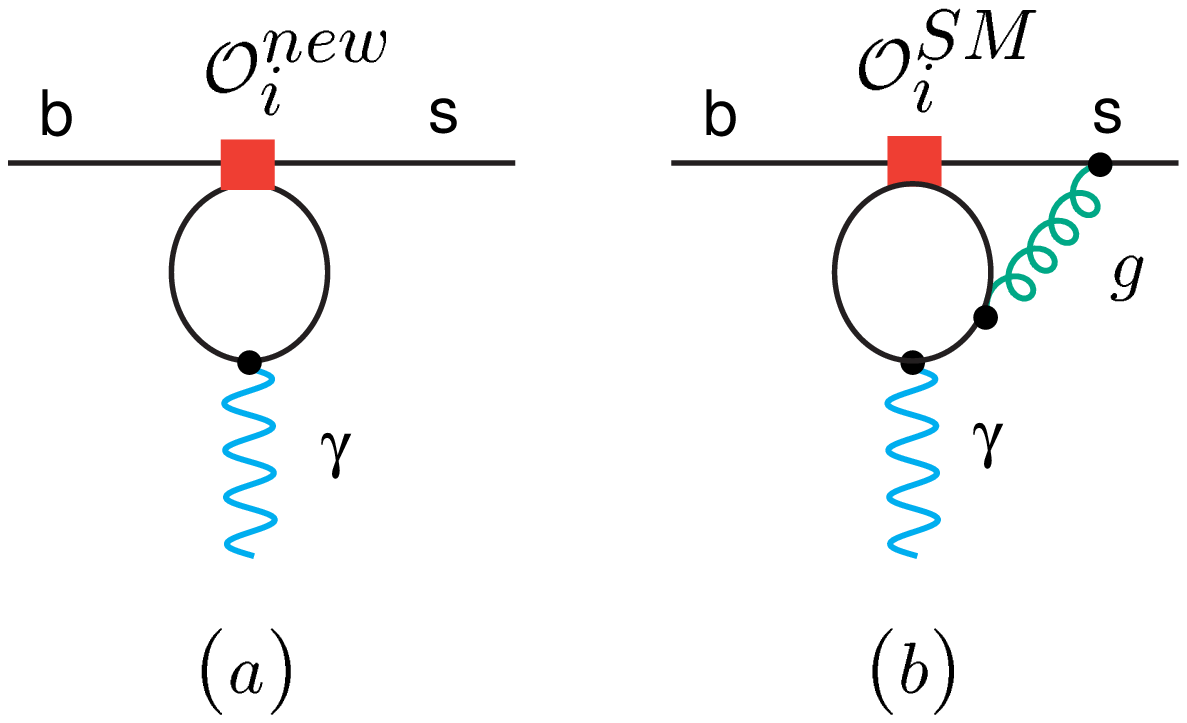,width=6.3cm}
\end{center}
\caption{
Left: matching of gluino--squark box on new scalar operators. Right:
mixing of new (scalar) operators at one loop (a) in constrast to the 
vectorial operators of the SM (b) which mix at two loop only.} 
\label{mixingsusy}
\end{figure}

The gluino contribution presents some peculiar features related
to the implementation of the QCD corrections.
In ref. \cite{OUR} this contribution to the decay $B \rightarrow X_s \gamma$
has been  investigated in great detail for
 supersymmetric models with generic soft terms. 
The gluino-induced contributions to the decay amplitude for $B \to X_s \gamma$
are of the following form:
\begin{equation}
\alpha_s(m_b) \, (\alpha_s(m_b) \log(m_b/M))^n \, \quad (LL),
\end{equation}
\begin{equation} 
\alpha^2_s(m_b) \,  (\alpha_s(m_b) \log(m_b/M))^n \, \quad (NLL).
\end{equation}
The relevant operator basis of the SM effective Hamiltonian gets enlarged
to contain magnetic and chromomagnetic operators with an extra factor of 
$\alpha_s$. Furthermore, 
one finds that  
gluino--squark boxes induce new scalar and tensorial 
four-quark operators, 
which are shown to 
mix into the magnetic operators without gluons already at one loop.  
On the other hand, the vectorial four-quark  operators 
mix only with an
additional gluon into magnetic ones (fig. \ref{mixingsusy}). 
Thus, they will contribute at NLL order only.
But from the numerical point
of view the contributions of the vectorial operators (although NLL) are
not necessarily suppressed w.r.t. the new four-quark contributions;
this is due to the expectation 
that the flavour-violation parameters
present in the Wilson coefficients of the new operators are expected
to be much smaller (or much more stringently constrained)
than the corresponding ones in the coefficients of the vectorial 
operators. This feature shows that a complete NLL calculation is important.

\medskip

To understand the sources of flavour violation that may be present in
supersymmetric models in addition to those enclosed in the CKM matrix,
one has to consider the contributions to the squark mass matrices
\begin{equation}
{\cal M}_f^2 \equiv  \left( \begin{array}{cc}
  m^2_{\,f,\,LL} +F_{f\,LL} +D_{f\,LL}           & 
                 \left(m_{\,f,\,LR}^2\right) + F_{f\,LR} 
                                                     \\[1.01ex]
 \left(m_{\,f,\,LR}^{2}\right)^{\dagger} + F_{f\,RL} &
             \ \ m^2_{\,f,\,RR} + F_{f\,RR} +D_{f\,RR}                
 \end{array} \right) \,,
\label{squarku}
\end{equation}
where $f$ stands for up- or down-type squarks.
In the super-CKM basis, where the quark mass matrices are diagonal 
and the squarks are rotated in parallel to their superpartners,
the $F$ terms  from the superpotential and the $D$ terms 
turn out to be diagonal 
$3 \times 3$ submatrices of the 
$6 \times 6$
mass matrices ${\cal M}^2_f$. This is in general not true 
for the additional terms $m^2_f$, originating from  the soft 
supersymmetric breaking potential. 
Because all neutral gaugino couplings are flavour diagonal
in the super CKM basis, the 
gluino contributions to the
decay $b \to s \gamma$ are induced by the off-diagonal
elements of the soft terms $m^2_{f,LL}$, $m^2_{f,RR}$, $m^2_{f,RL}$.

\medskip

As a first step, it is convenient to select {\it one} possible 
source of flavour violation in the squark sector at a time and
assume that all the remaining ones are vanishing. 
Following 
refs.~\cite{Massinsertion,MAS}, all diagonal entries in 
$m^2_{\,d,\,LL}$, $m^2_{\,d,\,RR}$, and $m^2_{\,u,\,RR}$
are set to be equal and their common value is denoted by
$m_{\tilde{q}}^2$.  The branching ratio can then be studied as a
function of \begin{equation} 
\delta_{LL,ij} = \frac{(m^2_{\,d,\,LL})_{ij}}{m^2_{\tilde{q}}}\,, 
\delta_{RR,ij} = \frac{(m^2_{\,d,\,RR})_{ij}}{m^2_{\tilde{q}}}\,, 
\delta_{LR,ij} = \frac{(m^2_{\,d,\,LR})_{ij}}{m^2_{\tilde{q}}}\, 
\, (i \ne j).
\label{deltadefa}
\end{equation}
Phenomenological analyses in the  so-called unconstrained  MSSM 
\cite{Donoghue,MAS,HAG}   neglected QCD corrections
and only used  the gluino contribution to saturate the experimental bounds. 
Moreover, no correlations between different sources of flavour 
violation were taken into account. 
In this way, one arrived at  `order-of-magnitude bounds' on the 
soft parameters~\cite{MAS,HAG}. The $B \rightarrow X_s \gamma$
decay is mainly sensitive to the off-diagonal elements 
$\delta_{LR,23}$ and $\delta_{RL,23}$ and constrains them to values of order 
$10^{-2}$. 
In~\cite{OUR}, the sensitivity of the bounds on the down squark
mass matrix to radiative QCD LL corrections was systematically analysed, 
including the SM and the gluino contributions. Some leading NLL contributions 
were considered in \cite{Okumura} and 
the large impact of the NLL corrections for
non-minimal models, in particular for large $\tan \beta$ was demonstrated.  

\medskip

A consistent analysis 
of the bounds on the sfermion mass matrix should also  include 
 interference effects between the various contributions.
In    \cite{NEWNEW}, the interplay  between the various 
sources of flavour violation and the interference effects 
of SM, gluino,
chargino, neutralino and charged Higgs boson contributions is
 analysed. 
New bounds on  simple combinations of
elements of the soft part of the squark mass matrices
are found to be, in general, one order of magnitude weaker 
than the bound on the single off-diagonal element $\delta_{LR,23}$, which 
was derived in previous work \cite{MAS,Masiero2001} by
neglecting any kind of interference effects.  
Thus, it turns out that --- at least 
within the decay $B \rightarrow X_s \gamma$ --- the flavour problem is less
severe than often stated.

\medskip

The measurement of the 
photon polarization within the decay $B \rightarrow X_s \gamma$ 
allows for another important SM test. Assuming that the decay is induced by
the magnetic dipole operator only, one starts with the effective
Hamiltonian 
\begin{equation}
{\cal H} = -\frac{4G_F}{\sqrt2} \lambda_t  \left( C_{7L}{\cal O}_{7L}
+  C_{7R}{\cal O}_{7R}\right),
\end{equation}
where ${\cal O}_{7L,R} \equiv \frac{e}{16\pi^2} \overline{m}_b 
\bar s\sigma_{\mu\nu}\frac{1 \pm \gamma_5}{2} b \, F^{\mu\nu}~$.
Then the photon polarization is defined by 
\begin{equation} 
\lambda_{\gamma} \equiv {|C_{7R}|^2 -|C_{7L}|^2\over |C_{7R}|^2 +|C_{7L}|^2}~.
\end{equation}
In the SM, one has $C_{7R}/C_{7L}=m_s/m_b \approx 0$ and therefore 
a mostly left-handed photon. But in many supersymmetric scenarios,  
and also in left--right-symmetric models,  the photon may have a 
large right--handed 
component. In \cite{Everett} the possibility of a strictly
right-handed photon within the framework of the MSSM was discussed.
Clearly, only in non-minimal models is such an extreme  deviation from the
SM prediction possible.

\medskip 

There are many suggestions for measuring $\lambda_\gamma$ in the literature
\cite{Atwood97,Melikhov98,Grossman00,Mannel97,Hiller01}. However, they all 
rely on very high statistics or on new experimental settings and will not 
 be possible in the near future. Quite recently, a new method has been 
 proposed, 
which can be realized with the present statistics
available at the $B$ factories. 
The photon polarization can be measured in radiative $B$ decays to excited 
kaons, using angular correlations among the three-body decay products of the 
excited kaons \cite{Gronau02}.  
It is essential for a helicity measurement to have a three-particle 
decay mode because $\lambda_\gamma$ is a parity-odd  
variable and there is no odd-momentum correlation in the two-body mode. 
In the decays 
$B^+\to (K^+_1(1400) \to K^0\pi^+ \pi^0)\gamma$ and $B^0\to (K^0_1(1400) \to 
K^+\pi^- \pi^0)\gamma$, however,  the up--down asymmetry of the photon 
momentum with respect to the $K\pi\pi$ decay plane in the frame
of the excited kaon 
measures the photon 
polarization rather efficiently. 
The up--down asymmetry can theoretically be predicted 
to be $A = (0.33 \pm 0.05) \times \lambda_\gamma$ 
in the case of the resonance $K_1 (1400)$ \cite{Gronau02}. Thus,
the method will definitely be sensitive for large deviations 
from the SM prediction already with the present luminosity at the $B$ 
factories.

\subsection{Constraints from $B \rightarrow X_s \ell^+ \ell^-$}

\begin{figure}
\begin{center}
\epsfig{figure=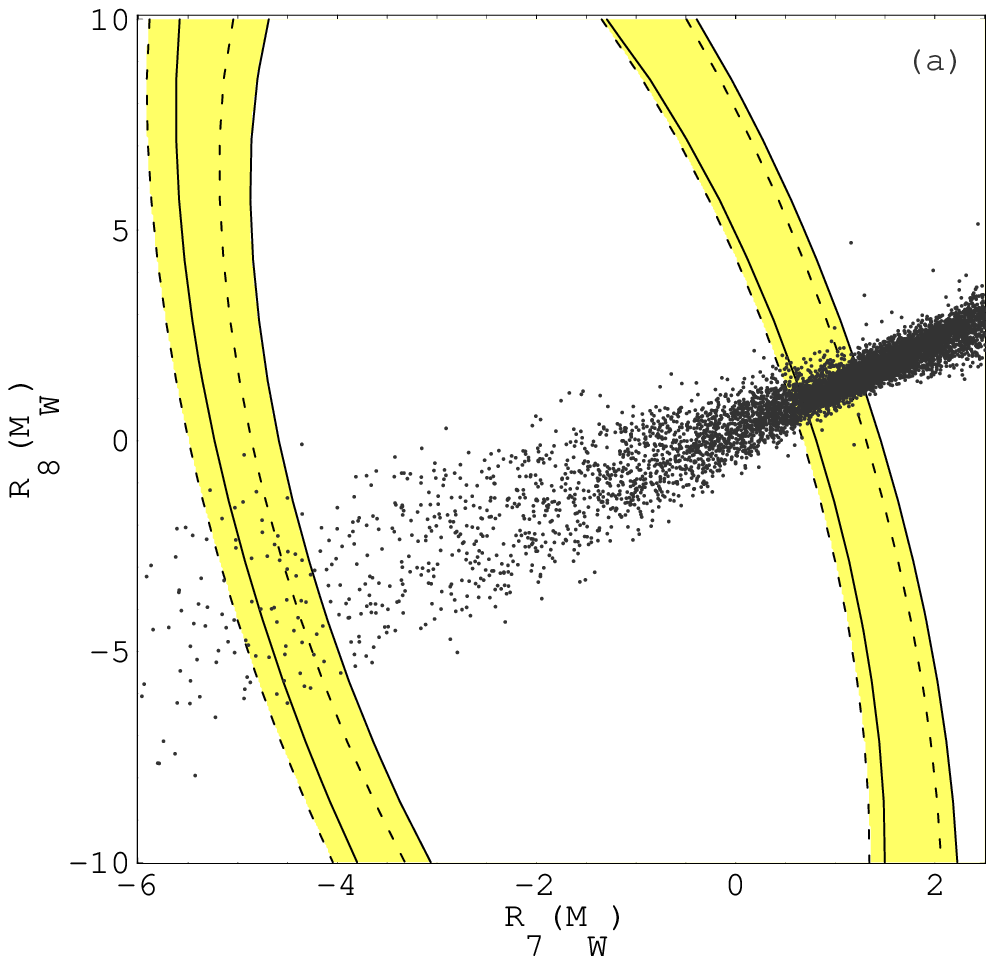,width=7cm}
\hspace*{.2cm}
\epsfig{figure=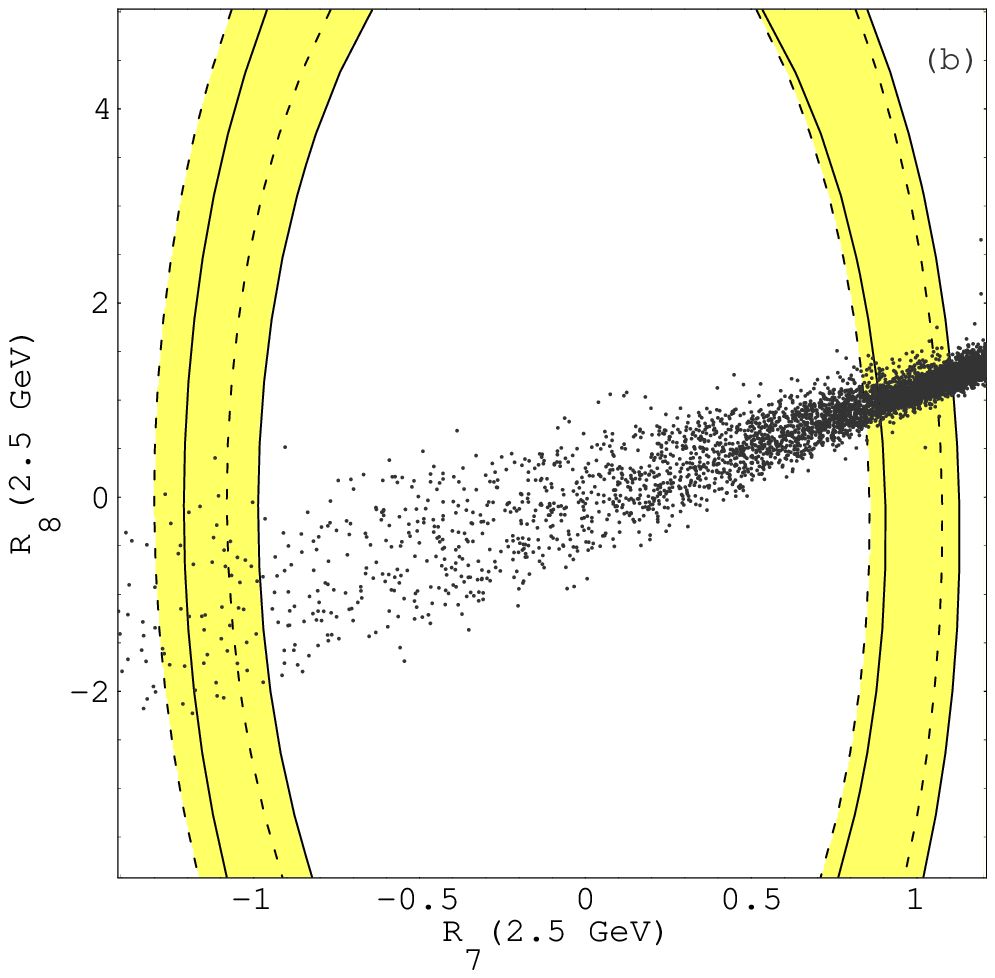,width=7cm}
\end{center}
\caption[]{
 $90 \%$ C.L. bounds in the $[R_7 (\mu), R_8(\mu)]$ plane
following from the measurement of the $B\to X_s \gamma$ branching ratio for
$\mu=m_W$ (left) and $\mu=2.5$ GeV (right), where 
$R_{7,8}= C_{7,8}^{total}/C_{7,8}^{SM}$. 
Theoretical uncertainties are taken into account. The solid and dashed
lines correspond to the $m_c = m_{c}^{pole}$ and $m_c =
m_{c}^{\overline{MS}} (\mu_b)$ cases respectively. The scatter points
correspond to the expectation in MFV models;
from \cite{LunghiGreub}.}
\label{C78bounds}
\end{figure}

The inclusive $B \rightarrow  X_s \ell^+ \ell^-$ decay is  another 
important tool to understand the nature of physics beyond the 
SM. 
In comparison to the decay $B \rightarrow X_s \gamma$,  it 
offers complementary information. For example one is able to resolve 
the sign ambiguity of the Wilson coefficient $C_7$, which is not fixed 
by the $B \rightarrow X_s \gamma$ constraint. The FB asymmetry, however, 
has terms proportional to $\Re (C_{10} C^{eff}_9 )$  and $\Re ( C_{10} C^{eff}_7 )$.  
As was first  advocated in \cite{AliMannel}, 
the invariant dilepton mass spectrum, 
the forward--backward charge asymmetry 
and the decay rate of $B \rightarrow X_s \gamma$ 
determine the magnitude and also the sign 
of the three Wilson coefficients $C_7,C_9,$ and $C_{10}$, and allow for a 
model-independent analysis of rare $B$ decays.

\medskip

There are several mSUGRA models and also several
model-independent analyses in the literature
\cite{francesca1,ChoMisiak,GotoOkada,HewettWells,Huangnew,KimKo,LunghiMasiero}.
It was always assumed that the operator basis is not 
enlarged in comparison to the SM. All the analyses found  
strong correlations between the decays $B \rightarrow
X_s \gamma$ and $B \rightarrow X_s \ell^+ \ell^-$ 

Within the mSUGRA model sizeable  deviations from the SM values
of the $B \rightarrow X_s \ell^+ \ell^-$ decay
are excluded through the severe  constraints on $C_7$ by the 
$B \rightarrow X_s \gamma$ measurement. But
 it  was also shown that in less restricted scenarios 
supersymmetric contributions  could potentially enhance the 
$B \rightarrow X_s \ell^+ \ell^-$ 
kinematic distributions, the dilepton mass spectrum and the 
FB asymmetry,  by more than $100 \%$ relative to the SM predictions.
One of the reasons of the enhancements is that the Wilson coefficient 
$C_7$ can change the sign with respect to the SM in some region of the 
parameter space. Within the mSUGRA model,  the experimental bounds have 
disfavoured the non-SM sign already (see fig. \ref{C78bounds}).

When the experimental uncertainties are  reduced soon, this fact 
will allow to discriminate between MFV models and
non-minimal models and will lead either to evidence of new physics 
or to very stringent constraints on the parameter space 
of such  models.

\begin{figure}
\begin{center}
\epsfig{figure=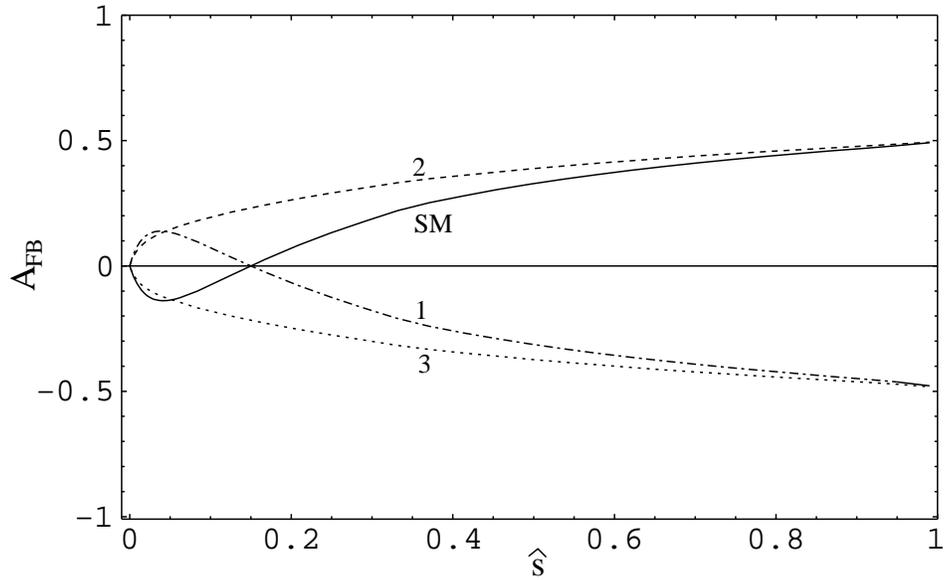,width=12.5cm}
\end{center}
\caption[]{Four different shapes of the normalized FB  asymmetry $\overline{A}_{FB}$
for the decay $B\to X_s \ell^+ \ell^-$. The four curves correspond to four
sample points of the Wilson coefficients that are compatible with
the present measurements of the integrated branching ratios;
from \cite{LunghiGreub}.}
\label{FBbeyond}
\end{figure}

\medskip 
 
Recently these analyses have been  updated in \cite{LunghiGreub}
based on the new experimental data of the semi-leptonic decays and
on partial results in the NNLL theoretical predictions.  
Within the analysis, 
the charm mass renormalization scheme 
ambiguity (see section \ref{phenobsg}) in the decay $B \rightarrow
X_s \gamma$ is  taken into account.  But also a too conservative  
error estimate regarding the charm mass dependence within the
decay $B \rightarrow X_s \ell^+ \ell^-$ is  assumed, which leads to a 
rather large error of $15 \%$ in the inclusive mode. 

It was found \cite{LunghiGreub} that with the 
present experimental knowledge the decay $B \rightarrow 
X_s \gamma$ still leads to  the most restrictive constraints.
Especially, the MFV scenarios are already highly constrained and only
small deviations to the SM rates and distributions are possible;
therefore no useful additional bounds from the semi-leptonic modes
beyond what are already known from the $B  \rightarrow  X_s \gamma$
can be deduced for the MFV models at the moment. 
But in  non-mimial models, 
additional constraints from the semi-leptonic mode already
emerge in some parts of the supersymmetric parameter space, namely 
for the off-diagonal elements within the squark mass matrix in the 
up-quark sector.  
 
Within the model-independent analysis, the impact of the partial 
NNLL contributions on the allowed ranges for the Wilson coefficients 
was already found to be significant. 
In this analysis, however, only the integrated branching ratios were used 
to derive constraints. It is clear that one needs measurements of the
kinematic distributions of the $B \rightarrow X_s \ell^+ \ell^-$,
the dilepton mass spectrum and the FB asymmetry in order to 
                                   determine the exact values and signs 
of the Wilson coefficients. 
In fig. \ref{FBbeyond}, the impact of these future measurements 
is illustrated. It shows the shape of the FB asymmetry for the SM and three
additional sample points, which are all still allowed by the present
measurements of the branching ratios; thus, 
even rather rough measurements of the FB asymmetry will either rule out 
large parts of the parameter space of extended  models or show
clear evidence for new physics beyond the SM.

\newpage

\setcounter{equation}{0}
\section{Direct CP violation in $b \rightarrow s$ transitions}
\label{CPsection}

The $B$ system provides us with an independent test of the CKM prescription 
of CP violation. Until recently, the neutral kaon system was the only 
environment where CP violation had been observed. Those effects 
in the kaon system are often plagued by large theoretical uncertainties
due to long-range QCD. So it was difficult to decide if the CKM description 
really accounts quantitatively for CP violation. 
In contrast, non-perturbative contributions are under control 
in the $B$ system thanks  to the heavy mass expansion. 
Moreover, there are gold-plated CP asymmetries
like  the one in the decay mode  $B \rightarrow \psi K_S$, which 
are theoretically very  clean, because the direct decay amplitude
is dominated by one single weak phase
and, thus, most of the hadronic uncertainties drop out
in the CP asymmetry. 

\medskip

The CKM prescription of CP violation with one single phase --- 
proposed in 1972 when the second family was not confirmed 
experimentally \cite{CKM72} --- is very predictive and has now
passed its first precision test in the golden $B$ mode, 
$B_d \rightarrow \psi K_S$, at the $10 \%$ level \cite{BELLECP,BABARCP}.  
Nevertheless, there is still room for non-standard CP phases, especially
in the FCNC $\Delta F = 1$ modes.
Actually, detailed measurements of CP asymmetries in rare $B$ decays 
will be possible in the near future. 

\medskip 

The  direct {\it normalized} CP asymmetries 
of the inclusive decay modes is given by~\footnote{There is a sign convention that is generally adopted 
in theory and experiment: on the partonic  level 
$\alpha_{CP}(b \rightarrow s \gamma) = 
(\Gamma(b \rightarrow s \gamma) - 
\Gamma(\bar b \rightarrow \bar s \gamma))/(\Gamma(b \rightarrow s \gamma) + \Gamma(\bar b \rightarrow \bar s \gamma))$; analogously 
$\alpha_{CP} \sim  (\Gamma(\bar B^0 \rightarrow \cdots) - \Gamma(B^0 \rightarrow \cdots))$ and  $\alpha_{CP} \sim  (\Gamma(B^- \rightarrow \cdots) - \Gamma(B^+ \rightarrow \cdots))$.}
\begin{equation}
\label{CPdirectdefinition} 
\nonumber 
\alpha_{CP}({B \rightarrow  X_{s/d} \, \gamma}) =
\frac{\Gamma(\bar B \rightarrow X_{s/d}\gamma)
     -\Gamma(B \rightarrow  X_{\bar s/\bar d}\gamma)}
     {\Gamma(\bar B \rightarrow  X_{s/d} \gamma)
     +\Gamma(B \rightarrow  X_{\bar s/\bar d}\gamma)}
\end{equation}
CLEO has already presented a measurement of the CP asymmetry in
the inclusive decay $B \to X_s \gamma$, actually a measurement of 
a weighted 
sum, $\alpha_{CP} = 0.965 \alpha_{CP}(B \rightarrow X_s \gamma)
+ 0.02 \alpha_{CP}(B \rightarrow X_d \gamma)$,   
yielding \cite{CleoCP}
\begin{equation}
\alpha_{CP} = (-0.079 \pm 0.108 \pm 0.022) \times (1.0 \pm 0.030) \,.
\end{equation}
The first error is statistical, the second and third errors  
additive and multiplicative systematic respectively. 
This measurement is based 
on $10^{7}$ $B \bar B$ events and implies that, 
at $90 \%$ confidence level, $\alpha_{CP}$ 
lies between $-0.27 < \alpha_{CP} < +0.10$; 
very large effects are thus already excluded.  
The same conclusion can be deduced  from the measurements of the CP asymmetry 
in the exclusive mode $B \rightarrow K^*(892) \gamma$ of CLEO
\cite{exclCPCLEO}, $\alpha_{CP} = +0.08 \pm 0.13_{stat} \pm 0.03_{syst}$,
of BABAR \cite{exclCPBABAR}, $\alpha_{CP} = -0.044 \pm 0.076 \pm 0.082$,
and of BELLE \cite{exclCPBELLE}, $\alpha_{CP} = -0.022 \pm 0.048 \pm 0.017$.
The preliminary measurement of BELLE is the best by far, based on $65.4 \times 10^6$ $B$ meson pairs and implies that, at $90 \%$ confidence level, 
$\alpha_{CP}$ in the exclusive $B \rightarrow K^* \gamma$ lies between 
$-0.106 < \alpha_{CP} < +0.062$.

\medskip

Theoretical NLL QCD predictions of 
the {\it normalized} CP asymmetries of
the  inclusive channels (see~\cite{AG7,KaganNeubert}) within  the SM 
can be expressed by the 
approximate formulae \cite{SoniWu}

\begin{equation}
\begin{array}{lll}                   
  \alpha_{CP}({B \rightarrow  X_s \gamma}) &\approx&  0.334 \times
\Im [\epsilon_s] \approx + 0.6 \% \, , \\
  \alpha_{CP}({B \rightarrow  X_d \gamma}) &\approx&  0.334 \times 
\Im [\epsilon_d] \approx - 16 \%  \label{SMnumbers}
\end{array}
\end{equation}
where 
\begin{equation}
\epsilon_s = \frac{V_{us}^*V_{ub}}{V_{ts}^*V_{tb}} \simeq 
-\lambda^2(\rho-i\eta), \quad \quad   
\epsilon_d = \frac{V_{ud}^*V_{ub}}{V_{td}^*V_{tb}} \simeq
        \frac{\rho -i\eta}{1-\rho+i\eta} .
\end{equation}
Numerically, the best-fit values of the CKM parameters are used
 \cite{SoniWu}. 
The two CP asymmetries are connected by the relative factor 
$\lambda^2 \, ((1-\rho)^2 + \eta^2)$. Moreover, 
the small SM prediction for the CP asymmetry in the decay 
$B \rightarrow X_s \gamma$ is a result of three suppression factors.
There is  an  $\alpha_s$ factor needed in order to have a strong phase;
moreover, there is a CKM suppression of order $\lambda^2$
and there is a GIM suppression of order $(m_c/m_b)^2$ reflecting the fact
that in the limit $m_c = m_u$ any CP asymmetry in the SM would vanish.

\medskip

An analysis for the leptonic counterparts  is presented in
\cite{Hiller}.
The normalized CP asymmetries may also be calculated for
exclusive decays: a
model calculation may be found in \cite{GSW}.
Theoretical predictions based on the QCD facorization approach 
\cite{Ali,Beneke,buchalla} are also affected by large 
uncertainties.  Only in the case of relatively large new physics effects 
will one be able to disentangle these effects 
 from the QCD uncertainties. 
But the available experimental data 
do not support this scenario in the 
$B \rightarrow K^* \gamma$  mode \cite{exclCPCLEO,exclCPBABAR,exclCPBELLE}.

\medskip

\begin{figure}
\begin{center}
\epsfig{figure=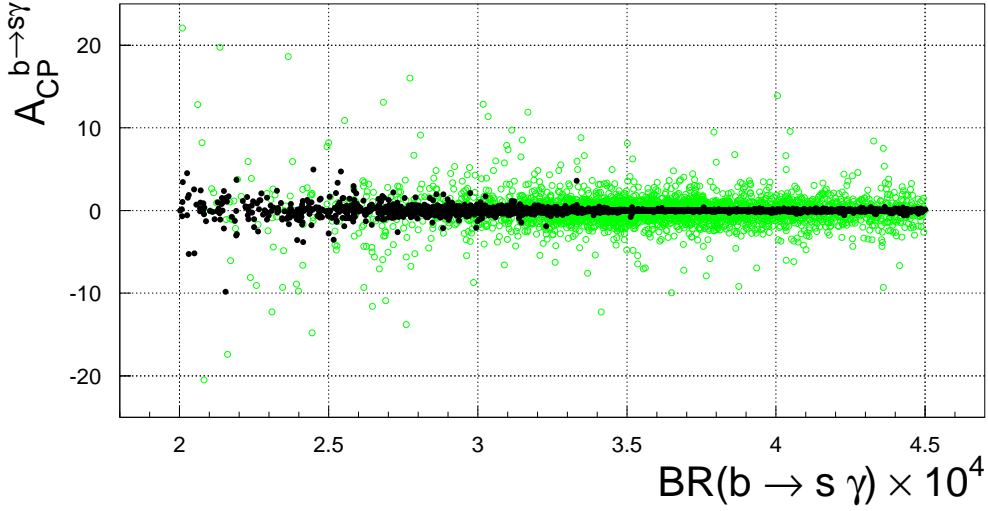,width=14cm}
\end{center}
\caption[]{$CP$ asymmetry vs `total' width of the decay $B \rightarrow X_s \gamma$.
The empty circles are computed without any restriction on the phases.
The filled black ones show the impact of 
the EDM's constraints; from \cite{Porod}.}
\label{CPflavourblind}
\end{figure}

Supersymmetric predictions for the CP asymmetries in 
$B \rightarrow X_{s/d} \gamma$ depend strongly on what is
assumed for the supersymmetry-breaking sector and are, thus,  
a rather model-dependent issue. 
The minimal supergravity model cannot account for large 
CP asymmetries beyond $2\%$ because of the constraints coming 
from the electron and neutron electric dipole moments
\cite{Goto}. This is generally true in models based on the MFV 
assumption (see also fig. \ref{CPflavourblind}). Non-minimal  
 models with squark mixing or models with R-parity violation
 allow for larger asymmetries, 
of the order of $10 \%$ or even larger \cite{Aoki,KaganNeubert}.
In \cite{KaganNeubert} it is argued that the asymmetry in case of non-minimal 
models could be even larger 
than $\pm 15 \%$  if the gluino mass is significantly lighter than the 
squark masses.
Recent studies of the $B \to X_d \gamma$ rate asymmetry  
in specific models led to asymmetries between $-40 \% $ and $+40\%$
\cite{Recksiegel} and  $-45 \%$ and $+ 21 \%$ \cite{Asatrian}.
In general, CP asymmetries may lead to clean 
evidence for  new physics by a significant deviation from the SM
prediction.  

From (\ref{SMnumbers}),  it is obvious  that a large 
CP asymmetry  in the 
$B \rightarrow X_s \gamma$ channel or a positive CP asymmetry in the 
inclusive $B \rightarrow X_d \gamma$ channel would be a clear signal
for new physics.

\medskip

The exclusive and inclusive decays of the form $b \to s \gamma$ and
$b \to d \gamma$, as well as their leptonic counterparts, provide a
stringent test, if the CKM matrix is indeed the only source of
CP violation. 
Using  U-spin, which is the $SU(2)$ subgroup of flavour $SU(3)$ relating
the $s$ and the $d$ quark and which is already a well-known tool 
in the context of non-leptonic decays \cite{Fleischer:1999pa,GronauBCP4},
one derives  relations between the CP asymmetries of the exclusive 
channels $B^- \to K^{*-} \gamma$ and $B^- \to \rho^- \gamma$ and 
of the inclusive channels $B \rightarrow X_s \gamma$ and 
$B \rightarrow X_d \gamma$.

Any CP violation in the SM has to be proportional to 
\begin{equation} \label{Jdet}
C = i \, J \,\,\, (m_u - m_c) (m_u - m_t) (m_c - m_t)
(m_d - m_s) (m_d - m_b) (m_s - m_b) \, , 
\end{equation}
where $J = {Im}[V_{ub} V_{cb}^* V_{cs} V_{us}^*]$ is the Jarlskog parameter.
Therefore, one should make use  of the U-spin symmetry only
with respect to the strong interactions. 
Defining the rate asymmetries ({not} the {\it normalized} CP asymmetries)
by  
\begin{equation} \label{ratediff}
\Delta \Gamma (B^- \to V^- \gamma) = \Gamma (B^- \to V^- \gamma) - \Gamma (B^+ \to V^+ \gamma)
\end{equation}
one arrives at the following relation \cite{mannelhurth}:
\begin{equation} 
\label{smsm}
 \Delta \Gamma (B^- \to K^{*-} \gamma) +
\Delta \Gamma (B^- \to \rho^- \gamma) = b_{exc} \Delta_{exc}, 
\end{equation}
where the right-hand side is written as a product of a relative 
U-spin breaking $b_{exc}$ and a typical size  $\Delta_{exc}$ 
of the CP violating rate difference. This is a direct consequence
of the unitarity of the CKM matrix and, thus, of the fact that 
the Jarlskog parameter is the only fourth-order quantity
that is invariant under rephasing of the quark fields within the SM.
The resulting relation  between $b \rightarrow s$ and $b \rightarrow d$ 
rate asymmetries due to 
\begin{equation}
J = {Im} (\lambda_u^{(s)} \lambda_c^{(s)*}) = - {Im} (\lambda_u^{(d)} \lambda_c^{(d)*})
\end{equation}
was first noticed in \cite{Soares}. 

\medskip 

In \cite{mannelhurth} 
the SM prediction for the
difference of branching ratios, based on model results in \cite{GSW}
and on a sum rule calculation of the form factors  \cite{Ali:1994vd},
was derived:
\begin{equation} \label{resexc1}
|\Delta {\cal B}(B^- \to K^{*-} \gamma) +
\Delta {\cal B}(B^- \to \rho^- \gamma)| \sim 4 \times 10^{-8}.
\end{equation}
Note that the right-hand side is model-dependent. 
The U-spin-breaking effects were also estimated 
in the QCD factorization approach \cite{buchalla}.
Within this approach, it was shown that the U-spin-breaking effect
essentially scales with the differences of the two form factors
$(F_{K^*}-F_\rho)$.
Using the form factors from the QCD sum rule calculation in 
\cite{ballbraun} and maximizing the CP asymmetries by a specific
choice of the CKM angle $\gamma$, the authors of \cite{buchalla} obtain
\begin{equation} 
 \Delta {\cal B}(B^- \to K^{*-} \gamma) +
\Delta {\cal B}(B^- \to \rho^- \gamma)  \sim - 3 \times 10^{-7},
\end{equation}
while for the separate asymmetries they obtain, $\Delta {\cal B}(B\to K^*\gamma) = - 7 \times 10^{-7}$ and 
$\Delta {\cal B}(B\to \rho\gamma) =  4 \times  10^{-7}$
, which explicitly shows the limitations of the relation
(\ref{smsm}) as a test of the SM. 

\medskip

The issue is much more attractive in the inclusive modes.
Because of  the heavy mass expansion for the inclusive process,
the leading contribution is
the free $b$-quark decay. In particular, there is no sensitivity to
the spectator quark and thus one arrives, within the partonic
contribution, at the following relation
for the CP rate asymmetries as the consequence of the CKM unitarity
\cite{Soares}:

\begin{equation} \label{resincg1}
\Delta \Gamma (B \to X_s \gamma) +
\Delta \Gamma (B \to X_d \gamma) = b_{inc} \Delta_{inc}. 
\end{equation}

In this framework one relies on
parton--hadron duality.  
So one can actually compute the breaking 
of U-spin by keeping a non-vanishing strange quark mass. 
The typical size of $b_{inc}$ can be roughly 
estimated to be of the order of 
$|b_{inc}| \sim m_s^2/m_b^2 \sim 5 \times 10^{-4}$;  
$|\Delta_{inc}|$  is again the average of the moduli of the two CP rate
asymmetries. These have been calculated (for vanishing strange quark mass),
e.g. in  \cite{AG7}, and one arrives at the following estimate 
within the partonic contribution \cite{mannelhurth}:
\begin{equation} \label{resinc3}
| \Delta {\cal B}(B \to X_s \gamma) +
\Delta {\cal B}(B \to X_d \gamma) | \sim 1 \times  10^{-9}. 
\end{equation}

\medskip

Going beyond the leading partonic contribution one has to check 
if the large suppression factor from the U-spin breaking is still
effective in addition to  the natural suppression factors 
already present  in the corresponding branching ratios.    
This question was addressed in \cite{mannelhurth2}. 
In the leading $1/m_b^2$ corrections, 
the U-spin-breaking effects  
also induce an additional overall factor $m_s^2/m_b^2$.
In the non-perturbative corrections from the charm quark
loop, which scale with  $1/m_c^2$, one finds again the same 
overall suppression factor, because
the operator $\tilde{\cal O}$ (see \ref{extraop})
does not  contain any information on the strange mass. 
The  corresponding long-distance contributions 
from up-quark loops, which scale with $\Lambda_{QCD}/m_b$
(see section \ref{sectionnonpert}), 
follow the same pattern \cite{mannelhurth2}. 
Thus, in the inclusive mode, the
right-hand side in (\ref{resinc3})  can be computed in a
model-independent way with the help  of the heavy mass expansion. 

\medskip 

Therefore, the  prediction (\ref{resinc3}) 
provides  a very clean  SM test, whether  generic 
new CP phases are active or not.
Any significant deviation from the estimate (\ref{resinc3}) 
would be a  direct  hint to non-CKM contributions to CP violation.

From the theoretical point of view the {\it sum} of the 
CP asymmetries in the inclusive $b \rightarrow s $ and $b \rightarrow d$
transitions turns out to be the favourable observable. 
This might be true also from the experimental point of view.

\newpage

\section{Further opportunities}

Generally, {\it exclusive} decay modes have large uncertainties due to
the hadronic form factors and it might thus be rather difficult 
to disentangle possible new physics contributions  from hadronic 
uncertainties in these modes --- 
at least in the absence of very large new effects. 
Therefore exclusive modes often can be used as QCD tests only.
However, there are exceptions to this rule. In specific 
ratios like CP asymmetries, hadronic uncertainties are reduced and  
 large  new physics effects might be  detectable.  
There are also exclusive modes that are as clean as inclusive 
modes because the corresponding  hadronic matrix elements can be 
determined  from experiment. The most important examples among them  
are the exclusive $B$ decay $B_s \rightarrow 
\mu^+ \mu^-$ and the exclusive rare kaon decays
$K_L \rightarrow \pi^0 \nu \bar \nu$ and 
$K^+ \rightarrow \pi^+ \nu \bar \nu$.
The  hadronic matrix elements of these FCNC (rare) processes 
can be related to well-known non-rare semi-leptonic decays.

As the inclusive decay $B \rightarrow X_s \nu \bar \nu$
(see section \ref{NNLLQCDbsll}), 
the exclusive decay $B_s \rightarrow \mu^+ \mu^-$ is
completely dominated by the top-quark contribution due
to the hard GIM mechanism.
QCD corrections within this  exclusive mode are already
calculated to NLL order. The remaining 
perturbative uncertainty is not larger than $\pm 1 \%$
 \cite{Buras93}.
The corresponding hadronic matrix element leads to the  
decay constant of the $B_s$ meson, 
$f_{B_s}$, which can be determined on the lattice. 
The related uncertainty represents the largest part of the   
theoretical error: $f_{B_s} = (238 \pm 31) MeV$ \cite{Latt}.  
The SM prediction for the branching ratio of the decay
$B_s \rightarrow \mu^+ \mu^-$
is of order $10^{-9}$.
Thus, this decay  will be accessible at LHC 
and also at $B$TeV.  
However, the branching ratio  can be much
larger within specific extensions of the SM. 
For example, the helicity-suppression of the SM contribution
leads to an enhanced sensitivity to the Higgs-mediated 
scalar FCNCs within the 2HDM and, especially  within the MSSM.
These non-standard contributions lead to a drastic 
enhancement in the large $\tan \beta$-limit
\cite{scalarFCNC}. 
Therefore, this decay might be even detectable at FERMILAB
before the LHC experiments  and the $B$TeV experiment
start to take data (see \cite{tanbetabmumu} for further discussions).

The other two important examples of theoretically clean {\it exclusive} 
modes, 
$K_L \rightarrow \pi^0\nu\bar{\nu}$ and $K^+ \rightarrow \pi^+ \nu 
\bar{\nu}$, are discussed in more detail in the following
section.

\setcounter{equation}{0}
\subsection{$K_L \rightarrow \pi^0\nu\bar{\nu}$ and $K^+ \rightarrow \pi^+ \nu 
\bar{\nu}$} 
\label{kaonsection}

\begin{figure}
\begin{center}
\epsfig{figure=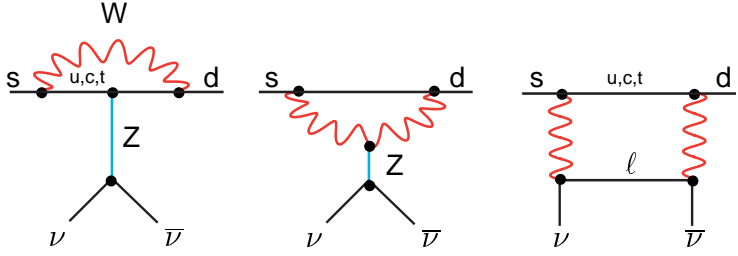,width=10cm}
\end{center}
\caption{One-loop diagrams contributing to $K \to\pi \nu\bar\nu$.}
\label{SMkaon}
\end{figure}

The rare decays $K_L\to\pi^0\nu\bar\nu$ and $K^+\to\pi^+\nu\bar\nu$ 
represent complementary opportunities for precision flavour physics.
They are also FCNC  processes induced at the one-loop
level via $Z^0$ penguin and box diagrams 
(see fig. \ref{SMkaon}) and are exceptionally clean processes.

As in the inclusive decay $B \rightarrow X_s \nu \bar \nu$
(see section \ref{NNLLQCDbsll}), the hard GIM mechanism is active:
 the short-distance top- and charm-contributions dominate the long-distance up-quark contribution within  
the charged mode $K^+\to\pi^+\nu\bar\nu$. The CKM factors of the charm 
contribution compensates the hard GIM suppression relative to the top
contribution in this specific case.  
  The short-distance amplitude is then governed 
by one single semi-leptonic operator, namely 
$(\bar s \gamma_\mu P_L d) (\bar \nu \gamma_\mu P_L \nu)$. 
Its hadronic 
matrix element can be determined experimentally
by  the semi-leptonic kaon decay.
In fact, the matrix element 
\mbox{$ \langle \pi^+  \mid \bar s \gamma_\mu P_L d \mid  K^+ \rangle$}
can be  related by isospin symmetry to the matrix element 
\mbox{$ \sqrt{2}\,   \langle \pi^0  \mid \bar s \gamma_\mu P_L u \mid  K^+ \rangle$} of the semi-leptonic decay \mbox{$K^+ \rightarrow \pi^0 e^+ \nu$}. 
The corresponding Wilson coefficient 
is already calculated to NLL QCD \cite{Buras93}
and the scale dependence is reduced  to $5 \%$ in the charged kaon mode. 

The situation is even more favourable in  the neutral mode,
which is dominated by the CP-violating part. There is no relative CKM 
enhancement of the charm contribution  and, thus, the amplitude
is completely dominated by the top contribution as  in the inclusive 
rare $B \rightarrow X_s \bar \nu \nu$  decay.
The NLL QCD calculation therefore leads to a $1 \%$ scale uncertainty
only \cite{Buras93}. 

\medskip

The validity of the OPE and the renormalization-group-improved 
perturbation theory in the charm contribution to the charged mode
has been  critically analysed:  
the separate scale dependence within the charm contribution of
$13 \%$ at NLL QCD is consistent with partial NNLL results \cite{Gerhard}. 
Moreover, subleading power corrections within the OPE of order 
$m_K^2/m_c^2$ --- which might lead to  $15 \%$ correction --- 
are estimated  to be at the level of  $5 \%$. However, for a reliable
determination of the latter  corrections, a lattice calculation of the
corresponding hadronic matrix elements will be 
indispensable \cite{Lewandowski}. 

\medskip 

The latest numerical SM predictions are  \cite{ginonew,Buras999}

\begin{equation}
\begin{array}{lll}
{\cal B}(K^+\to\pi^+\nu\bar\nu) &=& (7.2 \pm 2.1) \times 10^{-11}\\
{\cal B}(K_L\to\pi^0\nu\bar\nu) &=& (2.8\pm 1.1) \times 10^{-11}
\end{array}
\label{SMkaonkaon}
\end{equation}
The uncertainties of the present SM predictions are dominated by the  
current errors of the CKM parameters, while  the instrinsic error in 
the charged mode 
is about $6 \%$  (mainly from the charm contribution) and in the neutral 
mode about $2 \%$ only. 
This implies the important role of these decay modes for 
CKM phenomenology:
they play a unique role among $K$ decays, as does 
the $B_d\to\psi K_S$ mode among 
$B$ decays.
The measurements of the two kaon decay modes 
 allow for  a  measurement of the  angle $\beta$ of the unitarity triangle
                                        to a precision 
comparable to that obtained with 
the $B_d\to\psi K_S$ mode before the LHC era \cite{Buchalla:1994}.
The only necessary theoretical input is the internal charm contribution 
to $K^+\to\pi^+\nu\bar\nu$, which introduces some theoretical uncertainty
 (see above).

\medskip

The relation $(\sin 2\beta)_{\pi\nu\bar{\nu}} = 
(\sin 2\beta)_{\psi K_S}$ implies a very 
interesting connection between rare $K$ decays and $B$ physics,
which must be satisfied in the SM:
\begin{equation}
(\sin 2\beta)_{\pi\nu\bar{\nu}}  = (\sin 2\beta)_{\psi K_s} = 
-A_{CP}(\psi K_S)\, {1+x^2_d\o x_d}.
\label{golden}
\end{equation}  
$A_{CP}(\psi K_S)$ denotes 
the time-integrated CP-violating asymmetry in $B^0_d\to\psi K_S$ 
and $x_d=\Delta m/\Gamma$ gives the size of $B^0_d$--$\bar B^0_d$
mixing.  As was stressed 
in~\cite{Buchalla:1994}, besides the internal charm contribution 
to the charged kaon mode,  all quantities in (\ref{golden})
can be directly measured experimentally, and their relation is
almost independent of $V_{cb}$.

\medskip

\begin{figure}
\begin{center}
\epsfig{figure=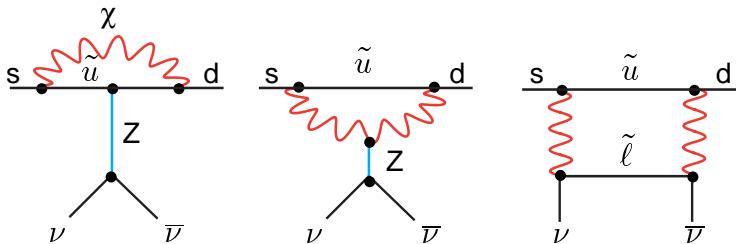,width=10cm}
\end{center}
\caption{Supersymmetric contributions to $K \to\pi \nu\bar\nu$.}
\label{SUSYkaon}
\end{figure}

Besides their  rich CKM phenomenology,  the decays  
$K_L \rightarrow \pi^0\nu\bar{\nu}$ and 
$K^+ \rightarrow \pi^+ \nu \bar{\nu}$ as loop-induced processes
 are  very sensitive to new physics beyond the SM.
In addition, the theoretical information is very clean
and the measurement of these decays thus leads  to 
very accurate constraints on any 
new physics model. 
Moreover, there is the possibility that these clean rare decay modes
themselves lead to first evidence of new physics when the measured 
decay rates are not compatible with the SM.

\medskip

New physics contributions in  
$K_L \rightarrow \pi^0\nu\bar{\nu}$ and 
$K^+ \rightarrow \pi^+ \nu \bar{\nu}$ can be parametrized 
in a model-independent way by two parameters
that  quantify the violation of the  
relation~(\ref{golden})~\cite{Nir98,Buras:1998}.
New effects in supersymmetric models can be induced
through new box- and penguin-diagram contributions which involve
new particles such as charged Higgs or  charginos and stops 
(fig.~\ref{SUSYkaon}) 
that replace the $W$ boson and the up-type quark 
of the SM (fig.~\ref{SMkaon}).

\medskip

Under the simplifying MFV assumption~\cite{ginonew} 
(see section \ref{generalities}), the relation~(\ref{golden}) 
is valid. Thus, the 
measurements of ${\cal B}(K_L \rightarrow \pi^0\nu\bar{\nu}$) and 
${\cal B}(K^+ \rightarrow \pi^+ \nu \bar{\nu}$) still directly 
determine the angle $\beta$, and 
a significant violation  of the relation (\ref{golden})  
would rule out this assumption.

\medskip

For the present experimental status of supersymmetry,
however,  a model-independent analysis that includes also a general
flavour change through the squark mass matrices is more suitable. 
If the  new sources of flavour change are parametrized by 
the mass-insertion approximation, an expansion 
of the squark mass matrices around their diagonal, 
it turns out that SUSY contributions in this more general setting
of the unconstrained MSSM 
allow  for a  significant violation of the relation (\ref{golden}). 
An enhancement of the branching ratios by an order of magnitude  
(in the case of $K^+ \rightarrow \pi^+ \nu \bar{\nu}$ by a factor
3)  with respect to the SM values is possible,
mostly thanks  to the chargino-induced Z-penguin 
contribution~\cite{Colangelo:1998}.
Recent analyses~\cite{Colangelo:1998,Burasb,all}
within the uMSSM
focused on the  correlation of rare decays and $\epsilon' / \epsilon$,
and led to  reasonable upper bounds for
the branching ratios: 
${\cal B}(K_L \rightarrow \pi^0\nu\bar{\nu}) \leq  1.2 \times 10^{-10}$
and ${\cal B}(K^+ \rightarrow \pi^+ \nu \bar{\nu}) \leq 1.7 \times 10^{-10}$, 
which should be compared with the latest numerical SM predictions
 (\ref{SMkaonkaon}).

\medskip

\begin{figure}
\begin{center}
\epsfig{figure=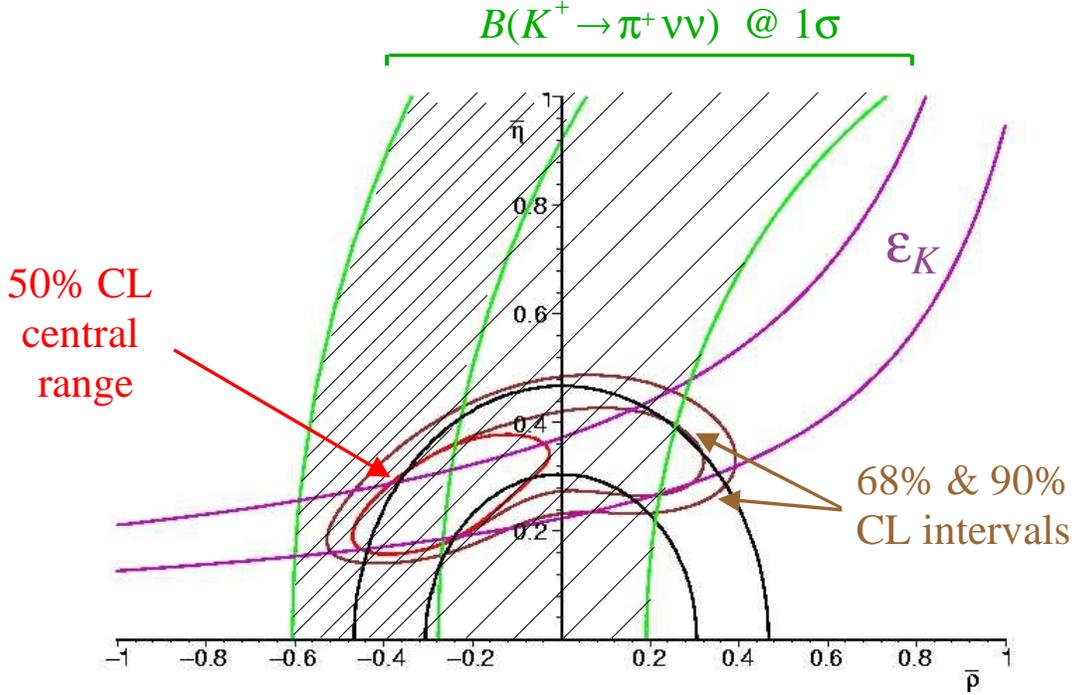,width=13cm}
\end{center}
\vspace{-6.4cm}
\caption[]{Allowed region in the  $\bar\rho$--$\bar \eta$ plane with the 
inclusion of  the latest $K^+\to\pi^+ \nu \bar\nu$  
and without $B_{s,d}$ data. The two external contours 
denote $68\%$ and $90\%$ confidence intervals; the inner 
one is the $68\%$ confidence interval {\it under the assumption} that 
the experimental error in the present measurement  is reduced by a 
factor 2, from \cite{ginonew}.}
\label{ginofigure}
\end{figure}

The rare decays $K^+ \rightarrow \pi^+ \nu \bar{\nu}$ and 
 $K_L \rightarrow \pi^0 \nu \bar{\nu}$ are  
specifically interesting in view of the suggested  experiments 
at the Brookhaven 
National Laboratory (USA) (\cite{brookhavennew}, \cite{BNLneutral})
and at FERMILAB  (\cite{fermilabnew}), and at  KEK  (\cite{keknew})
 (see \cite{Barker} for a review).

\medskip

For the neutral $K_L\to\pi^0\nu\bar\nu$ mode, the experimental situation
is not satisfactory yet; there is only an upper
bound available from KTeV \cite{ktevold}:
\begin{equation}
{\cal B}(K_L\to\pi^0\nu\bar\nu) < 5.9 \times 10^{-7},
\end{equation}
which is four orders of magnitude above the SM expectation. 
An indirect upper bound on ${\cal B}(K_L\to\pi^0\nu\bar\nu)$, 
using the current limit on 
${\cal B}(K^+\to\pi^+\nu\bar\nu)$ and isospin symmetry, can be
placed at $2.6 \times 10^{-9}$ \cite{Grossman98}. 
Future prospects are given by the  E391a experiment at KEK 
with
a sensitivity of $3 \times 10^{-10}$ (possible start 2003) \cite{keknew} 
and the E926  experiment 
(KOPIO) at Brookhaven which aims at a sensitivity of $10^{-13}$
\cite{BNLneutral}.

\medskip 

For the charged $K^+\to\pi^+\nu\bar\nu$  mode, the experimental situation
is more favourable.
The current Brookhaven experiment E787 has, to date,  
observed two clean 
candidate events for $K^+\to\pi^+\nu\bar\nu$.
The combined analysis including previous data \cite{brookhavenold}
leads to the following branching ratio \cite{brookhavennew2}:
\begin{equation}
{\cal B}(K^+\to\pi^+\nu\bar\nu)=(1.57^{+1.75}_{-0.82})\times 10^{-10}.
\end{equation}
The central value is more than twice the central value of the theoretical 
SM prediction, but the present measurement is still compatible with it, 
in view  of the large error bars on the experimental side.
Fig. \ref{ginofigure} illustrates
the possible future impact of more precise measurements --- to be expected 
from the Brookhaven experiment E949 with a sensitivity 
of $10^{-11}$/event (started 2001) \cite{brookhavennew}
and from the future
high-precision CKM experiment at FERMILAB 
with yet an order of magnitude higher sensitivity (starting 2007)
\cite{fermilabnew}. If the present central value is 
confirmed with a smaller error, this will clearly indicate a new-physics
contribution either in $B \bar B$ mixing or in the 
$K^+\to\pi^+\nu\bar\nu$ mode \cite{ginonew}.

\newpage


\section{Summary}
\label{sec:summary}
\setcounter{equation}{0}

In this paper we have reviewed the status of inclusive rare $B$ decays,
highlighting recent developments. 
These  decays  give special insight into the CKM matrix; 
moreover,  as FCNC processes,  they are
loop-induced and therefore  particularly sensitive to new physics.

Decay modes such as  
$B \rightarrow X_s \gamma$, $B \rightarrow X_s \nu \bar \nu$
and $B \rightarrow X_s \ell^+\ell^-$  (with specific kinematic cuts)
 are dominated by the partonic (perturbative) contributions and are, 
thus,  theoretically very clean   
 in contrast to the corresponding exclusive decay 
modes and represent laboratories to search for new physics. 
Non-perturbative contributions play a subdominant role and they  
are under control thanks to the heavy mass expansion. 
The inclusive rare $B$  decays  are or will be accessible at the present 
\mbox{$e^+$$e^-$} machines  (CLEO, BABAR, BELLE), with their low background 
 and their kinematic constraints, and will make 
precision flavour physics possible
in the near future.  

Significant theoretical  progress has been made during the last years.
Calculations of NLL (or even NNLL)  QCD corrections to these decay modes 
have been performed. The theoretical uncertainty has been significantly 
reduced. 
As was emphasized, the step from LL to NLL precision within the 
framework  of the 
renormalization-group-improved perturbation theory is not only a
quantitative,
but also a qualitative one, which tests the validity of the perturbative
approach in a  given problem. 

Within the theoretical prediction of $B \rightarrow X_s \gamma$,
the charm mass renormalization scheme ambiguity at NLL order 
represents the largest  uncertainty. In view of the precise experimental
data coming up from the $B$ factories in the near future, 
this uncertainty should be removed.

Inclusive rare $B$ decays allow for an indirect search for new physics,
a strategy complementary to the direct production of new (supersymmetric) 
particles,  which is reserved for the planned 
hadronic machines such as the LHC at CERN. 
But the indirect search at the 
$B$ factories already implies  significant restrictions for 
the parameter space of supersymmetric models and, thus, leads to important 
theoretically clean information  for the direct search of 
supersymmetric particles.

It is even possible that these rare processes give  first 
evidence of new physics outside the neutrino sector by a significant 
deviation from the SM prediction. 
But also in the long run,
after new physics has already been discovered, inclusive rare $B$  
decays will
play an important role in analysing in greater detail the 
underlying new dynamics.

Within supersymmetric models, the QCD calculation of the inclusive 
rare $B$ decays has not reached
the sophistication of the corresponding SM calculations. Nevertheless,
NLL analyses in specific scenarios already show that bounds on the 
parameter space of non-standard models are rather sensitive to NLL
QCD contributions. 

Detailed measurements of CP asymmetries in rare $B$ decays will also 
be possible in the near future. They will allow for a stringent and 
clean test if the CKM matrix is indeed the only source of CP violation. 
Moreover, a measurement of the photon polarization within the rare $B$ 
decays will be possible in order to check
the SM prediction of a left-handed photon.

The rare kaon decays, 
$K^+ \rightarrow \pi^+ \nu \bar{\nu}$ and 
 $K_L \rightarrow \pi^0 \nu \bar{\nu}$,  
offer complementary opportunities for precision flavour   
physics. Besides the current Brookhaven experiment, several 
more are planned or suggested to explore these 
theoretically clean decay modes.


\section*{Acknowledgements}
I thank Colin Jessop for many helpful discussions on the experimental 
aspects of \mbox{inclusive} rare $B$ decays and Ed Thorndike for useful 
comments. I am very grateful to Mikolaj Misiak 
for his careful reading of the manuscript
and for useful comments. \mbox{Discussions} with Thomas Besmer, 
Francesca Borzumati, Giancarlo  D'Ambrosio, Paolo Gambino, Adrian \mbox{Ghinculov}, 
Christoph Greub, Thomas Mannel, Gino Isidori, Daniel Wyler and 
York-Peng Yao are also gratefully acknowledged.



\end{document}